\documentclass[12pt]{JHEP3}
\usepackage{mathrsfs}
\usepackage{amsmath,amssymb}
\usepackage{epsfig}
\usepackage{relsize}
\usepackage{graphicx}
\input epsf

\def\x'{\mathaccent 19 x}
\def\y'{\mathaccent 19 y}
\def\n'{\mathaccent 19 n}
\def\u'{\mathaccent 19 u}

\def\et'{\mathaccent 19 \eta}
\def\th'{\mathaccent 19 \theta}
\def\lam'{\mathaccent 19 \lambda}
\def\varet'{\mathaccent 19 \vartheta}
\def\rh'{\mathaccent 19 \rho}
\def\ph'{\mathaccent 19 \phi}
\def\xb'{\mathaccent 19 {\bar{x}}}

\def\sp{{^{_{\mathscr P}}}}
\def\spg{{^{_{\cal P}}}}
\def\st{{^{_T}}}
\def\cex{{_{\cal C}}}

\def\bQ{\overline{Q}}
\def\bC{C^\dagger}

\def\hP{\hat{P}}

\def\l{{\lambda}}

\def\sfa{{\sf a}}
\def\sfb{{\sf b}}
\def\sfc{{\sf c}}
\def\sfd{{\sf d}}

\def\sl(2){\alg{sl}(2)}

\def\be{\begin{equation}}
\def\ee{\end{equation}}

\newcommand{\bea}{\begin{eqnarray}}
\newcommand{\eea}{\end{eqnarray}}

\newcommand{\bei}{\begin{itemize}}
\newcommand{\eei}{\end{itemize}}

\def\a {\alpha}
\def\b {\beta}
\def\s {\sigma}
\def\pa {\partial}

\def\g {\gamma}
\def\om {\omega}
\def\p{\phi}
\def\la{\label}
\def\e{\epsilon}
\def\ov{\over}

\def\S{\Sigma}

\def\hP{{\bf P}}

\def\bO{{\bf O}}
\def\bI{{\bf I}}
\def\bJ{{\bf J}}
\def\bJ{{\bf J}}
\def\bS{{\bf S}}
\def\bR{{\bf R}}
\def\bL{{\bf L}}
\def\bQ{{\bf Q}}
\def\tbQ{{\widetilde{ \bf Q}}}
\def\bC{{\bf C}}
\def\tbC{{\widetilde{ \bf C}}}

\def\bH{{\bf H}}
\def\tbH{{\widetilde {\bf H}}}

\def\hP{{\mathbb P}}

\def\bO{{\mathbb O}}
\def\bI{{\mathbb I}}
\def\bJ{{\mathbb J}}
\def\bJ{{\mathbb J}}
\def\bS{{\mathbb S}}
\def\bR{{\mathbb R}}
\def\bL{{\mathbb L}}
\def\bQ{{\mathbb Q}}
\def\tbQ{{\widetilde{ \mathbb Q}}}
\def\bC{{\mathbb C}}
\def\tbC{{\widetilde{ \mathbb C}}}

\def\bH{{\mathbb H}}
\def\tbH{{\widetilde {\mathbb H}}}

\newcommand{\alg}[1]{\mathfrak{#1}}
\newcommand{\su}{\alg{su}}

\newcommand{\AdS}{{\rm  AdS}_5\times {\rm S}^5}
\newcommand{\ads}{{\rm  AdS}_5\times {\rm S}^5}

\newcommand{\bem}{\left (\begin{matrix}}
\newcommand{\eem}{\end{matrix} \right )}

\def\S{{\cal S}}

\newcommand{\sn}{\mathop{\mathrm{sn}}\nolimits}
\newcommand{\cn}{\mathop{\mathrm{cn}}\nolimits}
\newcommand{\dn}{\mathop{\mathrm{dn}}\nolimits}
\def\am{\mathrm{am}}

\def\S{{\mathbb S}}
\def\V{{\mathscr V}}
\def\W{{\mathscr W}}
\def\D{{\mathscr D}}

\def\xmao{{x^{-}_{1}}}
\def\xpao{{x^{+}_{1}}}

\def\xmbt{{y^{-}_{2}}}
\def\xpbt{{y^{+}_{2}}}

\def\eps{\epsilon}

\author{Gleb Arutyunov$^a$\footnote{Email: G.Arutyunov@phys.uu.nl, frolovs@maths.tcd.ie} {}
\footnote{Correspondent fellow at Steklov Mathematical Institute,
Moscow.} \  and Sergey Frolov$^{b\, \dagger}$
 \\ $^{a}$ {\it Institute for Theoretical
Physics and Spinoza Institute,\\ ~~Utrecht University, 3508 TD
Utrecht, The Netherlands} \\ $^b$ {\it School of Mathematics,
Trinity College, Dublin 2, Ireland}}

\abstract{We find the S-matrix which describes the scattering of
two-particle bound states of the light-cone string sigma model on
$\AdS$. We realize the $M$-particle bound state representation of
the centrally extended $\su(2|2)$ algebra  on the space of
homogeneous (super)symmetric polynomials  of degree $M$ depending
on two bosonic and two fermionic variables. The scattering matrix $\S^{MN}$
 of $M$- and $N$-particle bound states is
a differential operator of degree $M+N$ acting on the product of
the corresponding polynomials. We require this operator to obey
the invariance condition and the Yang-Baxter equation, and we
determine it for the two cases $M=1,N=2$ and $M=N=2$. We show that
the S-matrices found satisfy {\it generalized physical unitarity,
CPT invariance, parity transformation rule and crossing symmetry.}
Although  the dressing factor as a function of {\it four}
parameters $x_1^+,x_1^-,x_2^+,x_2^-$  is universal for scattering
of any bound states, it obeys a crossing symmetry equation which
depends on $M$ and $N$.

}

\title{
The S-matrix of String Bound States}

\preprint{
          \smaller{\smaller{\smaller{ITP-UU-08-15}}}\\[-.5ex]
          \smaller{\smaller{\smaller{SPIN-08-14}}}\\[-.5ex]
          \smaller{\smaller{\smaller{TCDMATH 08-03}}}}

\begin{document}

\renewcommand{\thefootnote}{\arabic{footnote}}
\setcounter{footnote}{0}

\section{Introduction and summary}
It has been recognized in recent years that the S-matrix approach
provides a powerful tool to study the spectra of both the $\ads$
superstring and the dual gauge theory \cite{MZ}-\cite{B}. In the
physical (light-cone) gauge the Green-Schwarz string sigma-model \cite{MT} is equivalent
to a non-trivial massive integrable model of eight bosons and
eight fermions \cite{AF04,FPZ}. The corresponding scattering matrix, which arises
in the infinite-volume limit, is determined by global symmetries
almost uniquely \cite{B,AFZ}, up to an overall dressing phase
\cite{AFS} and the choice of a representation basis. A functional form of the dressing phase in terms of
local conserved charges has been bootstrapped \cite{AFS} from the
knowledge of the classical finite-gap solutions
\cite{Kazakov:2004qf}. Combining this form with the crossing
symmetry requirement \cite{Janik}, one is able to find an exact,
i.e. non-perturbative in the coupling constant, solution for the
dressing phase \cite{BHL} which exhibits remarkable interpolation
properties from strong to weak coupling \cite{BES}. The leading
finite-size corrections to the infinite-volume
spectrum are encoded into a set of asymptotic Bethe Ansatz
equations  \cite{BS}  based on this S-matrix.

\smallskip

In more detail, a residual symmetry algebra of the light-cone
Hamiltonian ${\mathbb H}$ factorizes into two copies of the
superalgebra $\su(2|2)$ centrally extended by two central charges,
the latter depend on the operator ${\mathbb P}$ of the world-sheet
momentum \cite{AFPZ}. Correspondingly, the S-matrix factorizes
into a product of two S-matrices $\S^{AA}$, each of them scatters
two fundamental supermultiplets $A$. Up to a phase, the S-matrix
$\S^{AA}$ is determined from the invariance condition which
schematically reads as \bea \nonumber \S^{AA}\cdot {\mathbb
J}_{12} ={\mathbb J}_{21}\cdot \S^{AA} \, ,\eea where ${\mathbb
J}_{12}$ is a symmetry generator in the two-particle
representation.

\smallskip

As in some quantum integrable models, it turns out that, in
addition to the supermultiplet of fundamental particles, the
string sigma-model contains an infinite number of bound states
\cite{D}. They manifest themselves as poles of the multi-particle
S-matrix built over $\S^{AA}$. The representation corresponding to
a bound state of $M$ fundamental particles constitutes a short
$4M$- dimensional multiplet of the centrally extended $\su(2|2)$
algebra \cite{Bn,CDO}. It can be obtained from the $M$-fold tensor
product of the fundamental multiplets by projecting it on the
totally symmetric component. A complete handle on the string
asymptotic spectrum and the associated Bethe equations requires,
therefore, the knowledge of the S-matrices which describe the
scattering of bound states. This is also a demanding problem for
understanding the finite-size string spectrum within the TBA
approach \cite{AFTBA}.
\smallskip

A well-known way to obtain the S-matrices in higher
representations from a fundamental S-matrix is to use the fusion
procedure \cite{Kulish:1981gi}. A starting point is a
multi-particle S-matrix
$$
\S_{a1}(p,p_1)\S_{a2}(p,p_2)\ldots \S_{aM}(p,p_M)\, ,
$$
where the (complex) momenta $p_1,\ldots ,p_{M}$ provide a solution
of the $M$-particle bound state equation and the index $a$ denotes
an auxiliary space being another copy of the fundamental irrep.
Next, one symmetrises the indices associated to the matrix spaces
$1,\ldots, M$ according to the Young  pattern of a desired
representation. The resulting object must give (up to
normalization) an S-matrix for scattering of fundamental particles
with $M$-particle bound states. A proof of the last statement uses
the fact that at a pole the residue of the S-matrix  is degenerate
and it coincides with a projector on a (anti-)symmetric irrep
corresponding to the two-particle bound state. This is, e.g., what
happens for the rational S/R-matrices based on the $\alg{gl}(n|m)$
superalgebras. The fusion procedure for the corresponding
transfer-matrices and the associated Baxter equations have been
recently worked out in \cite{Kazakov:2007fy,Kazakov:2007na}.

\smallskip

The fundamental S-matrix $\S^{AA}$ behaves, however, differently
and this difference can be traced back to the representation
theory of the centrally extended $\su(2|2)$. Let $\V^A$ be a
four-dimensional fundamental multiplet. It characterizes by the
values of the central charges which  are all functions of the
particle momentum $p$. For the generic values of $p_1$ and $p_2$
the tensor product $\V^A(p_1)\otimes \V^A(p_2)$ is an irreducible
long multiplet \cite{B,Bn}. In the special case when momenta $p_1$
and $p_2$ satisfy the two-particle bound state equation the long
multiplet becomes reducible. With the proper normalization of the
S-matrix the invariant subspace coincides with the null space
$$
\S^{AA}(p_1,p_2)\,\V^{\rm ker}=0\, , ~~~~\V^{\rm ker}\subset
\V^{A}\otimes\V^{A}\, .
$$
Indeed, if $v\in \V^{\rm ker}$, then ${\mathbb J}_{12}v \in
\V^{\rm ker}$ due to the intertwining property of the S-matrix:
$\S^{AA}\cdot {\mathbb J}_{12}v={\mathbb J}_{21}\cdot \S^{AA}v=0$.
Being reducible at the bound state point, the multiplet
$\V^A\otimes \V^A$ is, however, {\it indecomposable}. In
particular, there is no  invariant projector on either $\V^{\rm
ker}$ or on its orthogonal completion. The bound state
representation we are interested in corresponds to a factor
representation on the quotient space $\V^A/\V^{\rm ker}$. In this
respect, it is not clear how to generalize the known fusion
procedure to the present case.\footnote{The usual fusion procedure
works, however, for the rank one sectors of $\S^{AA}$, in which
case it becomes trivial. This will be discussed later on.}

\smallskip

The absence of apparent fusion rules motivates us to search for
other ways to determine the scattering matrices of bound states.
An obvious suggestion would be to follow the same invariance
argument for the bound state S-matrices as for the fundamental
S-matrix together with the requirement of factorized scattering,
the latter being equivalent to the Yang-Baxter (YB) equations. A
serious technical problem of this approach is, however, that the
dimension of the $M$-particle bound state representation grows as
$4M$, so that the S-matrix $\S^{MN}$ for scattering of the $M$-
and $N$-particle bound states will have rank $16MN$. Even for
small values of $M$ and $N$ working with such big matrices becomes
prohibitory complicated. Although, we are ultimately interested
not in the S-matrices themselves but rather in eigenvalues of the
associated transfer-matrices, we fist have to make sure that the
corresponding scattering matrices do exist and they satisfy the
expected properties.

\smallskip

The aim of this paper is to develop a new operator approach to
deal with the bound state representations. This approach provides
an efficient tool to solve both the invariance conditions and the
YB equations and, therefore, to determine the S-matrices
$\S^{MN}$, at least for sufficiently low values of $M$ and $N$.

\smallskip

Our construction relies on the observation that the $M$-particle
bound state representation $\V^M$ of the centrally extended
$\su(2|2)$ algebra can be realized on the space of homogeneous
(super)symmetric polynomials  of degree $M$ depending on two
bosonic and two fermionic variables, $w_a$ and $\theta_{\a}$,
respectively. Thus, the representation space is identical to an
irreducible short superfield $\Phi_M(w,\theta)$. In this
realization the algebra generators are represented by linear
differential operators ${\mathbb J}$ in $w_a$ and $\theta_{\a}$
 with coefficients depending on the representation
parameters (the particle momenta). More generally, we will
introduce a space $\D_M$ dual to $\V^M$, which can be realized as
the space of differential operators preserving the homogeneous
gradation of $\Phi_M(w,\theta)$. The S-matrix $\S^{MN}$ is then
defined as an element of
$${\rm End
}(\V^M\otimes \V^N)\approx\V^M\otimes \V^N\otimes \D_M\otimes
\D_N.$$ On the product of two superfields
$\Phi_M(w^1,\theta^1)\Phi_N(w^2,\theta^2)$ it  acts as a
differential operator of degree $M+N$. We require this operator to
obey the following intertwining property
$$
\S^{MN}\cdot {\mathbb J}_{12}= {\mathbb J}_{21}\cdot \S^{MN}\, ,
$$
which is the same invariance condition as before but now
implemented for the bound state representations. For two $\su(2)$
subgroups of $\su(2|2)$ this condition literally means the
invariance of the S-matrix, while for the supersymmetry generators
it involves the brading (non-local) factors to be discussed in the
main body of the paper. Thus, the S-matrix is an $\su(2)\oplus
\su(2)$-invariant element of ${\rm End }(\V^M\otimes \V^N)$ and it
can be expanded over a basis of invariant differential operators
$\Lambda_k$:
$$
\S^{MN}=\sum_k a_k \Lambda_k\, .
$$
As we will show, it is fairly easy to classify the differential
$\su(2)\oplus \su(2)$ invariants $\Lambda_k$. The coefficients
$a_k$ are then partially determined from the remaining invariance
conditions with the supersymmetry generators. It turns out,
however, that if the tensor product $\V^M\otimes \V^N$ has $m$
irreducible components then $m-1$ coefficients $a_k$ together with
an overall scale are left undetermined from the invariance
conditions and to find them one has to make use of the YB
equations. This completes the discussion of our general strategy
for a search of $\S^{MN}$. The operator formalism can be easily
implemented in the {\it Mathematica} program and it reduces
enormously a computational time, therefore, we give it a
preference in comparison to the matrix approach.

\smallskip

Having established a general framework, we will apply it to
construct explicitly the operators $\S^{AB}$ and $\S^{BB}$, which
are the S-matrices for scattering processes  involving the
fundamental multiplet $A$ and a multiplet $B$ corresponding to the
two-particle bound state. We will show that $\S^{AB}$ is expanded
over a basis of 19 invariant differential operators $\Lambda_k$
and all the corresponding coefficients $a_k$ up to an overall
normalization are determined from the invariance condition.
Construction of $\S^{BB}$ involves 48 operators $\Lambda_k$. This
time, two of $a_k$'s remain undetermined by the invariance
condition, one of them corresponding to an overall scale. As to
the second coefficient, we find it by solving the YB equations. It
is remarkable that with one coefficient we managed to satisfy two
YB equations: One involving $\S^{AB}$ and $\S^{BB}$, and the
second involving $\S^{BB}$ only. This gives an affirmative answer
to the question about the existence of the scattering matrices for
bound states.

\smallskip

Recall that the universal cover of the parameter space describing
the fundamental representation of the centrally extended
$\su(2|2)$ is an elliptic curve (a generalized rapidity torus
with real and imaginary periods $2\omega_1$ and $2\omega_2$,
respectively) \cite{Janik}. Since the particle energy and momentum
are elliptic functions with the modular parameter $-4g^2$, where
$g$ is the coupling constant, the fundamental S-matrix $\S^{AA}$
can be viewed as a function of two variables $z_1$ and $z_2$ with
values in the elliptic curve. Analogously, the $M$-particle bound
state representation can be uniformized by an elliptic curve but
with another modular parameter $-4g^2/M^2$. Correspondingly, in
general $\S^{MN}(z_1,z_2)$ is a function on a product of two tori
with {\it different} modular parameters.

\smallskip

In a physical theory, in addition to the YB equation, the
operators $\S$ must satisfy a number of important analytic
properties. Regarding $\S$ as a function of
generalized rapidity variables, we list them below.
\begin{itemize}
\item {\it Generalized Physical Unitarity}
\bea\nonumber \bS(z_1^*,z_2^*)^\dagger\cdot \bS (z_1,z_2)= \bI\,
\eea
\item {\it CPT Invariance}  \bea\nonumber \bS(z_1,z_2)^\st =
\bS(z_1,z_2)\, \eea
\item {\it Parity Transformation Rule}
$$ \S^{-1}(z_1,z_2) = \S(-z_1,-z_2)^\sp
$$
\item {\it Crossing Symmetry}
$$
\S^{c_1}(z_1,z_2)\, \S(z_1+\omega_2, z_2)= \bI\,,\quad
\S^{c_2}(z_1,z_2)\, \S(z_1, z_2-\omega_2)= \bI\,.
$$
\end{itemize}
For real values of momenta (or $z$-variables) $\S$ must be a
unitary operator. If $z'$s are complex the usual unitarity is
replaced by the generalized unitarity condition above. The CPT
invariance implies that the S-matrix is a symmetric operator. The
parity map $z\to -z$ supplied with a proper transformation
${\mathscr P}$ of the matrix indices gives an inverse matrix.
Finally, the crossing symmetry relates the
anti-particle-to-particle scattering matrix $\S^{c_1}$ to that of
particle-to-particle and it holds with the proper normalization of
$\S$ only. As we will see, although the dressing factor as a
function of {\it four} parameters $x_1^+,x_1^-,x_2^+,x_2^-$  is
universal for scattering of any bound states, as a consequence of
the crossing symmetry, it obeys an equation which explicitly
depends on $M$ and $N$.

\smallskip

It appears that fulfillment of some of these  physical properties
crucially depends on the choice of a local basis in representation
spaces scattered by the S-matrix, a freedom is encoded in the
so-called $\eta$-phases. As was shown in \cite{AFTBA}, there
exists a unique \footnote{Up to momentum-independent unitary
transformations.} basis corresponding to a special choice of
$\eta$'s in which the matrix realization of $\S^{AA}$ enjoys all
the physicality requirements, including the generalized unitarity.

\smallskip

In this paper we will find a genuine operator realization of all
the above-listed properties. We will also  show that they do hold
for the operators $\S^{AA}$, $\S^{AB}$ and $\S^{BB}$. Most
importantly, the generalized unitarity, the CPT invariance and the
parity transformation for these S-matrices will take place with
our ``physical" choice of the $\eta$-phases only! In addition,
this choice will also render the canonically normalized S-matrices
to be meromorphic functions on the product of two tori.

\smallskip

Let us now discuss possible applications of our results and the
related problems. First of all, the S-matrices can be used to
derive asymptotic Bethe equations for
two-particle bound states generalizing the considerations in \cite{B,MM,Le}. Then, it is not clear to us if and how
the approach we adopted here can be extended for arbitrary values
of $M$ and $N$, perhaps, with an exception of the cases $M=1,2$
and $N$ arbitrary. In this respect,  the understanding of the fusion
procedure is an important problem to be solved. One should find an
efficient way of dealing with the multi-particle fundamental
S-matrix on factor representations corresponding to bound states.
In any case, the S-matrices obtained through the fusion procedure
must necessarily agree with our findings. In particular, one can
use $\S^{AB}$ and $\S^{BB}$ to check the conjectures existing in
the literature \cite{Bn,Belitsky:2008wj} about the eigenvalues of
the transfer-matrices in higher representations.

\smallskip

Another interesting point concerns the Yangian
symmetry\footnote{See also appendix 8.3 of \cite{AFZ} for the
discussion of higher symmetries of the fundamental
\mbox{S-matrix}. } \cite{Beisert:2007ds}-\cite{Spill:2008tp} and
the issue of the universal R-matrix. The fundamental S-matrix has
been shown \cite{Beisert:2007ds} to commute with the Yangian for
the centrally extended $\su(2|2)$ and we expect the same property
to hold for the S-matrices we found. In particular, it would be
interesting to see whether the Yangian symmetry is powerful enough
to fix all the coefficients of the S-matrix $\S^{MN}$ without
appealing to the YB equations. Concerning the universal R-matrix,
would it exist, one could use it to deduce all the bound states
S-matrices $\S^{MN}$ and the fusion procedure would not be
required. Due to non-invertibility of the Cartan matrix for
$\su(2|2)$, existence of the universal R-matrix remains an open
problem. On the other hand, at the classical level \cite{sualg}
one is able to identify a universal analogue of the classical
$r$-matrix \cite{Moriyama:2007jt,Beisert:2007ty}. It is of
interest to verify if the semi-classical limit of our S-matrices
agrees with this universal classical $r$-matrix \cite{Leeuw}.

\smallskip

As was explained in \cite{AFTBA}, to develop the TBA approach for
$\ads$ superstring one has to find the scattering matrices for
bound states of the accompanying mirror theory. The bound states
of this theory are ``mirrors" of those for the original string
model. Moreover, the scattering matrix of the fundamental mirror
particles is obtained from $\S^{AA}$ by the double Wick rotation.
The same rotation must also relate the S-matrices of the string
and mirror bound states, which on the rapidity tori should
correspond to shifts by the imaginary quarter-periods.

\smallskip

The leading finite-size corrections
\cite{magnon}-\cite{Minahan:2008re} to the dispersion relation for
fundamental particles (giant magnons \cite{HM}) and bound states can be also
derived by applying the perturbative L\"uscher approach \cite{Luscher}, which
requires the knowledge of the fundamental S-matrix
\cite{AJK}-\cite{Heller:2008at}. It is quite interesting
to understand the meaning of the bound state S-matrices in the
L\"uscher approach and use them, e.g., to compute the corrections
to the dispersion relations corresponding to string bound states.

\smallskip

Let us mention that our new S-matrices might also have an
interesting physical interpretation outside the framework of
string theory and the AdS/CFT correspondence \cite{M}. Up to normalization, $\S^{AA}$ coincides \cite{Bn,MM}
with the Shastry R-matrix \cite{Shastry} for the one-dimensional
Hubbard model. The operators $\S^{AB}$ and $\S^{BB}$ might have a
similar meaning for higher Hubbard-like models describing the
coupling of the Hubbard electrons to matter fields. Also, the
representation of the S-matrices in the space of symmetric
polynomials we found provides a convenient starting point for a
search of possible $q$-deformations \cite{Beisert:2008tw}. The
space of symmetric polynomials admits a natural $q$-deformation
with the corresponding symmetry algebra realized by difference
operators. It would be interesting to find the $q$-deformed
versions of $\S^{AB}$ and $\S^{BB}$ along these lines.

\section{Bound state representations}
In this section we discuss the atypical totally symmetric
representations of the centrally extended $\su(2|2)$ algebra which
are necessary to describe bound states of the light-cone string
theory  on $\AdS$. These representations are $2M|2M$-dimensional,
and they are parameterized by four parameters $a,b,c,d$ which are
meromorphic functions on a rapidity torus.

\subsection{Centrally extended $\su(2|2)$ algebra}

The centrally extended $\su(2|2)$ algebra which we will denote
$\su(2|2)_\cex$ was introduced in \cite{B}. It is generated by the
bosonic rotation generators $\bL_a{}^b\,,\ \bR_\a{}^\b$, the
supersymmetry generators $\bQ_\a{}^a,\,\ \bQ_a^{\dagger}{}^\a$,
and three central elements $\bH$, $\bC$ and $\bC^\dagger$ subject
to the following relations \bea \label{su22} &&
\left[\bL_a{}^b,\bJ_c\right]=\delta_c^b \bJ_a - {1\ov 2}\delta_a^b
\bJ_c\,,\qquad~~~~ \left[\bR_\a{}^\b,\bJ_\g\right]=\delta^\b_\g
\bJ_\a - {1\ov 2}\delta^\b_\a \bJ_\g\,,
 \nonumber \\
&& \left[\bL_a{}^b,\bJ^c\right]=-\delta_a^c \bJ^b + {1\ov 2}\delta_a^b \bJ^c\,,\qquad
~~\left[\bR_\a{}^\b,\bJ^\g\right]=-\delta_\a^\g \bJ^\b + {1\ov 2}\delta_\a^\b \bJ^\g\,, \nonumber \\
&& \{ \bQ_\a{}^a, \bQ_b^\dagger{}^\b\} = \delta_b^a \bR_\a{}^\b + \delta_\a^\b \bL_b{}^a +{1\ov 2}\delta_b^a\delta^\b_\a  \bH\,, \nonumber \\
&& \{ \bQ_\a{}^a, \bQ_\b{}^b\} =
\epsilon_{\a\b}\epsilon^{ab}~\bC\, , ~~~~~~~~ \{
\bQ_a^\dagger{}^\a, \bQ_b^\dagger{}^\b\} =
\epsilon_{ab}\epsilon^{\a\b}~\bC^\dagger \,. \eea Here the first
two lines show how the indices $c$ and $\gamma$ of any Lie algebra
generator transform under the action of $\bL_a{}^b$ and
$\bR_\a{}^\b$. For the $\AdS$ string model the central element
$\bH$ is hermitian and is identified with the world-sheet
light-cone Hamiltonian, and the supersymmetry generators
$\bQ_\a{}^a$ and $\bQ_a^{\dagger}{}^\a$. The central elements
$\bC$ and $\bC^\dagger$ are hermitian conjugate to each other:
$\left( \bQ_\a{}^a\right)^\dagger=\bQ_a^{\dagger}{}^\a$.  It was
shown in \cite{AFPZ} that the central elements $\bC$ and
$\bC^\dagger$  are expressed through the world-sheet momentum
$\hP$ as follows \bea \label{Cc} \bC={i\ov
2}g\,(e^{i\hP}-1)e^{2i\xi}\,,\quad \bC^\dagger=-{i\ov
2}g\,(e^{-i\hP}-1)e^{-2i\xi}\,,\quad  g ={\sqrt\l\ov 2 \pi}\, .
\eea The phase $\xi$ is an arbitrary function of the central
elements. It reflects an external ${\rm U(1)}$  automorphism of
the algebra (\ref{su22}): $\bQ\to e^{i\xi}\bQ\,,\ \bC\to
e^{2i\xi}\bC$. In our  paper \cite{AFZ} the choice $\xi=0$ has
been made in order to match with the gauge theory spin chain
convention by \cite{B} and to facilitate a comparison with the
perturbative string theory computation of the S-matrix performed
in \cite{KMRZ}. We use the same choice of $\xi$ in the present
paper too.

\smallskip

Without imposing the hermiticity conditions on the generators of
the algebra the U(1) automorphism of the algebra can be extended
to the external $\sl(2)$ automorphism \cite{Bn}, which acts on the
supersymmetry generators as follows \bea\la{nq}
\begin{aligned}
&&\tbQ_\a{}^a =u_1\,  \bQ_\a{}^a +
u_2\,\epsilon^{ac}\,\bQ_c^\dagger{}^\g\, \epsilon_{\g\a}\,,\quad
\tbQ^\dagger_a{}^\a = v_1\, \bQ^\dagger_a{}^\a +
v_2\,\epsilon^{\a\b}\,\bQ_\b{}^b\, \epsilon_{ba}\,,\\
&&\bQ_\a{}^a =v_1\,  \tbQ_\a{}^a -
u_2\,\epsilon^{ac}\,\tbQ_c^\dagger{}^\g\, \epsilon_{\g\a}\,,\quad
\bQ^\dagger_a{}^\a = u_1\, \tbQ^\dagger_a{}^\a -
v_2\,\epsilon^{\a\b}\,\tbQ_\b{}^b\, \epsilon_{ba}\,,
\end{aligned}\eea where the coefficients may depend on the central
charges and must satisfy the $\sl(2)$ condition
$$
u_1 v_1 - u_2 v_2 =1\,.
$$
Then, by using the commutation relations (\ref{su22}),  we find
that the transformed generators satisfy the same relations
(\ref{su22}) with the following new central charges \bea
\label{chargesn}
\begin{aligned}
\tbH &= (1 + 2 u_2 v_2) \bH -2u_1 v_2 \bC - 2 u_2 v_1\bC^\dagger\,,\\
\tbC&= u_1^2 \bC +u_2^2 \bC^\dagger - u_1u_2\bH \, ,\quad
\tbC^\dagger=v_1^2 \bC^\dagger +v_2^2 \bC - v_1v_2\bH  \,.
\end{aligned} \eea If we now require
that the new central charges $\tbC$ and $\tbC^\dagger$ vanish,
while the transformed supercharges $\tbQ$ and $\tbQ^\dagger$ are
hermitian conjugate to each other, we find
\bea\la{uv}\begin{aligned} & u_1 = v_1 = \frac{1}{\sqrt{2}}
\sqrt{1 + \frac{\bH}{\sqrt{\bH^2-4
   \bC \bC^\dagger}}}\,,\quad \\
   &
 u_2 = {\bC\ov \sqrt{\bH^2-4
   \bC \bC^\dagger}}{1\ov u_1}\,,\quad\quad
 v_2 = {\bC^\dagger\ov \sqrt{\bH^2-4
   \bC \bC^\dagger}}\,  \frac{1}{v_1}\,.
   \end{aligned}
\eea With this choice of the parameters $u_i,v_i$, the new
Hamiltonian takes the following simple form \bea\nonumber \tbH =
\sqrt{\bH^2-4\bC \bC^\dagger}\,. \eea We see that any irreducible
representation of the centrally-extended algebra with $\tbH\neq 0$
can be obtained from a representation of the usual $\su(2|2)$
algebra with zero central charges $\bC=\bC^\dagger=0$. This will
play an important role in our derivation of the bound state
scattering matrices.

\subsection{The atypical  totally symmetric  representation}
The atypical  totally symmetric  representation\footnote{Denoted
by $\langle M-1,0;\Vec{C}\rangle$ in \cite{Bn}.} of
$\su(2|2)_\cex$ which describes $M$-particle bound states of the
light-cone string theory on $\AdS$  has dimension $2M|2M$ and it
can be realized on the graded vector space with the following
basis \bei
\item a  tensor symmetric  in $a_i$: $|e_{a_1...a_M}\rangle$, where $a_i = 1,2$ are bosonic indices which gives $M+1$
bosonic states
\item a  symmetric in $a_i$ and skew-symmetric in $\a_i$:
$\ |e_{a_1...a_{M-2}\a_1\a_2}\rangle$, where $\a_i = 3,4$ are fermionic indices which gives $M-1$ bosonic
states.
The total number of bosonic states is $2M$.
\item a tensor symmetric in $a_i:\  |e_{a_1...a_{M-1}\a}\rangle$  which gives
$2M$ fermionic states.
\eei

\smallskip

\noindent We denote the corresponding vector space as
$\V^M(p,\zeta)$ (or just $\V^M$ if the values of $p$ and $\zeta$
are not important), where $\zeta=e^{2i\xi}$. For non-unitary
representations the parameters $p$ and $\zeta$ are arbitrary
complex numbers which parameterize the values of the central
elements (charges) on this representation: $\bH  |e_i\rangle= H
|e_i\rangle\,,\ \bC |e_i\rangle =C |e_i\rangle\,,\ \bC^\dagger
|e_i\rangle =\overline{C} |e_i\rangle$, where $|e_i\rangle$ stands
for any of the basis vectors. The bosonic generators act in the
space in the canonical way {\small \bea\nonumber &&\bL_a{}^b
|e_{c_1c_2\cdots c_M}\rangle =\delta_{c_1}^b|e_{ac_2\cdots
c_M}\rangle + \delta_{c_2}^b|e_{c_1ac_3\cdots c_M}\rangle +\cdots
-  {M\ov 2}\delta_a^b|e_{c_1c_2\cdots
c_M}\rangle\hspace{10cm}\\\nonumber &&\bL_a{}^b |e_{c_1\cdots
c_{M-2}\g_1\g_2}\rangle =\delta_{c_1}^b|e_{ac_2\cdots
c_{M-2}\g_1\g_2}\rangle + \delta_{c_2}^b|e_{c_1ac_3\cdots
c_{M-2}\g_1\g_2}\rangle +\cdots  - {M-2\ov
2}\delta_a^b|e_{c_1\cdots
c_{M-2}\g_1\g_2}\rangle~~~~~~~\hspace{10cm}\\\nonumber &&\bL_a{}^b
|e_{c_1c_2\cdots c_{M-1}\g}\rangle =\delta_{c_1}^b|e_{ac_2\cdots
c_{M-1}\g}\rangle + \delta_{c_2}^b|e_{c_1ac_3\cdots
c_{M-1}\g}\rangle +\cdots  -  {M-1\ov 2}\delta_a^b|e_{c_1\cdots
c_{M-1}\g}\rangle\,,\hspace{10cm} \eea } and similar formulas for
$\bR_\a{}^\b$ \bea\nonumber &&\bR_\a{}^\b |e_{c_1c_2\cdots
c_M}\rangle =0\\\nonumber &&\bR_\a{}^\b |e_{c_1c_2\cdots
c_{M-2}\g_1\g_2}\rangle =\delta_{\g_1}^\b|e_{c_1\cdots
c_{M-2}\a\g_2}\rangle + \delta_{\g_2}^\b|e_{c_1\cdots
c_{M-2}\g_1\a}\rangle   -  \delta_\a^\b|e_{c_1\cdots
c_{M-2}\g_1\g_2}\rangle\\\nonumber &&\bR_\a{}^\b |e_{c_1c_2\cdots
c_{M-1}\g}\rangle =\delta_{\g}^\b|e_{c_1\cdots c_{M-1}\a}\rangle -
{1\ov 2}\delta_\a^\b|e_{c_1\cdots c_{M-1}\g}\rangle\,. \eea Then
the most general action of supersymmetry  generators compatible
with the $\su(2)$ symmetry is of the form \bea \nonumber
&&\bQ_\a{}^a |e_{c_1c_2\cdots c_M}\rangle = a_1\left(
\delta_{c_1}^a|e_{c_2\cdots c_M\a}\rangle  +
\delta_{c_2}^a|e_{c_1\cdots c_M\a}\rangle+\cdots \right)
\,,\\\nonumber &&\bQ_\a{}^a  |e_{c_1c_2\cdots
c_{M-2}\g_1\g_2}\rangle  =b_2\e^{ac_{M-1}}\left(\e_{\a\g_1}
|e_{c_1\cdots c_{M-1}\g_2}\rangle  -\e_{\a\g_2} |e_{c_1\cdots
c_{M-1}\g_1}\rangle \right) \,,\ \ \ \ \ \\\nonumber &&\bQ_\a{}^a
|e_{c_1c_2\cdots c_{M-1}\g}\rangle  =b_1\e^{ac_{M}}\e_{\a\g}
|e_{c_1\cdots c_{M}}\rangle  +a_2\left(
\delta_{c_1}^a|e_{c_2\cdots c_{M-1}\a\g}\rangle  +
\delta_{c_2}^a|e_{c_1\cdots c_{M-1}\a\g}\rangle+\cdots \right)
\,,\ \ \ \ \ \ \eea where the constants $a_1,a_2,b_1,b_2$ are
functions of $g$, $p$ and $\zeta$.  The action of
$\bQ_a^\dagger{}^\a$ is given by similar formulas \bea \nonumber
&&\bQ_a^\dagger{}^\a |e_{c_1c_2\cdots c_M}\rangle =
c_1\e^{\a\g}\left(\e_{ac_1} |e_{c_2\cdots c_M\g}\rangle  +
\e_{ac_2}|e_{c_1\cdots c_M\g}\rangle+\cdots \right) \,,\\\nonumber
&&\bQ_a^\dagger{}^\a |e_{c_1c_2\cdots c_{M-2}\g_1\g_2}\rangle
=d_2\left(\delta^{\a}_{\g_1}|e_{c_1\cdots c_{M-2}a\g_2}\rangle  -
\delta^{\a}_{\g_2}|e_{c_1\cdots c_{M-2}a\g_1}\rangle\right)
\,,\\\nonumber &&\bQ_a^\dagger{}^\a  |e_{c_1c_2\cdots
c_{M-1}\g}\rangle  =d_1\delta^\a_\g |e_{c_1\cdots c_{M-1}a}\rangle
+c_2\e^{\a\rho}\left(\e_{ac_1} |e_{c_2\cdots c_{M-1}\rho\g}\rangle
+  \e_{ac_2}|e_{c_1\cdots c_{M-1}\rho\g}\rangle+\cdots \right).\ \
\ \ \ \ \eea The familiar fundamental representation \cite{B,Bn}
corresponds to $M=1$.

\smallskip

The constants $a_i,b_i,c_i,d_i$ are not arbitrary. They obey the
constraints which follow from the requirement that the formulae
above give a representation of $\su(2|2)_\cex$ \bea\nonumber
&&a_1d_1-b_1c_1=1\,,\quad a_2d_2-b_2c_2=1\\\nonumber
&&b_1d_2=b_2d_1\,,\quad c_1 a_2=a_1 c_2\,. \eea These relations
show that one can always rescale the basis vectors in such a way
that the parameters with the subscript 2 would be equal to those
with the subscript 1
$$
a_2=a_1\equiv \sfa\,,\quad b_2=b_1\equiv \sfb\,,\quad
c_2=c_1\equiv \sfc\,,\quad d_2=d_1\equiv \sfd\, .
$$
It is this choice we will make till the end of the paper. Thus, we
have four parameters subject to the following universal
$M$-independent constraint \bea\la{constr}
 \sfa \sfd -  \sfb  \sfc =1\,.
\eea The values of central charges, however, depend on $M$, and
they are given by \bea\la{hcc} {H\ov M} =  \sfa \sfd +  \sfb
\sfc\,,\quad {C\ov M}= \sfa \sfb\,,\quad {\overline{C}\ov M} =
\sfc \sfd \,. \eea We see that if we replace $H/M\to H$ and
$C/M\to C$ we just obtain the relations of the fundamental
representation. In terms of the parameters $g\,, x^\pm$ \cite{Bn}
this replacement is equivalent to the substitution $g\to g/M$ in
the defining relation for $x^\pm$. As a result, we obtain the
following convenient parametrization of $ \sfa, \sfb,  \sfc ,
\sfd$ in terms of\footnote{The parameter $\zeta$ in \cite{Bn}
should be rescaled as $\zeta\to -i\zeta$ to match our definition.}
$g,x^+,x^-,\zeta,\eta$  {\small \bea\nonumber \sfa = \sqrt{{g \ov
2M}}\eta\,,\    \sfb = \sqrt{g\ov 2M}{i\zeta\ov \eta}\left(
{x^+\ov x^-}-1\right)\,,\ \sfc = -\sqrt{g\ov 2M}{\eta\ov \zeta
x^+}\,,\   \sfd = \sqrt{g\ov 2M}{x^+\ov i\eta}\left( 1-{x^-\ov
x^+}\right)\,.\\\hspace{-10cm}\la{abcd} \eea } Here the parameters
$x^\pm$ satisfy the $M$-dependent constraint \bea\la{xpm} x^+
+{1\ov x^+} -x^--{1\ov x^-}={2M\ov g}i\,, \eea which follows from
$\sfa \sfd -  \sfb  \sfc=1$, and they are related to the momentum
$p$ as \bea\nonumber  {x^+\ov x^-} = e^{ip}\,. \eea The values of
the central charges can be found by using eq.(\ref{hcc})
\bea\la{hc2}
\begin{aligned}
 H &= M + {ig\ov x^+}-
{ig\ov x^-} = ig x^- - ig x^+ -M\,,\quad H^2 = M^2 + 4
g^2\sin^2{p\ov 2}\, ,\\ C&={i\ov 2}g\zeta\left( {x^+\ov x^-}-1\right) ={i\ov 2}g
\zeta\left( e^{ip}-1\right)\,,\quad
 \overline{C}={g\ov 2 i\zeta}\left( {x^-\ov x^+}-1\right) = {g\ov 2i\zeta}\left( e^{-ip}-1\right)\,.~~~~~
\end{aligned}\eea Let us stress that according to eq.(\ref{Cc}) the central
charges $C$ and $\overline{C}$ are functions of the string tension
$g$, and the world-sheet momenta $p$ is independent of $M$. This
explains the rescaling $g\to g/M$ in eqs.(\ref{abcd}).

\smallskip

The totally symmetric representation is completely determined by
the parameters $g,x^+,x^-,\zeta$, and $M$. The parameter $\eta$
simply reflects a freedom in the choice of the basis vectors
$|e_i\rangle$. For a non-unitary representation it can be set to
unity by a proper rescaling of $|e_i\rangle$. As was shown in
\cite{AFZ,AFTBA}, the string theory singles out the following
choice of $\eta$ and $\zeta$ \bea\la{urep} \zeta
=e^{2i\xi}\,,\quad \eta =e^{i\xi}\, e^{{i\ov 4}p}\, \sqrt{i x^- -
i x^+}\, , \eea where the parameter $\xi$ should be
real for unitary representations. For a single symmetric
representation the parameter $\zeta$ is equal to 1. The S-matrix,
however, acts in the tensor product $\V^M(p_1,e^{ip_2})\otimes
\V^N(p_2,1) \sim \V^M(p_1,1)\otimes \V^N(p_2,e^{ip_1})$ of the
representations, see \cite{AFZ} for a detailed discussion.

\smallskip
The string choice guarantees that the S-matrix satisfies the usual (non-twisted) YB equation and the generalized unitarity condition \cite{AFTBA}.

\subsection{The rapidity torus}

The other important property of the string choice for the phase $\eta$ is that it is only with this choice the parameters  $\sfa, \sfb, \sfc , \sfd$ are meromorphic functions of the torus rapidity variable $z$  \cite{Janik}. Let us recall that the dispersion
formula
 \bea\la{dispn}
  \left( {H\ov M} \right)^2 - 4\left( {g\ov M} \right)^2\sin^2{p\ov 2} = 1
 \eea
suggests the following natural parametrization of the energy and momentum in terms
of Jacobi elliptic functions
\bea\la{pez}
p=2\,{\am\,z}\,,~~\quad~~ \sin{p\ov 2} = \sn(z,k)\,,~~\quad~~ H =
M\dn(z,k)\, , \eea where we introduced the elliptic
modulus\footnote{Our convention for the elliptic modulus is the
same as accepted in the {\it Mathematica} program, e.g., ${\rm
sn}(z,k)={\rm JacobiSN}[z, k]$.   }
$k=-4g^2/M^2<0$. The corresponding torus has two periods $2\omega_1$ and $2\omega_2$, the
first one is real and the second one is imaginary \bea\nonumber
2\omega_1=4{\rm K}(k)\, , ~~~~~~~~~ 2\omega_2=4i{\rm K}(1-k)-4{\rm
K}(k)\, ,
 \eea
where ${\rm K}(k)$ stands for the complete elliptic integral of
the first kind.  The rapidity torus
 is an analog of the
rapidity plane in two-dimensional relativistic models. Note, however, that the elliptic modulus and the periods of the torus depend on the dimension of the symmetric representation.

\smallskip

In this parametrization the real $z$-axis is the
physical one  because for real
values of $z$ the energy is positive and the momentum is real due
to \bea\nonumber 1 \leq\dn(z,k)\leq \sqrt{k'}\, ,~~~~~~~~z\in
{\mathbb R}\, , \eea where $k'\equiv 1 -k$ is the complementary
modulus.
\smallskip

The representation parameters $x^{\pm}$,
which are subject to the constraint (\ref{xpm}) can be also expressed
in terms of Jacobi elliptic functions as
\bea \la{xpxmz}
x^{\pm}=\frac{M}{2g}\Big(\frac{\cn z}{\sn z} \pm i \Big)(1+\dn z)
\, .
\eea
Then, as was mentioned above, the S-matrix acts in the tensor product of two symmetric representations with parameters $\zeta=e^{2i\xi}$ equal to either  $e^{ip_2}$ or $e^{ip_1}$, and, therefore, the factor $e^{i\xi}$ which appears in the expression (\ref{urep}) for $\eta$ is a meromorphic function of the torus rapidity variable $z$. Thus,
if the parameter $\eta$ is a meromorphic function of $z$ then the parameters
$\sfa, \sfb,  \sfc, \sfd$ of the symmetric representation also are meromorphic functions of $z$. Indeed, as was shown in \cite{AFTBA}, one can
resolve the branch cut ambiguities of $\eta$ by means of the following relation
\bea\la{etaz}
e^{\frac{i}{4}p}\, \sqrt{ix^-(p)-ix^+(p)}=
\frac{\sqrt{2M}}{\sqrt{g}}\frac{{\rm dn}\, \frac{z}{2}\big({\rm
cn}\, \frac{z}{2}+i \, {\rm sn}\, \frac{z}{2}{\rm dn}\,
\frac{z}{2}\big)}{1+{4g^2\ov M^2}\, {{\rm sn}^4\frac{z}{2}}}\equiv \eta(z,M)
\eea
valid in the region $-\frac{\om_1}{2}<{\rm Re}\,
z<\frac{\om_1}{2}$ and $i\,\om_2<{\rm Im}\, z<-i\,\om_2$. We conclude, therefore, that with the choice of $\eta$ made in \cite{AFTBA}  the parameters
$\sfa, \sfb,  \sfc, \sfd$ of the symmetric representation are meromorphic functions of the torus rapidity variable $z$. As a consequence, the S-matrix is also a meromorphic function of $z_1, z_2$ (up to a non-meromorphic scalar factor).

\section{S-matrix in the superspace}
In this section we identify the totally symmetric representations
with the $2M|2M$-dimensional graded vector space of monomials  of degree $M$ of two bosonic and two fermionic variables. The generators of
the centrally extended $\su(2|2)$ algebra are realized as differential operators acting in this space. The S-matrix is naturally realized in this framework as
a differential $\su(2) \oplus \su(2)$ invariant operator in the tensor product of two representations.

\subsection{Operator realization of $\su(2|2)_\cex$}
It is well-known that the $\su(2)$ algebra can be realized by differential operators acting in the vector space of analytic functions of two bosonic variables $w_1,w_2$. An irreducible representation of spin $j$ is then identified with the vector subspace of
monomials of degree $2j$.

\smallskip

A similar realization also exists for the centrally extended
$\su(2|2)$ algebra. To this end we introduce a vector space of
analytic functions of two bosonic variables $w_a$, and two
fermionic variables $\theta_\a$. Since any such a function can be
expanded into a sum $$
 \Phi(w, \theta) =\sum_{M=0}^\infty \, \Phi_M(w, \theta)\,
$$ of homogeneous symmetric polynomials of degree $M$ \bea
\la{supf}   \Phi_M(w, \theta) &=&\p^{c_1\ldots c_M}w_{c_1}\cdots
w_{c_M}+\p^{c_1\ldots c_{M-1}\gamma}w_{c_1}\cdots
w_{c_{M-1}}\theta_{\gamma}+
\\&&\hskip 4.5cm +\, \p^{c_1\ldots
c_{M-2}\gamma_1\gamma_2}w_{c_1}\cdots
w_{c_{M-2}}\theta_{\gamma_1}\theta_{\gamma_2} \nonumber\eea the
vector space is in fact isomorphic to the direct sum of all
totally symmetric representations of $\su(2|2)_\cex$.

\smallskip

Then, the bosonic and fermionic generators of $\su(2|2)_\cex$ can
be identified with the following differential operators acting in
the space \bea \nonumber  \bL_a{}^b &=& w_a \frac{\pa}{\pa w_b} -
{1\ov 2}\delta_a^b w_c \frac{\pa}{\pa w_c}\,, \quad\qquad
\bR_\a{}^\b = \theta_\a \frac{\pa}{\pa \theta_\b} -
{1\ov 2}\delta^\b_\a \theta_\g \frac{\pa}{\pa \theta_\g}\\
\nonumber \bQ_{\a}{}^a &=& \sfa\, \theta_{\a}\frac{\pa}{\pa
w_a}+\sfb\, \e^{ab}\e_{\a\b}w_b\frac{\pa}{\pa\theta_{\b}} \,,\quad
\bQ_a^\dagger{}^\a = \sfd\,w_a\frac{\pa}{\pa\theta_{\a}}+\sfc\,
\e_{ab}e^{\a\b}\theta_{\b}\frac{\pa}{\pa w_b}~~~~~\\ \nonumber
\bC&=& \sfa\sfb\Big(w_a\frac{\pa}{\pa
w_a}+\theta_{\a}\frac{\pa}{\pa \theta_\a}\Big) \,, \quad\qquad
\bC^\dagger= \sfc\sfd\Big(w_a\frac{\pa}{\pa
w_a}+\theta_{\a}\frac{\pa}{\pa \theta_\a}\Big) \\\la{operalg}
&&\hskip 1.5cm \bH=(\sfa\sfd + \sfb\sfc)\Big(w_a\frac{\pa}{\pa
w_a}+\theta_{\a}\frac{\pa}{\pa \theta_\a}\Big)\,. \eea Here we use
the conventions \bea\nonumber
\frac{\pa}{\pa\theta_{\a}}\theta_{\b}+\theta_{\b}\frac{\pa}{\pa\theta_{\a}}=\delta^{\a}_{\b}\,
, ~~~~~~\e_{\a\gamma}\e^{\b\gamma}=\delta_{\a}^{\b}\, ,
~~~~~\e^{\a\b}\e_{\rho\delta}=\delta_{\rho}^{\a}\delta_{\delta}^{\b}-\delta_{\rho}^{\b}\delta^{\a}_{\delta}\,,
\eea and the constants $\sfa,\sfb, \sfc,\sfd$ satisfy the only
condition
$$
\sfa\sfd-\sfb\sfc=1\,,
$$
and are obviously identified with the parameters of the totally symmetric representation we discussed in the previous section.

\smallskip
The representation carried by $\Phi(w, \theta)$ is reducible, and
to single out an irreducible component one should restrict oneself
to an irreducible superfield $ \Phi_M(w, \theta)$ (\ref{supf}).
Then the basis vectors $ |e_{c_1c_2\cdots c_{M}}\rangle$, $
|e_{c_1c_2\cdots c_{M-1}\g}\rangle$ and $ |e_{c_1c_2\cdots
c_{M-2}\g_1\g_2}\rangle$ of a totally symmetric representation
should be identified with the monomials $w_{c_1}\cdots w_{c_{M}}$,
$w_{c_1}\cdots w_{c_{M-1}}\theta_{\gamma}$, and $w_{c_1}\cdots
w_{c_{M-2}}\theta_{\gamma_1}\theta_{\gamma_2}$, respectively.

\smallskip
One can define a natural scalar product on the vector space. One
introduces the following basis of monomials \bea\la{basismon}
|m,n,\mu,\nu\rangle = N_{mn\mu\nu}\,w_1^{m}\,w_2^{n}\,
\theta_3^{\mu}\,\theta_4^{\nu}\,,\ \   \eea where $m,n\ge 0\,,\
\mu,\nu=0,1\,,\ m+n+\mu+\nu=M$. This basis is assumed to be
orthonormal \bea\nonumber \langle  a,b,\a,\b |m,n,\mu,\nu\rangle =
\delta_{am}\delta_{bn}\delta_{\a\mu}\delta_{\b\nu}\,. \eea The
normalization constants $N_{mn\mu\nu}$ can then be determined from
the requirement of the unitarity of the representation, see e.g.
\cite{ArSok} for the $\su(2)$ case. In particular the hermiticity
condition for the generators $\bL_a{}^b:\, \
\left(\bL_a{}^b\right)^\dagger = \bL_b{}^a$ leads to the relation
\cite{ArSok} \bea\nonumber N_{mn\mu\nu} = \left({1\ov m!\, n!
}\right)^{1/2}\, N(m+n)\,. \eea Further, the condition that the
generators $\bQ_{\a}{}^a$ and $ \bQ_a^\dagger{}^\a$ are hermitian
conjugate to each other fixes the normalization constants $N(m+n)$
to coincide\bea\nonumber N(M-2)=N(M-1)=N(M)\,. \eea The overall
normalization constant can be set to any number, and we choose it
so that the normalization constant $N_{M-2,0,1,1}$ be equal to 1.
This gives \bea\nonumber N_{mn\mu\nu} = \left({(M-2)!\ov m!\, n!
}\right)^{1/2}\,. \eea

\smallskip

Having defined the scalar product, one can easily check that both
the transposed and hermitian conjugate operators are obtained by
using the following rules \bea\la{hermc}
&&\left(w_a\right)^\dagger = {\pa\ov\pa w_a}\,,\quad \left(
{\pa\ov\pa w_a}\right)^\dagger = w_a\,,\quad
\left(\theta_\a\right)^\dagger = {\pa\ov\pa \theta_\a}\,,\quad
\left( {\pa\ov\pa  \theta_\a}\right)^\dagger =
\theta_\a\\\nonumber &&\left(w_a\right)^t = {\pa\ov\pa
w_a}\,,\quad \left( {\pa\ov\pa w_a}\right)^t = w_a\,,\quad
~\left(\theta_\a\right)^t = {\pa\ov\pa \theta_\a}\,,\quad \left(
{\pa\ov\pa  \theta_\a}\right)^t =  \theta_\a \,, \eea that means
that $w_a\,,\, {\pa\ov\pa w_a}\,,\,   \theta_\a\,,\, {\pa\ov\pa
\theta_\a}$ are considered to be real. For the product of several
operators the usual rules are applied: $\left({\mathbb A} {\mathbb
B}\right)^\dagger ={\mathbb B}^\dagger {\mathbb A}^\dagger$ and
$\left({\mathbb A} {\mathbb B}\right)^t = {\mathbb B}^t {\mathbb
A}^t$. In particular, by using these rules one can readily verify
that the central element $\bH$ is hermitian,  and the
supersymmetry generators $\bQ_\a{}^a$ and $\bQ_a^{\dagger}{}^\a$,
and the central elements $\bC$ and $\bC^\dagger$ are hermitian
conjugate to each other, provided the parameters
$\sfa,\sfb,\sfc,\sfd$ satisfy the following relations: $\sfd^* =
\sfa\,,\, \sfc^* = \sfb$.

\smallskip

The basis of monomials (\ref{basismon}) can be also used to
 find the matrix form of the algebra generators.
 Denoting the basis vectors of the $M$-particle bound state representation as  $ |e_{i}\rangle$,
 where $i=1,\ldots ,4M$,  we can define the matrix elements of any differential operator $\bO$ acting in the vector space of monomials by the formula
\bea\la{matrelem0}
\bO\cdot  |e_{i}\rangle = O_{i}^{k}(z) \, |e_{k}\rangle,
\eea
and use them to construct the matrix form of $\bO$
\bea\la{Om}
O(z)= \sum_{ik}
O_{i}^{k}(z)\, E_k{}^i \,,
\eea
where $E_k{}^i \equiv  E_{ki}$ are the usual matrix unities.

\bigskip

 Let us finally note that the $\sl(2)$ external  automorphism of $\su(2|2)_\cex$
 just redefines the constants $\sfa,\sfb,\sfc,\sfd$ as follows
\bea \la{tQc}
 \tilde{\sfa} = \sfa\, u_1 - \sfc\, u_2\,,\quad  \tilde{\sfb} = \sfb\, u_1 - \sfd\, u_2\,;\qquad
  \tilde{\sfc} = \sfc\, v_1 - \sfa\, v_2\,,\quad  \tilde{\sfd} = \sfd\, v_1 - \sfb\, v_2\,,
\eea
and can be used to
set, e.g. $\sfb=0=\sfc$. This  simplifies  the
derivation of the S-matrix.

\subsection{S-matrix operator}

It is clear that the graded tensor product
$\V^{M_1}(p_1,\zeta_1)\otimes \V^{M_2}(p_2,\zeta_2)$ of two
totally symmetric representations  can be identified with   a
product of two superfields $\Phi_{M_1}$ and $\Phi_{M_2}$ depending
on different sets of coordinates. An algebra generator acting in
the product is given by the sum of two differential operators each
acting on its own superfield and depending on its own set of
parameters $\sfa,\sfb, \sfc,\sfd$ or, equivalently, on $p$ and
$\zeta$. Given any two sets of parameters, one obtains a
representation of $\su(2|2)_\cex$ (generically reducible)  with
central charges $H,C,\overline{C}$ equal to the sum of central
charges of the symmetric representations: $C=C_1+C_2$, and so on.
The same tensor product representation can be obtained from two
other symmetric representations $\V^{M_1}(p_1,\tilde{\zeta_1})\,,
\V^{M_2}(p_2,\tilde{\zeta_2})$ if the parameters $\tilde{\zeta_i}$
satisfy certain shortening conditions \cite{Bn}.

\smallskip

As was extensively discussed in \cite{AFZ}, in string theory the
Hilbert space of two-particle in-states is identified with the
tensor product $\V^{M_1}(p_1,e^{ip_2})\otimes \V^{M_2}(p_2,1)$,
and the Hilbert space of two-particle out-states is identified
with  $\V^{M_1}(p_1,1)\otimes \V^{M_2}(p_2,e^{ip_1})$. The two
Hilbert spaces are isomorphic to the product of two superfields
$\Phi_{M_1}$ and $\Phi_{M_2}$, and the S-matrix $\bS$  is an
intertwining operator \bea\la{Sop} \bS (p_1,p_2):
~~~~~\V^{M_1}(p_1,e^{ip_2})\otimes \V^{M_2}(p_2,1)\to
\V^{M_1}(p_1,1)\otimes \V^{M_2}(p_2,e^{ip_1}) \, .~~~~~\eea It is
realized as a differential operator acting on $\Phi_{M_1}(w_a^1,
\theta_\a^1)\Phi_{M_2}(w_a^2, \theta_\a^2)$, and satisfying the
following invariance condition \bea\la{invcon} \bS(p_1,p_2)\cdot
(\bJ(p_1,e^{ip_2}) + \bJ(p_2,1)) = (\bJ(p_1,1) + \bJ(p_2,e^{ip_1})
)\cdot \bS(p_1,p_2)\, \eea for any of the symmetry generators.
Here $\bJ(p_i,\zeta)$ stands for any of the generators
(\ref{operalg}) realized as a differential operator acting on
functions of $w_a^i, \theta_\a^i$. The parameters $\sfa,\sfb,
\sfc,\sfd$ of the operator coefficients are expressed through
$p_i$ and $\zeta$ by means of formulae (\ref{abcd}) with $\eta$
given by eq.(\ref{urep}).

\smallskip

Since $\Phi_{M_1}\Phi_{M_2}$ is isomorphic to the graded tensor
product of two symmetric representations, the S-matrix operator
(\ref{Sop}) satisfying eq.(\ref{invcon}) is analogous to the
so-called fermionic $R$-operator, see e.g. \cite{kor}. This
analogy becomes even more manifest when one considers the action
of $\S$ on the tensor product of several symmetric representation.
For this reason $\S$ could be also called the fermionic
$\bS$-operator. In particular, in absence of interactions, i.e. in
the limit $g\to\infty$, it must reduce to the identity operator.
\smallskip

It is clear that for bosonic generators $\bL_a{}^b \,,
\bR_\a{}^\b$ generating the $\su(2) \oplus \su(2)$ subalgebra, the
invariance condition (\ref{invcon}) takes the standard form
\bea\la{suLR} \left[\, \bS(p_1,p_2)\,,\, \bL_a{}^b\, \right] =0 =
\left[\, \bS(p_1,p_2)\,,\, \bR_\a{}^\b\, \right] \, . \eea
Therefore, the S-matrix is a  $\su(2) \oplus \su(2)$-invariant
differential operator which maps $\Phi_{M_1}(w_a^1,
\theta_\a^1)\Phi_{M_2}(w_a^2, \theta_\a^2)$ into a linear
combination of the products of two homogeneous polynomials of
degree $M_1$ and $M_2$.

\smallskip
Any differential operator acting in $\V^{M_1}\otimes \V^{M_2}$ can
be viewed as an element of
$${\rm End}(\V^{M_1}\otimes
\V^{M_2})\approx \V^{M_1}\otimes \V^{M_2} \, \otimes \,
\D_{M_1}\otimes\D_{M_2}\, ,$$ where $\D_M$ is the vector space
dual to $\V^M$. The dual space is realized as the space of
polynomials of degree $M$ in the derivative operators ${\pa\ov \pa
w_a}\,,\, {\pa\ov\pa\theta_\a}$. A natural pairing between $\D_M$
and $\V^M$ is induced by ${\pa\ov \pa w_a} w_b = \delta_a^b\,,\,
{\pa\ov \pa \theta_\a} \theta_\b = \delta_\a^\b$. Therefore,
eq.(\ref{suLR}) means that the S-matrix is a $\su(2)\oplus \su(2)$
singlet (invariant) component  in the tensor product decomposition
of \mbox{$\V^{M_1}\otimes \V^{M_2} \, \otimes \,
\D_{M_1}\otimes\D_{M_2}$}.

\smallskip

Thus, the S-matrix can be naturally represented as
\bea\la{smatl}
 \bS(p_1,p_2)= \sum_i \, a_i(p_1,p_2)\, \Lambda_i\,,
\eea where $\Lambda_i$ span a  complete basis of differential
$\su(2) \oplus \su(2)$-invariant operators (obviously independent
of $p_1,p_2$), and  $a_i$ are  coefficients of the S-matrix which
could be determined from the remaining invariance conditions
(\ref{invcon}), and some additional requirements such as the YB
equation.

\smallskip

The basis of the operators  $\Lambda_i$ can be easily found by
first branching the symmetric representations $\V^{M_1} , \V^{M_2}
$ into their $\su(2)\oplus \su(2)$ components, see eq.(\ref{supf})
\bea\nonumber \V^{M_i} = V^{M_i\ov 2}\times V^{0}_{(0)}\, +\,
V^{{M_i-1\ov 2}}\times V^{1/2}_{(1)} \,+ \,V^{{M_i-2\ov 2}}\times
V^{0}_{(2)} = \sum_{k=0}^2V^{{M_i-k\ov 2}}\times V^{j_k}_{(k)}\, .
\eea Here $V^j$ denotes a spin $j$ representation of $\su(2)$
realized in the space of homogeneous symmetric polynomials of
degree $2j$ of two bosonic variables, while $V^{j_k}_{(k)}$ stands
for a spin $j_k$ representation of $\su(2)$ realized in the space
of homogeneous degree $k$ polynomials of two fermionic variables,
so that $k$ can take only 3 values: $k=0,1,2$, and $j_0=0\,, \,
j_1=1/2\,,\, j_2=0$.

\smallskip
The tensor product $\V^{M_1}\otimes \V^{M_2}$ is then decomposed
into a sum of irreducible $\su(2)\oplus \su(2)$ components as
follows \bea\la{decom} \V^{M_1}\otimes \V^{M_2}
=\sum_{k_1,k_2=0}^2\,\sum_{j_B={|M_2-M_1-k_2+k_1|\ov
2}}^{{M_2+M_1-k_2-k_1\ov 2}} V^{j_B}_{(M_1-k_1,M_2-k_2)}\times
\sum_{j_F=|j_{k_1}-j_{k_2}|}^{j_{k_1}+j_{k_2}} V^{j_F}_{(k_1,k_2)}
\,.~~~~~~~ \eea Here $ V^{j_B}_{(M,N)}$ denotes a spin $j_B$
representation of $\su(2)$, which is realized in the space spanned
by products of degree $M$ with degree $N$ symmetric polynomials,
the first one in two bosonic variables $w_a^1$ and the second in
$w_a^2$. Analogously, $ V^{j_F}_{(k_1,k_2)}$ stands for a spin
$j_F$ representation of $\su(2)$ realized in the space of products
of degree $k_1$ polynomials in two fermionic variables
$\theta_\a^1$ with degree $k_2$ polynomials  in two other
fermionic variables $\theta_\a^2$.

\smallskip
Finally, the dual space $\D_{M_1}\otimes\D_{M_2}$ has a similar
decomposition (\ref{decom}) in terms of the vector spaces
$D^j_{(n_1,n_2)}$ dual to $V^j_{(n_1,n_2)}$. Since $V^j\otimes
D^j$ contains the singlet component (the invariant), the
$\su(2)\oplus \su(2)$-invariant differential operators simply
correspond to the singlet components of the spaces $V^j\times V^k
\otimes D^j\times D^k$ in the tensor product decomposition of
$\V^{M_1}\otimes \V^{M_2} \, \otimes \, \D_{M_1}\otimes\D_{M_2}$.

\smallskip

In principle, it is not difficult to count the number of
components in eq.(\ref{decom}) as well as the number of invariant
operators for arbitrary $M_1,M_2$. We consider here\footnote{The
case $M_1=M_2=2$ will be discussed in detail in section
\ref{secs22}. } only the special case of $M_1=1\,,\, M_2=M$ which
corresponds to the scattering of a fundamental particle with a
$M$-particle bound state. One has to consider separately three
different cases: $M=1$, $M=2$ and $M\ge 3$.

\smallskip

For $M=1$, by using (\ref{decom}), one gets \bea\la{decom11}
\V^{1}\otimes \V^{1} &=&V^{1}_{(1,1)}\times
V^{0}_{(0,0)}+V^{0}_{(0,0)}\times V^{1}_{(1,1)} \\\nonumber
&+&V^{0}_{(1,1)}\times V^{0}_{(0,0)}+V^{0}_{(0,0)}\times
V^{0}_{(1,1)} + V^{1/2}_{(1,0)}\times
V^{1/2}_{(0,1)}+V^{1/2}_{(0,1)}\times V^{1/2}_{(1,0)}\,.~~~~~~~
\eea We see that there are six components in this decomposition
and they give rise to six {\it diagonal invariants} of the
symbolic form $V^{j_B}_{(N_1,N_2)}\times V^{j_F}_{(k_1,k_2)}\cdot
D^{j_B}_{(N_1,N_2)}\times D^{j_F}_{(k_1,k_2)}$. Then the
representations $V^{0}\times V^{0}$ and $V^{1/2}\times V^{1/2}$
come with multiplicity 2, and that gives additional four {\it
off-diagonal invariants} of the form $V^{j_B}_{(N_1,N_2)}\times
V^{j_F}_{(k_1,k_2)}\cdot D^{j_B}_{(K_1,K_2)}\times
D^{j_F}_{(n_1,n_2)}$ with $\{N_i, k_i\}\neq \{K_i,n_i\}$. Thus,
the total number of invariant operators in the $M_1=M_2=1$ case is
10.

\smallskip
In the case $M_1=1, M_2=2$ we get the following decomposition with the nine components
\bea\la{decom12}
\V^{1}\otimes \V^{2} &=&V^{3/2}_{(1,2)}\times V^{0}_{(0,0)}+V^{1/2}_{(0,1)}\times V^{1}_{(1,1)}
+ V^{1}_{(1,1)}\times V^{1/2}_{(0,1)} + V^{1}_{(0,2)}\times V^{1/2}_{(1,0)}
 \\\nonumber
&&\hspace{-1cm} + V^{0}_{(1,1)}\times
V^{1/2}_{(0,1)}+V^{0}_{(0,0)}\times V^{1/2}_{(1,2)} +
V^{1/2}_{(1,2)}\times V^{0}_{(0,0)} +V^{1/2}_{(1,0)}\times
V^{0}_{(0,2)}+V^{1/2}_{(0,1)}\times V^{0}_{(1,1)}\, \eea giving
rise to 9 diagonal invariants. Representations $V^{1}\times
V^{1/2}$ and $V^{0}\times V^{1/2}$ come with multiplicity 2, and
$V^{1/2}\times V^{0}$ occurs with multiplicity 3, and that gives
additional 10 off-diagonal invariants Thus, the total number of
invariant operators in the $M_1=M_2=1$ case is 19.

\smallskip
Finally, in the case $M_1=1, M_2\equiv M \ge 3$ we get the
following decomposition over the ten components \bea\la{decom13}
\V^{1}\otimes \V^{M} &=&V^{M+1\ov 2}_{(1,M)}\times
V^{0}_{(0,0)}+V^{M-1\ov 2}_{(0,M-1)}\times V^{1}_{(1,1)}+V^{M-3\ov
2}_{(1,M-2)}\times V^{0}_{(0,2)}\\\nonumber &+& V^{M\ov
2}_{(1,M-1)}\times V^{1/2}_{(0,1)} + V^{M\ov 2}_{(0,M)}\times
V^{1/2}_{(1,0)} +V^{M-2\ov 2}_{(1,M-1)}\times
V^{1/2}_{(0,1)}+V^{M-2\ov 2}_{(0,M-2)}\times V^{1/2}_{(1,2)}
 \\\nonumber
&+&
 V^{M-1\ov 2}_{(1,M)}\times V^{0}_{(0,0)}
+V^{M-1\ov 2}_{(1,M-2)}\times V^{0}_{(0,2)}+V^{M-1\ov
2}_{(0,M-1)}\times V^{0}_{(1,1)}\, , \eea which produces 10
diagonal and 10 off-diagonal invariants. So, the total number of
invariant operators in the $M_1=1\,,\, M_2\ge 3$ case is 20, and
it is independent of $M_2$.

\smallskip
We will fix the normalization of diagonal invariants such that
they provide the orthogonal decomposition of the identity operator
\bea\nonumber {\mathbb I} = \sum\, \Lambda_i^{\rm diag}\,. \eea In
what follows we will always denote the diagonal invariant operator
corresponding to the component $V^{M_1+M_2\ov 2}_{(M_1,M_2)}\times
V^{0}_{(0,0)}$ with the maximum spin $j_B = (M_1+M_2)/2$ as
$\Lambda_1$, and we will set the coefficient $a_1$ of the S-matrix
(\ref{smatl}) to be equal to one.\footnote{With this choice the
string S-matrix can be obtained from  (\ref{smatl}) by multiplying
it by the scalar S-matrix of the $\su(2)$ sector, see section
\ref{seccross} for details.}

\smallskip

Having found all invariant operators, one can use the invariance
condition (\ref{invcon}) to determine some of the coefficients
$a_i$. A natural question is how many coefficients will be left
undetermined or, in other words, how many different (up to an
overall factor) $\su(2|2)$ invariant S-matrices exist. To answer
to this question, one should decompose the tensor product
$\V^{M_1}\otimes\V^{M_2}$ into the sum of irreducible $\su(2|2)$
components. Then the number of different $\su(2|2)$ invariant
S-matrices is equal to the number of components in the
decomposition. The tensor product decomposition was studied in
\cite{Bn} with the  result\bea\la{dec2} \V^{M_1}\otimes\V^{M_2} =
\sum_{n = |M_1-M_2| +2}^{M_1+M_2} \W^{n}\, , \eea where $\W^n$ is
a typical (or long) irreducible supermultiplet of dimension
$16(n-1)$ denoted as $\{ n-2,0; \Vec{C}\}$ in \cite{Bn}.

\smallskip
Assuming that $M_1\le M_2$, we find  $M_1$ irreducible components
in eq.(\ref{dec2}), and, therefore, there exists $M_1$ $\su(2|2)$
invariant S-matrices solving eq.(\ref{invcon}). Setting $a_1=1$
leaves $M_1-1$ coefficients undetermined. They could be fixed by
imposing  additional equations such as the unitarity condition and
the YB equation.

\smallskip
In the simplest case $M_1=1\,,\, M_2\equiv M$ there is only one
component in the decomposition (\ref{dec2}) \bea\la{dec1m}
\V^{1}\otimes\V^{M} = \W^{M+1}\,, \eea and, as the consequence,
the $\su(2|2)$ invariant S-matrix is determined uniquely up to an
overall factor.

\smallskip
The first nontrivial case corresponds to $M_1=2\,,\, M_2\equiv
M\ge 2$, where eq.(\ref{dec2}) has two components \bea\la{dec2m}
\V^{2}\otimes\V^{M} =  \W^{M+2}+\W^M \,. \eea A coefficient of the
S-matrix  which cannot be determined by the invariance condition
is fixed by the YB equation as we will demonstrate in section
\ref{secs22}.

\subsection{Matrix form of the S-matrix} \la{matrform}

The operator formalism provides a very efficient way of solving
the invariance conditions (\ref{invcon}), and checking the YB
equation. It is convenient, however, to know also the usual matrix
form of the S-matrix operator. It can be used to check such
properties of the S-matrix as  unitarity  and crossing symmetry.

\smallskip
To find the matrix form of the S-matrix
 realized the invariant differential operator, we use the basis of monomials (\ref{basismon}).
Denoting the basis vectors of the $M_1$- and $M_2$-particle bound
state representations as $ |e_{i}\rangle$ and $ |e_{I}\rangle$,
respectively, and the basis vectors of the tensor product of these
representations as  $ |e_{iJ}\rangle$  where $i=1,\ldots ,4M_1$
and $J=1,\ldots ,4M_2$ ,  we define the matrix elements of any
operator $\bO$ acting in the tensor product by the formula
\bea\la{matrelem} \bO\cdot  |e_{iJ}\rangle = O_{iJ}^{kL}(z_1,z_2)
\, |e_{kL}\rangle. \eea The elements $O_{iJ}^{kL}$ can be now used
to construct various matrix forms of $\bO$. In particular, the
matrix form of the S-matrix $\bS$ which satisfies the {\it graded}
YB equation  is given by \bea\la{Smg} S^g_{12}(z_1,z_2)=
\sum_{ikJL} S_{iJ}^{kL}(z_1,z_2)\, E_k{}^i \otimes E_L{}^J\,, \eea
where $E_k{}^i \equiv  E_{ki}$ are the usual matrix unities.

\smallskip
On the other hand, the matrix form of the S-matrix $\bS$ which
satisfies the {\it usual} YB equation, see section \ref{secs11},
is obtained by multiplying $S^g_{12}$ by the graded identity
$I_{12}^g = (-1)^{\epsilon_i\epsilon_J} E^i_i\otimes E^J_J$, where
$\epsilon_i\,, \epsilon_J$ are equal to 0 for bosonic and to 1 for
fermionic states. It is  given by \bea\la{Sm}
S_{12}(z_1,z_2)=I_{12}^gS^g_{12}= \sum_{ikJL}
S_{iJ}^{kL}(z_1,z_2)\,  (-1)^{\epsilon_k\epsilon_L}\, E_k{}^i
\otimes E_L{}^J\,. \eea

\smallskip
For reader's convenience we list in appendices the matrix form of
the  $\su(2)\oplus\su(2)$ invariant differential operators
expressing them as sums over symbols $E_{kiLJ}$, the latter can be
equal to either $E_k{}^i \otimes E_L{}^J$ or to
$(-1)^{\epsilon_k\epsilon_L}\, E_k{}^i \otimes E_L{}^J$ (or
anything else one wants).

\subsection{General properties of  S-matrix}

In a physical theory the S-matrix should satisfy some general
properties.

\medskip

{\it Physical Unitarity.} First of all, for real values of the
momenta $p_1,p_2$ or, equivalently, for the real torus rapidity
variables $z_1,z_2$,  the S-matrix must be a  unitary operator.
Moreover, in a relativistic theory  the S-matrix satisfies a more
general condition $\bS(\theta^*)^\dagger\cdot \bS (\theta)= \bI$
called the generalized unitarity. For a non-relativistic theory
with a S-matrix depending on two parameters the condition takes
the following form \bea\la{guc} \bS(z_1^*,z_2^*)^\dagger\cdot \bS
(z_1,z_2)= \bI\,. \eea To analyze the generalized unitarity
condition, we notice that the diagonal invariant operators can be
normalized to be hermitian, and it is convenient to normalize and
order the off-diagonal invariant operators in such a way that they
come in hermitian-conjugate pairs \bea\la{Lhc}
\left(\Lambda_i^{\rm diag}\right)^\dagger = \Lambda_i^{\rm
diag}\,,\qquad \left(\Lambda_{2k}^{\rm off-diag}\right)^\dagger =
\Lambda_{2k+1}^{\rm off-diag}\,, \eea where $k$ takes integer
values if the number of diagonal operators is odd, and
half-integer values if the number of diagonal operators is even.

\smallskip

It is clear that the generalized unitarity condition imposes
severe  restrictions on the coefficients $a_i$ of the S-matrix. As
was shown in  \cite{AFTBA}, the S-matrix of fundamental particles
of the light-cone string theory satisfies the generalized
unitarity only for a special choice of the parameters $\eta_i$ of
the fundamental representations. Then, the only degree of
arbitrariness left is a constant unitary rotation of the basis of
the fundamental representation. We will show in section
\ref{secsmatr} that the bound state S-matrices also satisfy the
generalized unitarity condition.

\medskip

{\it CPT invariance.} In a CPT-invariant relativistic field theory
the S-matrix must be a symmetric operator, and we impose the same
condition on the scattering matrix for any bound states of the
light-cone string theory \bea\la{symc} \bS(z_1,z_2)^\st =
\bS(z_1,z_2)\,. \eea Taking into account that, due to the
definition (\ref{hermc}), the transposition acts  on the invariant
operators in the same way as the hermitian conjugation (\ref{Lhc})
\bea\nonumber \left(\Lambda_i^{\rm diag}\right)^\st =
\Lambda_i^{\rm diag}\,,\qquad \left(\Lambda_{2k}^{\rm
off-diag}\right)^\st = \Lambda_{2k+1}^{\rm off-diag}\,, \eea we
find that for a symmetric S-matrix no restriction arises on the
coefficients $a_i$ corresponding to the diagonal invariant
operators, while the off-diagonal coefficients must satisfy the
following non-trivial relations \bea\la{rel2} a_{2k}^{\rm
off-diag} =a_{2k+1}^{\rm off-diag}\,.~~~~~~~~ \eea We will see in
section \ref{secsmatr} that these relations do hold but again only
with the choice of $\eta_i$  as made in \cite{AFTBA}.

\medskip

{\it Parity Transformation.} The S-matrix is an  invertible
operator, and one can ask whether  the inverse S-matrix  is
related
 to the S-matrix itself but not through the inversion procedure.  One relation has been already
described above, where the inverse S-matrix is just the hermitian
conjugate one. Another quite natural relation arises due to the
parity transformation ${\mathscr P}$ of the world-sheet coordinate
$\s$, which changes the sign of  the torus rapidity variables
$z_i\to -z_i$. To describe the action of the parity transformation
on the invariant operators, one notices that since they are
homogeneous polynomials in fermions $\theta_\a^i$ and derivatives
$\pa/\pa\theta_\a^i$, one can introduce the following natural
grading on the space of invariant operators \bea\la{grading}
\eps_\Lambda = {1\ov 2}\left(n_{\theta} -
n_{\pa/\pa\theta}\right)\,, \eea where $n_{\theta}$ and
$n_{\pa/\pa\theta}$ are the numbers of $\theta$'s and
$\pa/\pa\theta$'s occurring in an invariant operator $\Lambda$,
respectively. It is not difficult to see that the degree of any
invariant operator is an integer. Furthermore, the vector space of
invariant operators can be supplied with the structure of a graded
algebra because \bea\nonumber \eps_{\Lambda_1\cdot \Lambda_2}
=\eps_{\Lambda_1}+ \eps_{ \Lambda_2} \,. \eea An action of the
parity transformation on any invariant operator can be defined in
the following way \bea\la{parL} \Lambda^\sp =
(-1)^{\eps_\Lambda}\, \Lambda\,. \eea Then, the inverse operator
$\S^{-1}$  must coincide with the parity-transformed operator
\bea\la{Sm1p} \S^{-1}(z_1,z_2) = \S(-z_1,-z_2)^\sp = \sum_k
a_k(-z_1,-z_2)(-1)^{\eps_{\Lambda_k}}\, \Lambda_k\,. \eea This
gives a very simple way of finding the inverse S-matrix. Comparing
eqs.(\ref{guc}), (\ref{rel2}) and (\ref{Sm1p}), we obtain the
following relations between  the coefficients of the S-matrix
\bea\la{rel3} &&a_k(z_1^*,z_2^*)^* = (-1)^{\eps_{\Lambda_k}}\,
a_k(-z_1,-z_2)\, .~~~ \eea Coefficients of the bound state
S-matrices we study in section \ref{secsmatr} do indeed satisfy
these relations.

\medskip

{\it Unitarity and Hermitian Analyticity.} To discuss the next
property, we need to distinguish the S-matrix $\S^{MN}$ which
describes the scattering of a $M$-particle bound state with a
$N$-particle one and acts as  an intertwiner \bea \la{smn}
\S^{MN}(z_1,z_2)=\sum_{k} a_k(z_1,z_2) \Lambda_k\,:~~~~\V^M\otimes
\V^N ~~\to~~ \V^M\otimes \V^N \,, \eea from the S-matrix $\S^{NM}$
which describes the scattering of a $N$-particle bound state with
a $M$-particle one
 \bea \la{snm}
 \S^{NM}(z_1,z_2)=\sum_{k} b_k(z_1,z_2) \Lambda_k^{NM}\,:~~~~\V^N\otimes
\V^M ~~\to~~ \V^N\otimes \V^M \,. \eea It is clear that the number
of invariant operators  in $\V^M\otimes \V^N$ and $\V^N\otimes
\V^M$ is the same, and, moreover, if $\Lambda_k$ acts on the
products $\Phi_{M}(w_a^1, \theta_\a^1)\Phi_{N}(w_a^2,
\theta_\a^2)$ then $\Lambda_k^{NM}$ acts on the products
$\Phi_{N}(w_a^1, \theta_\a^1)\Phi_{M}(w_a^2, \theta_\a^2)$ and it
is obtained from $\Lambda_k$ by means of exchange $w_a^1
\leftrightarrow w_a^2\,,\, \theta_\a^1 \leftrightarrow
\theta_\a^2$. Note also that in eq.(\ref{smn}) the variable $z_1$
is the torus rapidity variable of the $M$-particle bound state,
while in eq.(\ref{snm}) $z_1$ is the torus rapidity variable of
the $N$-particle bound state.

The coefficients $b_k$ of the S-matrix $\S^{NM}$  can be used to
construct an invariant operator which acts in $\V^M\otimes \V^N$
\bea \la{smn2} \S^{NM}_{21}(z_2,z_1)=\sum_{k} b_k(z_2,z_1)
\Lambda_k\,:~~~~\V^M\otimes \V^N ~~\to~~ \V^M\otimes \V^N \,, \eea
where the subscript 21 indicates that the order of $\Phi_{N}$ and
$\Phi_{M}$ was exchanged. We also exchanged $z_1$ and $z_2$ to
attach $z_1$ to the $M$-particle bound state. It is clear that the
combined exchange is just the graded permutation of the spaces
$\V^N$ and $\V^M$.

\smallskip
Then, the unitarity condition\footnote{Not to be confused with the
generalized physical unitarity (\ref{guc}).} states that the
operator $\S^{NM}_{21}(z_2,z_1)$ is the operator  inverse to the
S-matrix operator $\S^{MN}(z_1,z_2)$ \bea\la{uuc}
\S^{NM}_{21}(z_2,z_1)\cdot \S^{MN}(z_1,z_2) ={\mathbb I}\,. \eea
Comparing this formula with eq.(\ref{Sm1p}) for the inverse
S-matrix operator and eq.(\ref{rel3}), we get that the
coefficients $b_k$ are related to $a_k$ in the following way
\bea\la{rel4} b_k(z_2,z_1) = (-1)^{\eps_{\Lambda_k}}\,
a_k(-z_1,-z_2) = a_k(z_1^*,z_2^*)^* \,. \eea As the consequence of
eq.(\ref{uuc}) and eq.(\ref{guc}) one finds another relation
\bea\la{hera} \bS^{MN}(z_1^*,z_2^*)^\dagger =
\S^{NM}_{21}(z_2,z_1)\,, \eea which is the operator form of the
hermitian analyticity condition.

\smallskip

In the case of the scattering of $M$-particle bound states between
themselves, that is for $M=N$, there is one more relation between
the S-matrix and its inverse. To find it, we notice that for $M=N$
the action of the graded permutation ${\cal P}$ on the invariant
operators $\Lambda_i$ can be realized by means of exchange $w_a^1
\leftrightarrow w_a^2\,,\, \theta_\a^1 \leftrightarrow
\theta_\a^2$. Under this exchange an invariant operator transforms
to another invariant operator, the latter, due to $M=N$,  acts in
the same space \bea\nonumber \Lambda_k\to \Lambda_k^\spg\,. \eea
Then, the operator $\S^{MM}_{21}(z_2,z_1)\equiv\S_{21}(z_2,z_1)$
can be obtained from $\S^{MM}(z_1,z_2)\equiv\S(z_1,z_2)$ by
exchanging $z_1 \leftrightarrow z_2$ in the coefficients $a_k$ and
replacing $\Lambda_k$ by $ \Lambda_k^\spg$ \bea \la{sinv2}
\S_{21}(z_2,z_1)=\bS(z_2,z_1)^\spg=\sum_{k} a_k(z_2,z_1)
\Lambda_k^\spg \, . \eea Thus, in the operator form the unitarity
condition looks as \bea\la{nuc} \bS(z_2,z_1)^\spg\cdot \bS
(z_1,z_2)= \bI\, , \eea while the hermitian analyticity condition
takes the form \bea\la{hera2} \bS(z_1^*,z_2^*)^\dagger =
\bS(z_2,z_1)^\spg\,. \eea The last condition leads to additional
non-trivial relations between the coefficients $a_k$ which, as we
show in section \ref{secsmatr}, are satisfied by our bound state
S-matrices.

\medskip

{\it Crossing Symmetry.} The S-matrix also satisfies the crossing
symmetry relations. We discuss them in detail in section
\ref{seccross}.

\medskip

{\it Yang-Baxter Equation.} The $\su(2|2)$-invariant S-matrix for the full asymptotic
spectrum, the latter includes both the fundamental particles and
all the bound states, can be schematically represented as the
following block-diagonal matrix
\bea\nonumber {\mathbb S}=\left(
\begin{array}{ccccccc}
{\mathbb S}^{\bf 11} ~&~~&~~&~&~~&~~&   \\
~&{\mathbb S}^{\bf 12} ~&~&~&~~&~~& \\
~&~&~{\mathbb S}^{\bf 21}  ~&~&~&~~&~~ \\

\vspace{-0.1cm}

~&~&~& {\mathbb S}^{\bf 22} ~&~~&~~&    \\

\vspace{-0.3cm}

 &   & & &\cdot ~&~~&~~ \\
 \vspace{-0.3cm}
  &   & & & &\cdot&~~ \\
   &   & & &~&~~&\cdot
\end{array}
 \right) \, .
 \eea
Here ${\mathbb S}^{\bf 11}$ is the scattering matrix for the
fundamental multiplets, ${\mathbb S}^{\bf 12}$ is the scattering
matrix for the fundamental multiplet with the two-particle bound
state multiplet and so on. One should view ${\mathbb S}$ as an
operator acting in the direct sum
$$
{\mathscr V}={\mathscr V}^{\bf 1}\oplus {\mathscr V}^{\bf 2}\oplus
{\mathscr V}^{\bf 3}\oplus \ldots ,
$$
where  ${\mathscr V}^{M}$ is a carrier space for the irreducible
representation of $\su(2|2)_\cex$ corresponding to the
$M$-particle bound state. The condition of the factorised
scattering is equivalent to the Yang-Baxter equation
 \bea\la{ybe}
 \S_{23}\S_{13}\S_{12}=\S_{12}\S_{13}\S_{23} \, .\eea
Each side of the last equality is understood as an operator acting
in $\V^{\otimes 3}$.

\smallskip

The YB equation (\ref{ybe}) is equivalent to an infinite  number
of equations for the individual S-matrices. They are obtained by
restricting $\V^{\otimes 3}$ to the tensor product of three
irreducible representations
$\V^{M_1}\otimes\V^{M_2}\otimes\V^{M_3}$
 \bea\la{ybe2}
 \S_{23}^{M_2M_3}\S_{13}^{M_1M_3}\S_{12}^{M_1M_2}=\S_{12}^{M_1M_2}\S_{13}^{M_1M_3}\S_{23}^{M_2M_3} \, .\eea

\smallskip

In particular, in the next section we will consider a {\it consistent truncation} of
the theory which amounts to keeping the fundamental particles and two-particle
bound states only. The corresponding truncated S-matrix is \bea\nonumber
\check{\S}=\left( \begin{array}{cccc}
{\mathbb S}^{AA} ~&~& &\\
& {\mathbb S}^{AA}& &   \\
& & {\mathbb S}^{BA}  ~&~ \\
& & & {\mathbb S}^{BB}
\end{array}
 \right) \, .\eea
Here and in what follows for the sake of clarify we identify $\S^{\bf 11}\equiv
\S^{AA}$, etc., and we use the notation $\V^1\equiv\V^A$ for the fundamental representation, and $\V^2\equiv\V^B$ for the two-particle bound state representation.

\smallskip
Consistency of the truncation means that the
scattering of the lower-particle bound states (one- and two- in
our present context) between themselves should factorise
independently of the presence/absence of the higher-particle bound
states. Thus, we require $\check{\S}$ to satisfy the Yang-Baxter
equation (\ref{ybe}) which results in the following set of inequivalent equations for the
individual S-matrices \bea\nonumber \S^{AA}_{23}\S^{AA}_{13}\S^{AA }_{12} & = &\S^{AA }_{12}\S^{AA}_{13}\S^{AA}_{23} \, ,\\\nonumber
\S^{AB}_{23}\S^{AB}_{13}\S^{AA }_{12} &= &\S^{AA
}_{12}\S^{AB}_{13}\S^{AB}_{23} \, , \\\nonumber
\S^{BB}_{23}\S^{AB}_{13}\S^{AB}_{12} &= &\S^{AB
}_{12}\S^{AB}_{13}\S^{BB}_{23} \, , \\\nonumber
\S^{BB}_{23}\S^{BB}_{13}\S^{BB}_{12} &=&\S^{BB
}_{12}\S^{BB}_{13}\S^{BB}_{23} \, . \eea Here the first
and the last equations are the standard Yang-Baxter equations for
the S-matrices corresponding to scattering of the fundamental and
the two-particle bound states respectively. The second and the
third equations are defined in the spaces $\V^{A}\otimes
\V^{A}\otimes \V^{B}$ and $\V^{A}\otimes \V^{B}\otimes \V^{B}$ correspondingly.

\bigskip

In the matrix language the properties of the S-matrix operator we discussed above take the following form.

\medskip

{\it Physical Unitarity.}
\bea\nonumber
S_{12}^g(z_1^*,z_2^*)^\dagger \, S_{12}^g(z_1,z_2)=I\ &\Longleftrightarrow&\ S_{12}(z_1^*,z_2^*)^\dagger \, S_{12}(z_1,z_2)=I\,,
\eea
where $S_{12}^g$ and $S_{12}$ are defined in (\ref{Smg}) and  (\ref{Sm}).

\medskip

{\it CPT invariance.} \bea\nonumber S_{12}^g(z_1,z_2)^\st =
S_{12}^g(z_1,z_2)\ &\Longleftrightarrow&\ S_{12}(z_1,z_2)^\st =
I_{12}^g S_{12}(z_1,z_2)I_{12}^g\,, \eea where the superscript
$``T"$ denotes the usual matrix  transposition. Thus, the CPT
invariance just means that the graded S-matrix is symmetric.

\medskip

{\it Parity Transformation.} \bea\nonumber S_{12}^g(z_1,z_2)^{-1}
= I_{12}^gS_{12}^g(-z_1,-z_2)I_{12}^g\ \Longleftrightarrow\
S_{12}(z_1,z_2)^{-1} = S_{12}(-z_1,-z_2)\,, \eea where the parity
transformation acts on the S-matrix  and any invariant operator as
$\left(S_{12}^g\right)^\sp = I_{12}^gS_{12}^gI_{12}^g$. This also
means that in matrix language the grading of the tensor product of
matrix unities  $E_k{}^i \otimes E_L{}^J$ is defined to be
$(-1)^{\eps_k \eps_L +\eps_i\eps_J}$.

\medskip

{\it Unitarity and Hermitian Analyticity.}

In the general case $M\neq N$
\bea\nonumber
S^{g,NM}_{21}(z_2,z_1)S^{g,MN}_{12}(z_1,z_2) = I\ &\Longleftrightarrow&\ S^{NM}_{21}(z_2,z_1)S^{MN}_{12}(z_1,z_2) = I\,,
\eea
where
$$
S_{21}^{g,NM}(z_2,z_1)= P_{12}^gS_{12}^{g,NM}(z_2,z_1)P_{12}^g=\sum_{IKjl}\,
S^{NM}{}_{Ij}^{Kl}(z_2,z_1)\, (-1)^{\epsilon_K\epsilon_l+\epsilon_I\epsilon_j}\, E_l{}^j \otimes E_K{}^I \,,
$$
and $P_{12}^g$ is the graded permutation matrix. The hermitian
analyticity condition takes the form \bea\nonumber
S^{g,MN}_{12}(z_1^*,z_2^*)^\dagger = S_{21}^{g,NM}(z_2,z_1)\
&\Longleftrightarrow&\ S^{MN}_{12}(z_1^*,z_2^*)^\dagger =
S_{21}^{NM}(z_2,z_1)\,. \eea In the  case $M= N$, we get
$\S^{MM}(z_1,z_2)\to S_{12}(z_1,z_2)$, and  the unitarity and
hermitian analyticity conditions take the following form
\bea\nonumber
S_{21}^g(z_2,z_1) \, S_{12}^g(z_1,z_2)=I\ &\Longleftrightarrow&\ S_{21}(z_2,z_1) \, S_{12}(z_1,z_2)=I\,,\\
\nonumber
S^{g}_{12}(z_1^*,z_2^*)^\dagger = S_{21}^{g}(z_2,z_1)\ &\Longleftrightarrow&\ S_{12}(z_1^*,z_2^*)^\dagger = S_{21}(z_2,z_1)\,,
\eea
where
$$
S_{21}^g(z_2,z_1) = P_{12}^gS_{12}^g(z_2,z_1)P_{12}^g=\sum_{ikjl}
S_{ij}^{kl}(z_2,z_1) (-1)^{\epsilon_k\epsilon_l+\epsilon_i\epsilon_j}\, E_l{}^j \otimes E_k{}^i \,,
$$
and this means that the action of the graded permutation operator ${\cal P}$ is just given by the conjugation by the graded permutation matrix.

\medskip

{\it Yang-Baxter Equation.}
The operator YB equation
(\ref{ybe}) has the same matrix form only for the S-matrix $S_{12}$ defined by (\ref{Sm}):
 $$
  S_{23}S_{13}S_{12}=S_{12}S_{13}S_{23}\,,$$
where to simplify the notation we omit the explicit dependence of
the S-matrices on $z_i$ and $M_i$.
  To show this, we notice that the matrix form of the operators $\S_{12}, \S_{13}$ and $\S_{23}$ is
  \bea\nonumber
S_{12}^{\rm matr} = S_{12}^g\,, \quad S_{13}^{\rm matr} = I_{23}^gS_{13}^g I_{23}^g\,, \quad S_{23}^{\rm matr} =  I_{12}^gI_{13}^gS_{23}^g I_{13}^g I_{12}^g\,,
\eea
where $S_{ij}^g$ denotes the usual embedding of the matrix $S^g$ into the product of three spaces.
We see that the graded form of the S-matrix satisfies the following graded YB equation
\bea\nonumber
 I_{12}^gI_{13}^gS_{23}^g I_{13}^g I_{12}^g I_{23}^gS_{13}^g I_{23}^gS_{12}^g=S_{12}^g I_{23}^gS_{13}^g I_{23}^g I_{12}^gI_{13}^gS_{23}^g I_{13}^g I_{12}^g \,.
\eea In terms of the S-matrix $S_{12}=I_{12}^gS_{12}^G$ the
previous equation takes the form \bea\nonumber
 I_{12}^gI_{13}^gI_{23}^gS_{23}I_{12}^g I_{23}^gS_{13} I_{23}^g I_{12}^gS_{12}= I_{12}^gS_{12} I_{23}^g I_{13}^gS_{13} I_{12}^gI_{13}^g S_{23} I_{13}^g I_{12}^g \,.
\eea Taking into account the identities \bea\nonumber I_{12}^g
I_{23}^gS_{13} = S_{13} I_{12}^g I_{23}^g\,,\quad I_{12}^gI_{13}^g
S_{23} =S_{23}I_{13}^g  I_{12}^g\,, \eea we conclude that $S_{12}$
satisfies the usual YB equation. This completes our discussion of
the properties of the  S-matrix.

\section{S-matrices}\la{secsmatr}
In this section we discuss the explicit construction and properties of the three  S-matrices describing the scattering of fundamental particles and two-particle bound states.

\subsection{The S-matrix $\S^{AA}$} \la{secs11}

As the warm-up exercise, here we will explain how to obtain the
known scattering matrix $\S^{AA}$ for fundamental particles
\cite{B,AFZ} in the framework of our new superspace approach.

\smallskip

According to our general discussion in section 3, the scattering
matrix $\S^{AA}$ is an $\su(2)\oplus \su(2)$ invariant
differential operator which acts on the tensor product of two
fundamental multiplets \bea\nonumber \S^{AA}:~~~~~ \V^A\otimes \V^A ~~\to
~~\V^A\otimes \V^A\, . \eea Recall that
$$
\V^A\otimes \V^A =\W^2\, ,
$$
where $\W^2$ is a long irreducible supermultiplet of dimension 16
\cite{Bn}. Under the action of $\su(2)\oplus \su(2)$ this
supermultiplet branches as follows \bea\label{decW2}
\W^2&=&V^0\times V^0 +V^0\times V^0+V^1\times V^0+\\
\nonumber &&~~~~~~~~~~~~~~~~+V^0\times V^1+V^{1/2}\times
V^{1/2}+V^{1/2}\times V^{1/2}. \eea The 16-dimensional basis of
the tensor product space $\V^A\otimes \V^A$
$$
~~~~~~~~~~~~~~\hskip 2.5cm\V^A\otimes \V^A={\rm Span
}\Big\{w_a^1w_b^2\, , w_a^1 \theta_{\a}^2\, , w_a^1
\theta_{\a}^2\, , \theta_{\a}^1\theta_{\beta}^2\Big\},
~~~~~~~~a=1,2; ~~~\a=3,4
$$
can be easily adopted to the decomposition (\ref{decW2}):
\bea
\label{W2}\begin{aligned} w_a^1w_b^2&=
\frac{1}{2}(w_a^1w_b^2-w_b^1w_a^2)+\frac{1}{2}(w_a^1w_b^2+w_b^1w_a^2)
~~\rightarrow~~ (V^0+V^1)\times V^0 \\
w_a^1 \theta_{\a}^2 &~~\rightarrow~~  V^{1/2}\times
V^{1/2} \\
w_a^2 \theta_{\a}^1 &~~\rightarrow~~   V^{1/2}\times
V^{1/2}\\
\theta_{\a}^1\theta_{\beta}^2 &=
\frac{1}{2}(\theta_{\a}^1\theta_{\b}^2-\theta_{\b}^1\theta_{\a}^2)+
\frac{1}{2}(\theta_{\a}^1\theta_{\b}^2+\theta_{\b}^1\theta_{\a}^2)~~\rightarrow~~
V^0\times (V^0 + V^1) \end{aligned} \eea The space $\D_2$ of the
second order differential operators dual to $\W^2$ branches under
$\su(2)\oplus \su(2)$ in the same way: \bea\label{decdualW2}
\D_2&=&D^0\times D^0 +D^0\times D^0+D^1\times D^0+\\
\nonumber &&~~~~~~~~~~~~~~~~+D^0\times D^1+D^{1/2}\times
D^{1/2}+D^{1/2}\times D^{1/2}. \eea
 A basis adopted to this
decomposition is formed by the following differential operators
 \bea \label{dW2} \begin{aligned}
\e^{cd}{\pa\ov\pa w_1^c}{\pa \ov \pa
w_2^d} & ~~\rightarrow~~ D^0 \times D^0\,,\quad
\eps_{\a\b}{\pa\ov\pa \theta_1^\a}{\pa\ov\pa \theta_2^\b} ~~\rightarrow~~
D^0 \times D^0\\
\frac{\pa}{\pa w_a^2}\frac{\pa}{\pa \theta_{\a}^1} &
~~\rightarrow~~D^{1/2} \times D^{1/2}\,,\quad
\frac{\pa}{\pa w_a^1}\frac{\pa}{\pa \theta_{\a}^2}  ~~\rightarrow~~D^{1/2} \times D^{1/2}\\
{\pa\ov\pa w_1^a}{\pa\ov\pa w_2^b}+{\pa\ov\pa w_1^b}{\pa\ov\pa
w_2^a} &~~\rightarrow~~D^1 \times D^0\,,\quad
{\pa\ov\pa \theta_1^\a}{\pa\ov\pa \theta_2^\b}+{\pa\ov\pa
\theta_1^\b}{\pa\ov\pa \theta_2^\a} ~~\rightarrow~~ D^0
\times D^1
\end{aligned}
\eea

\smallskip
The S-matrix we are looking for is an element of the space
$\W^2\otimes \D_2$ which is invariant under the action of
$\su(2)\oplus \su(2)$. We obviously have {\small \bea \nonumber
\W^2\otimes \D_2 &=& (V^0\times V^0 +V^0\times V^0+V^1\times
V^0+V^0\times V^1+V^{1/2}\times V^{1/2}+V^{1/2}\times
V^{1/2})\\
&\otimes & \nonumber (D^0\times D^0 +D^0\times D^0+D^1\times
D^0+D^0\times D^1+D^{\frac{1}{2}}\times D^{1/2}+D^{1/2}\times
D^{1/2}) \eea}
 The invariants of $\su(2)\oplus \su(2)$ arising in
this tensor product decomposition are easy to count. Indeed, since
$V^j\otimes D^j$ contains the singlet component (invariant), the
$\su(2)\oplus \su(2)$ invariants simply correspond to the singlet
components of the spaces $V^j\otimes D^j\times V^k\otimes D^k$ in
the tensor product decomposition of $\W^2\otimes \D_2$. Thus, the
invariant subspace is generated by \bea \nonumber {\rm
Inv}(\W^2\otimes \D_2)&=& 4(V^0\otimes D^0)\times (V^0\otimes
D^0)+4 (V^{1/2}\otimes D^{1/2})\times (V^{1/2}\otimes
D^{1/2})+\\\nonumber &~&+ (V^1\otimes D^1)\times (V^0\otimes D^0)+
(V^0\otimes D^0)\times (V^1\otimes D^1) , \eea where integers in
front of the spaces in the r.h.s of the last formulas indicate the
corresponding multiplicities. Therefore, in the present case we
find 10 invariant elements $\Lambda_k$ in the space $\W^2\otimes
\D_2$. Their explicit form can be easily found by using the bases
(\ref{W2}) and (\ref{dW2}): \bea \label{L11}
\begin{aligned}
&\Lambda_1=\frac{1}{2}(w_a^1w_b^2+w_b^1w_a^2)\frac{\pa^2}{\pa
w_a^1 \pa w_b^2} \,, \qquad\, \qquad\Lambda_7=\eps^{ab}w_a^1w_b^2\eps_{\a\b}\frac{\pa^2}{\pa\theta_{\b}^2\pa\theta_{\a}^1}~~~~~\\
&\Lambda_2=\frac{1}{2}(w_a^1w_b^2-w_b^1w_a^2)\frac{\pa^2}{\pa
w_a^1 \pa w_b^2}\,, \qquad\,  \qquad\Lambda_8=\eps^{\a\b}\theta_{\a}^1\theta_{\b}^2
\eps_{ab}\frac{\pa^2}{\pa w_a^1 \pa w_b^2} \\
&\Lambda_3=\frac{1}{2}(\theta_{\a}^1\theta_{\b}^2+\theta_{\b}^1\theta_{\a}^2)\frac{\pa^2}{ \pa {\theta}_\b^2\pa
\theta_{\a}^1}  \,, \qquad \qquad \quad~ \Lambda_9=w_a^1\theta_{\a}^2\frac{\pa^2}{\pa w_a^2\pa
\theta_{\a}^1}\\
&\Lambda_4=\frac{1}{2}(\theta_{\a}^1\theta_{\b}^2-\theta_{\b}^1\theta_{\a}^2)\frac{\pa^2}{\pa {\theta}_\b^2 \pa
\theta_{\a}^1}\,, \qquad \qquad \quad~  \Lambda_{10}=w_a^2\theta_{\a}^1\frac{\pa^2}{\pa w_a^1\pa
\theta_{\a}^2} \\
&\Lambda_5=w_a^1\theta_{\a}^2\frac{\pa^2}{\pa
w_a^1\pa\theta_{\a}^2}\,,\\
&\Lambda_6=w_a^2\theta_{\a}^1\frac{\pa^2}{\pa
w_a^2\pa\theta_{\a}^1}\,.
\end{aligned}
\eea
Thus, the S-matrix is the
following differential operator in the space $\V^A\otimes \V^A$:
\bea\nonumber \S^{AA}(z_1,z_2)=\sum_{k=1}^{10}a_k\Lambda_k \, \eea The
differential operators $\Lambda_1,\ldots \Lambda_6$ correspond to
the diagonal invariants in $\W^2\otimes \D_2$. Their
normalization has been chosen in such a way that for
$a_1=a_2=\ldots =a_6=1$ and  $a_7=a_8=a_9=a_{10}=0$ the operator
$\S^{AA}$ coincides with an identity operator.

\smallskip
The unknown coefficients $a_k$ can be now determined from the
permutation relations of $\S^{AA}$ with the supersymmetry
generators. We find
 \bea
\nonumber
~\qquad\qquad  a_1=1\,\hspace{13cm}
\eea

 \vspace{-0.6cm}

\bea
\nonumber
~\qquad\qquad
a_2=2\,\frac{(x^+_1-x^+_2) (x^-_1
   x^+_2-1)x^-_2}{(x^+_1 - x^-_2)(x^-_1 x^-_2-1)
    x^+_2}-1\,,\hspace{13cm}
 \eea

 \vspace{-0.6cm}

\bea
\nonumber
~\qquad\qquad
a_3=\frac{x^+_2-x^-_1}{x^-_2-x^+_1}
\frac{\tilde{\eta}_1
  \tilde{\eta}_2}{ \eta_1\eta_2} ,\hspace{13cm}
  \eea

 \vspace{-0.6cm}

\bea
\nonumber
~\qquad \qquad
 a_4=\frac{(x^-_1-x^+_2)}{(x^-_2-x^+_1)}\frac{\tilde{\eta}_1
  \tilde{\eta}_2}{ \eta_1\eta_2} - 2\,\frac{
   (x^-_2 x^+_1-1) (x^+_1-x^+_2)
  x^-_1 }{(x^-_1 x^-_2-1)
   (x^-_2-x^+_1) x^+_1} \frac{\tilde{\eta}_1
  \tilde{\eta}_2}{ \eta_1\eta_2} ,\hspace{13cm}
  \eea

 \vspace{-0.6cm}

\bea
\nonumber
~\qquad \qquad
a_5=\frac{x^-_2-x^-_1}{x^-_2-x^+_1 } \frac{\tilde{\eta}_2}{\eta _2}  ,\hspace{13cm}
\eea

 \vspace{-0.6cm}

\bea
\nonumber
~\qquad \qquad
a_6=\frac{x^+_1-x^+_2}{x^+_1-x^-_2}\frac{\tilde{\eta}_1}{\eta_1},\hspace{13cm}
 \eea

 \vspace{-0.6cm}

 \bea
 \nonumber
~\qquad \qquad
 a_7 = -\frac{i (x^-_1-x^+_1)
   (x^-_2-x^+_2)
   (x^+_1-x^+_2)}{(x^-_1
   x^-_2-1) (x^-_2-x^+_1) }\frac{1}{\eta
   _1 \eta _2},\hspace{13cm}
\eea

 \vspace{-0.6cm}

\bea
\nonumber
~\qquad \qquad
a_8 =\frac{i x^-_1 x^-_2
   (x^+_1-x^+_2)}{(x^-_1 x^-_2-1)
   (x^-_2-x^+_1) x^+_1
   x^+_2}\tilde{\eta}_1\tilde{\eta}_2,\hspace{13cm} \eea

 \vspace{-0.6cm}

\bea
\nonumber
~\qquad \qquad
a_9 =\frac{x^+_1-x^-_1}{x^+_1-x^-_2}\frac{\tilde{\eta}_2}{\eta_1},\hspace{13cm} \eea

 \vspace{-0.6cm}

\bea
\nonumber
~\qquad \qquad
a_{10}=\frac{x^-_2-x^+_2}{x^-_2-x^+_1}\frac{\tilde{\eta}_1}{\eta_2}\,,\hspace{13cm}
 \eea
The coefficients $a_k$ are determined up to an overall scaling
factor, and we normalize them in a canonical way by setting
$a_1=1$.  The parameters $\eta_k$ are not fixed by the invariance
condition. They are determined by imposing the generalized
unitarity condition and the YB equation \cite{AFTBA}, and are
given by the following formulas \bea\la{etak} \eta_1 =
e^{ip_2/2}\eta(z_1,1)\,,\quad \eta_2 =\eta(z_2,1)\,,\quad
\tilde{\eta}_1 = \eta(z_1,1)\,,\quad \tilde{\eta}_2 =
e^{ip_1/2}\eta(z_2,1)\,, \eea where $\eta(z,M)$ is defined by
(\ref{etaz}).

\smallskip

The S-matrix satisfies all the properties we discussed in the previous section. First,  the physical unitarity condition $\bS^\dagger\cdot \bS = \bI$ can be easily checked by using the explicit form of the coefficients $a_i$, and the hermitian conjugation conditions
\bea\nonumber
\left(\Lambda_i\right)^\dagger = \Lambda_i\,,\quad i=1,\ldots ,6\,;\qquad \left(\Lambda_7\right)^\dagger = \Lambda_8\,,\ \ \left(\Lambda_9\right)^\dagger = \Lambda_{10}\,.
\eea
Moreover, with the choice of $\eta_i$ (\ref{etak}) made in \cite{AFTBA}, the S-matrix also satisfies the generalized unitarity condition
$
\bS(z_1^*,z_2^*)^\dagger\cdot \bS (z_1,z_2)= \bI\,.
$

\smallskip

Second, the S-matrix is a symmetric operator
$
\bS(z_1,z_2)^\st = \bS(z_1,z_2)\,,
$
and the coefficients $a_i$  satisfy the following relations
\bea\la{rel112}
&&a_7(z_1,z_2) = a_8(z_1,z_2)\,,\quad a_9(z_1,z_2)=a_{10}(z_1,z_2)\,,~~~~~~~~
\eea
if one uses $\eta_i$ given by  (\ref{etak}).

\smallskip

Third, by using the action of the parity transformation on an invariant operator defined in (\ref{parL}), we find
\bea\nonumber
\Lambda_i^\sp  = \Lambda_i \,,\ i=1,\ldots ,6; 9,10\,;\quad \Lambda_7^\sp  = - \Lambda_7\,,\
\Lambda_8^\sp = - \Lambda_8\,,~~~~~
\eea
and check that
the inverse operator $\S^{-1}$  is equal to the parity-transformed S-matrix operator
$
\S^{-1}(z_1,z_2) = \S(-z_1,-z_2)^\sp \,,
$
and the relations (\ref{rel3}) between  on the coefficients of the S-matrix are satisfied.

\smallskip

Fourth, as was discussed in the previous section, in the operator formalism the action of the graded permutation on the invariant operators $\Lambda_i$ can be realized by means of the exchange $w_a^1 \leftrightarrow w_a^2\,,\, \theta_\a^1 \leftrightarrow \theta_\a^2$. By using the explicit form of $\Lambda_i$ one finds
\bea\la{permL}
\Lambda_i^\spg =\Lambda_i \,,\ i=1,2,3 ,4\,;\quad \Lambda_5^\spg = \Lambda_6\,,\
\Lambda_7^\spg = - \Lambda_7\,,\
\Lambda_8^\spg = - \Lambda_8\,,\ \Lambda_9^\spg = \Lambda_{10}\,,~~~~~
\eea
and verifies  the unitarity condition $\bS(z_2,z_1)^\spg\cdot \bS (z_1,z_2)= \bI\,,
$
and the hermitian analyticity condition
$
\bS(z_1^*,z_2^*)^\dagger = \bS(z_2,z_1)^\spg\,,$
which
implies the following relations between the coefficients $a_i$ of the S-matrix
\bea\la{rel1}
&&a_i(z_2,z_1)^* = a_i(z_1^*,z_2^*)\,,\ i=1,2,3 ,4\,;\quad a_5(z_2,z_1)^* = a_6(z_1^*,z_2^*)\,,\ \\ \nonumber &&a_7(z_2,z_1)^* = - a_8(z_1^*,z_2^*)\,,\  \ a_9(z_2,z_1)^* = a_{9}(z_1^*,z_2^*)\,,\ a_{10}(z_2,z_1)^* =  a_{10}(z_1^*,z_2^*)\,.~~~~~~~~
\eea
Comparing relations (\ref{rel1}) and  (\ref{rel112}), we conclude that the coefficients $a_i$ also satisfy the following relations
\bea\nonumber
&&a_7(z_2,z_1)^* = -a_7(z_1^*,z_2^*)\,,\ a_8(z_2,z_1)^* = - a_8(z_1^*,z_2^*)\,,\ \\ \nonumber &&a_9(z_2,z_1)^* = a_{10}(z_1^*,z_2^*)\,,\ a_{10}(z_2,z_1)^* =  a_{9}(z_1^*,z_2^*)\,.~~~~~~~~
\eea

\smallskip

Finally, given the $\eta$'s, one can check
the fulfilment of the Yang-Baxter equation
 \bea\la{YBs11}
  \S_{23}^{AA}\S_{13}^{AA}\S_{12}^{AA}=\S_{12}^{AA}\S_{13}^{AA}\S_{23}^{AA} \, ,\eea
where each side of the last equality is understood as an operator
acting in $\V^A{^{\otimes 3}}$,  the latter being identified with
the product of three superfields $\Phi_{1}(w_a^1, \theta_\a^1)$,
$\Phi_{2}(w_a^2, \theta_\a^2)$ and $\Phi_{3}(w_a^3, \theta_\a^3)$.
This completes the construction of the graded fermionic
$\bS$-operator which describes the scattering of two fundamental
multiplets.

\smallskip

By using the general procedure outlined in section 3, we can also
construct the matrix representation for the $\su(2)\oplus \su(2)$
invariant differential operators acting in the tensor product of
fundamental particle representations. For reader's convenience we
present them below
\bea\nonumber~\qquad \qquad  \Lambda_1 &=& E_{1111}+\frac{1}{2
   }
   E_{1122}+\frac{1
   }{2}
   E_{1221}+\frac{1
   }{2}
   E_{2112}+\frac{1
   }{2}
   E_{2211}+E_{2222}\hspace{13cm}
   \eea

 \vspace{-0.8cm}

\bea
\nonumber~\qquad \qquad
\Lambda_2 &=& \frac{1}{2}
   E_{1122}-\frac{1
   }{2}
   E_{1221}-\frac{1
   }{2}
   E_{2112}+\frac{1
   }{2} E_{2211}\hspace{13cm}
\eea

 \vspace{-0.8cm}

\bea
\nonumber~\qquad \qquad
\Lambda_3 &=& E_{3333}+\frac{1}{
   2}
   E_{3344}+\frac{1
   }{2}
   E_{3443}+\frac{1
   }{2}
   E_{4334}+\frac{1
   }{2}
   E_{4433}+E_{4444}\hspace{13cm}
\nonumber
\eea

 \vspace{-0.8cm}

\bea \nonumber~\qquad \qquad  \Lambda_4 &=& \frac{1}{2}
   E_{3344}-\frac{1
   }{2}
   E_{3443}-\frac{1
   }{2}
   E_{4334}+\frac{1
   }{2} E_{4433}\hspace{13cm}
\eea

 \vspace{-0.8cm}

\bea
\nonumber~\qquad \qquad
\Lambda_5 &=& E_{1133}+E_{1144}+E_{2233}+E_{224
   4}\hspace{13cm}
\eea

 \vspace{-0.8cm}

\bea
\nonumber~\qquad \qquad
\Lambda_6 &=& E_{3311}+E_{3322}+E_{4411}+E_{442
   2}\hspace{13cm}
\eea

 \vspace{-0.8cm}

\bea
\nonumber~\qquad \qquad
\Lambda_7 &=& E_{1324}-E_{1423}-E_{2
   314}+E_{2413}\hspace{13cm}
\eea

 \vspace{-0.8cm}

\bea
\nonumber~\qquad \qquad
\Lambda_8 &=& E_{3142}-E_{3241}-E_{4132}+E_{4231}\hspace{13cm}
\eea

 \vspace{-0.8cm}

\bea
\nonumber~\qquad \qquad
\Lambda_9 &=& E_{1331}+E_{1441}+E_{2332}+E_{2442}\hspace{13cm}
\eea

 \vspace{-0.8cm}

\bea
\nonumber~\qquad \qquad
\Lambda_{10} &=& E_{3113}+E_{3223}+E_{4114}+E_{4224} \hspace{13cm}
\eea
Here the symbols $E_{kilj}$ can be equal to either $E_k{}^i \otimes E_l{}^j$ or to $(-1)^{\epsilon_k\epsilon_l}\, E_k{}^i \otimes E_l{}^j$, where $E_k{}^i \equiv  E_{ki}$ are the
standard $4\times 4$ matrix unities. Computing the matrix
representation for $\S^{AA}$ by using $E_{kilj}=(-1)^{\epsilon_k\epsilon_l}\, E_k{}^i \otimes E_l{}^j$ we find that it precisely coincides
with the $\su(2|2)$-invariant S-matrix found in \cite{AFZ,AFTBA}.

In the matrix language the above-described properties of the
S-matrix operator take the form discussed in the previous section.

\subsection{The S-matrix $\S^{AB}$} \la{secs12}
The S-matrices $\S^{AB}$ and $\S^{BA}$ for scattering of a
fundamental particle $A$ with the two-particle bound state $B$ are
$\su(2)\oplus \su(2)$ invariant third-order differential operators
which act as intertwiners \bea \nonumber &&\S^{AB}:~~~~\V^A\otimes
\V^B ~~\to~~ \V^A\otimes \V^B \\
\nonumber &&\S^{BA}:~~~~\V^B\otimes \V^A ~~\to~~ \V^B\otimes
\V^A\, \eea
Below we discuss the S-matrix $\S^{AB}$ in detail and
comment on its relation to $\S^{BA}$.

\smallskip

According to section 3.2, the tensor product $\V^A\otimes \V^B$ is
isomorphic to a long supermultiplet $\W^3$ of dimension 32. The
total number of invariant differential operators $\Lambda_k$ for
this case is 19 and, therefore, \bea \label{SABa}
\S^{AB}=\sum_{k=1}^{19}a_k \Lambda_k\, . \eea Representations for
$\Lambda_k$ by third-order differential operators and by rank 32
matrices are listed in appendix \ref{ASAB1}. The invariants
$\Lambda_1,\ldots ,\Lambda_9$ are diagonal and they are normalized
to provide the orthogonal decomposition of unity.

\smallskip

Since the tensor product $\V^A\otimes \V^B$ is irreducible, all
the coefficients $a_k$ in (\ref{SABa}) can be determined, up to an
overall scaling factor and parameters $\eta_k$, from the invariance conditions involving
the generators ${\mathbb Q}$ and ${\mathbb Q}^{\dagger}$. The
corresponding solution for the coefficients $a_k$ is presented in
appendix \ref{ASAB2}.

\smallskip

Now we are ready to discuss the properties of $\S^{AB}$. First,
under the hermitian conjugation the operators $\Lambda_k$
transform as follows \bea\nonumber (\Lambda_i)^{\dagger}=\Lambda_i,
~~~i=1,\ldots, 9;~~~~~~~
(\Lambda_{10+2k})^{\dagger}=\Lambda_{11+2k}\, ,~~~~~ k=0,\ldots, 4
. \eea These transformation rules allow one to check the
generalized unitarity condition. Second, with our choice of
$\eta$'s the operator $\S^{AB}$ is symmetric, i.e.
$(\S^{AB})^\st=\S^{AB}$. Consequently, the coefficients $a_k$
satisfy the following relations
$$
a_{10+2k}(z_1,z_2)=a_{11+2k}(z_1,z_2)\,
$$
for $k=0,\ldots, 4$.

\smallskip

Third,  the parity transformation  (\ref{parL}) acts on the invariant operators as follows
\bea\nonumber
\Lambda_i^\sp  = \Lambda_i \,,\ i=1,\ldots ,9;14,15,18,19\,;\quad \Lambda_k^\sp  = - \Lambda_k\,,\ k=10,11,12,13,16,17\,,~~~~~
\eea
and
the inverse S-matrix  is equal to the parity-transformed S-matrix, and the relations (\ref{rel3}) between  on the coefficients of the S-matrix are satisfied.

\smallskip

Fourth,  the unitarity condition (\ref{uuc}) $
\S^{BA}_{21}(z_2,z_1)\cdot \S^{AB}(z_1,z_2) ={\mathbb I}
$
allows us to find the coefficients of the S-matrix $S^{BA}$ by using
the relations (\ref{rel4})
\bea\nonumber
b_k(z_1,z_2) = (-1)^{\eps_{\Lambda_k}}\, a_k(-z_2,-z_1) = a_k(z_2^*,z_1^*)^* \,,
\eea
and check that the resulting S-matrix $S^{BA}$ satisfies the invariance conditions
(\ref{invcon}).

\smallskip

\begin{figure}[t]
\begin{center}
\hskip -2.5cm
\includegraphics*[width=0.9\textwidth]{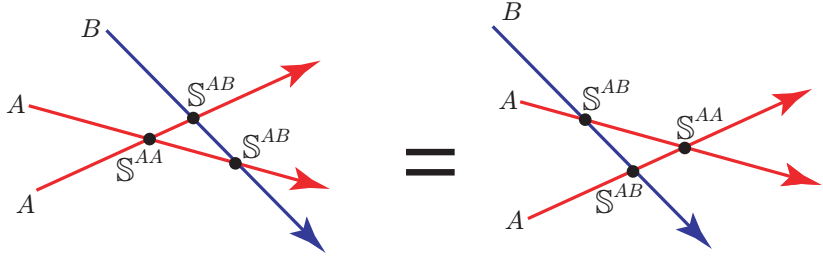}~~~~~~~~
\end{center}
\vskip -5.5cm \caption{Factorisation of the three-particle
S-matrix for the scattering process involving two fundamental
particles and one two-particle bound state.
 } \label{torus3}
\end{figure}

Having found the S-matrices $\S^{AB}$ and $\S^{BB}$ one can check
the factorization property of the three-particle S-matrix
involving two fundamental particles and one two-particle bound
state, see Fig. 1. The corresponding Yang-Baxter equation \bea\nonumber
\S^{AB }_{23}\S^{AB }_{13}\S^{AA}_{12} &= &\S^{AA }_{12}\S^{AB
}_{13}\S^{AB }_{23} \,  \eea is an operator identity on the triple
tensor product $\V^A\otimes \V^A\otimes \V^B$, which we have shown
to hold.

Finally, one can check that on solutions of the equation \bea
\la{boundSAB} x_1^-=y_2^+ \eea the coefficients
$a_4,a_7,a_9,a_{16},a_{17}$ vanish. As the consequence, the
S-matrix $\S^{AB}$ degenerates and has  rank 12. This precisely
corresponds to forming a three-particle bound state of dimension
12. Under the condition (\ref{boundSAB}) the 32-dim tensor product
representation $\V^A\otimes\V^B=\W^3$ becomes reducible \cite{Bn}.
However, as was discussed in the Introduction, this representation
is indecomposable. The invariant subspace is a 20-dim short
representation, while  the three-particle bound state
representation should be understood as a factor representation.

\subsection{The S-matrix $\S^{BB}$}\la{secs22}
The S-matrix $\S^{BB}$ is the following differential operator in
the space $\V^B\otimes \V^B$: \bea\nonumber
\S^{BB}(z_1,z_2)=\sum_{k=1}^{48}a_k\Lambda_k \, ,\eea where  48
invariant operators $\Lambda_k$ and the corresponding matrices are
listed in the appendix \ref{ASBB}. In particular,
$\Lambda_1,\ldots \Lambda_{16}$ correspond to the diagonal
invariants. Their normalization has been chosen in such a way that
for $a_1=a_2=\ldots =a_{16}=1$ and $a_{17}=\ldots =0$ the operator
$\S^{BB}$ coincides with the identity operator.

\smallskip

\begin{figure}[t]
\begin{center}
\includegraphics*[width=1.0\textwidth]{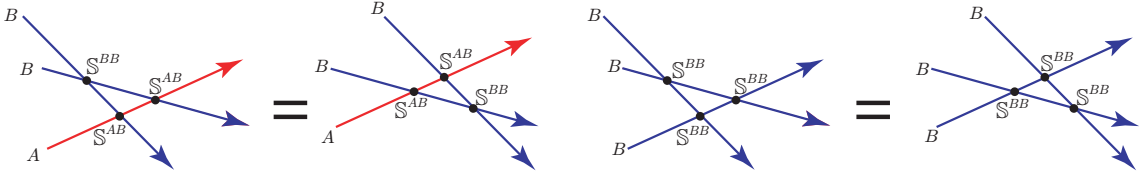}~~~~~~~~
\end{center}
\vskip -3cm\caption{The Yang-Baxter equation corresponding to the
scattering process of a fundamental particle $A$ with two
two-particle bound states $B$.
 } \label{torus3b}
\end{figure}

The operator $\S^{BB}$ provides the first example of an invariant
differential operator for which the coefficients $a_k$ cannot be
fully determined from the commutation relations with supersymmetry
generators.  As was already mentioned in section 3.2, the tensor
product decomposition $\V^B\otimes \V^B$ comprises two long
multiplets $\W^2$ and $\W^4$. As a consequence, a solution of
the invariance condition is a two-parametric family
$$
a_k\equiv a_k(a_1,a_2;\sfa,\sfb,\sfc,\sfd)\, ,~~~~ k>2\, ,
$$
where we have chosen the independent coefficients $a_1$ and $a_2$
to parametrize our solution. In addition, $a_k$ depend in a very
complicated way on the parameters $\sfa,\sfb,\sfc,\sfd$ which
describe four representations of the centrally extended $\su(2|2)$
involved in the invariance condition. It appears convenient to
explicitly distinguish between two independent solutions by
introducing \bea a_1^f &=&1\, , ~~~~a_2^f=0\, , ~~~a_k^f\equiv
a_k(1,0;\sfa,\sfb,\sfc,\sfd)\, ,~~~k>2, \nonumber \\
a_1^s&=&0\, , ~~~~a_2^s=1\, , ~~~~a_k^s\equiv
a_s(0,1;\sfa,\sfb,\sfc,\sfd)\, ,~~~k>2, \nonumber
 \eea
so that a general solution is a linear combination of these two.
Consequently, we introduce two differential operators \bea\nonumber
\S_1^{BB}&=&\sum_{k=1}^{48}a_k^{ f}\Lambda_k\, ,\\\nonumber
\S_2^{BB}&=&\sum_{k=1}^{48}a_k^{s}\Lambda_k \, .\eea Thus, the
S-matrix $\S^{BB}$ with the canonical normalization $a_1=a_1^f=1$
can be now written as \bea \la{SBBfs}
\S^{BB}=\S_1^{BB}+q\,\S_2^{BB}\, , \eea where $q\equiv a_2$ is a
single parameter which cannot be determined from the invariance
condition. Below we will find it by solving the YB equations
involving $\S^{AB}$ and $\S^{BB}$.

\smallskip

First we consider the YB equation involving one fundamental
particle and two bound states \bea
\S^{BB}_{23}\S^{AB}_{13}\S^{AB}_{12} &= &\S^{AB
}_{12}\S^{AB}_{13}\S^{BB}_{23} \, , \la{YB1} \eea  where $\S^{AB}$
is given by eq.(\ref{SBBfs}). Evaluated on basis elements of the
tensor product $\V^{A}\otimes \V^{B}\otimes V^{B}$ this operator
equation is equivalent to 256 relations between the coefficients
of the S-matrices involved. Analysis of these relations allows one
to extract the following form of the unknown coefficient $q$ \bea
q=-\frac{ 3a_{11}^{AB}(a_1^{f}+a_3^{f}-2a_{13}^{f}) -8(2a^{
AB}_1+a^{AB}_2-3a^{AB}_6)a_{24}^{f}}
{3a_{11}^{AB}(b_1^{s}+b_3^{s}-2b_{13}^{s})-8(2a^{AB}_1+a^{AB}_2-3a^{AB}_6)b_{24}^{s}}\,
,\label{solq} \eea where by $a_k^{AB}$ we denoted the coefficients
of $\S^{AB}$ to distinguish them from $a_k$ of $\S^{BB}$.

\smallskip

This result for $q$ might seem rather odd. Indeed, the coefficient
$q$ must also guarantee the fulfilment of another YB equation,
which involves the bound states only \bea
\S^{BB}_{23}\S^{BB}_{13}\S^{BB}_{12} &= &\S^{BB
}_{12}\S^{BB}_{13}\S^{BB}_{23} \, . \la{YB2} \eea This relation
implies that $q$ should be a function of the variables $z_1,z_2$
living on the rapidity tori associated to two bound states, while
solution (\ref{solq}) involves the rapidity torus of a fundamental
particles. This issue can be resolved by noting that the following
important identity takes place \bea
2a^{AB}_1+a^{AB}_2-3a^{AB}_6=3\frac{(1-\xpbt\xmbt)}{\sqrt{\xpbt\xmbt}}\frac{(\xpbt+\xmbt)}{(\xpbt-\xmbt)}a^{AB}_{11}\,
. \label{identity}\eea Due to this identity the coefficients
$a^{AB}_k$ drop out from the formula for $q$ and we find \bea\nonumber
q=-\frac{ a_1^{f}+a_3^{f}-2a_{13}^f
-8\frac{(1-\xpbt\xmbt)}{\sqrt{\xpbt\xmbt}}\frac{(\xpbt+\xmbt)}{(\xpbt-\xmbt)}a_{24}^{f}}
{b_1^{s}+b_3^{s}-2b_{13}^{s}-8\frac{(1-\xpbt\xmbt)}{\sqrt{\xpbt\xmbt}}\frac{(\xpbt+\xmbt)}{(\xpbt-\xmbt)}b_{24}^{s}}\,
. \eea Thus, $q$ depends on the coefficients of the operator
$\S^{BB}$ only.

\smallskip

Having found the coefficients $a_k^f$, $a_k^s$ and $q$, we can now
reconstruct all the coefficients $a_k=a_k^f+q a_k^s$. The
coefficients $\sfa,\sfb,\sfc,\sfd$ of the bound state
representations are parametrized in terms of the variables
$y^{\pm}_1$ and $y^{\pm}_2$ satisfying the constraints \bea\la{sc}
y_1^++\frac{1}{y_1^+}-y_1^--\frac{1}{y_1^-}=\frac{4i}{g}\, , ~~~~
y_2^++\frac{1}{y_2^+}-y_2^--\frac{1}{y_2^-}=\frac{4i}{g}\, .\eea
In practice, the formulae arising for $a_k$ as the functions of
$y^{\pm}_1,y^{\pm}_2$ are very involved, but they can be
drastically simplified by using constraints (\ref{sc}). Most
efficiently, this can be done by using the external $\sl(2)$
automorphism of the centrally extended algebra $\su(2|2)$
discussed in section 2.1. Indeed, in the invariance condition
(\ref{invcon}) the supersymmetry generators $\bQ$ and
$\bQ^{\dagger}$ can be replaced by $\tilde{\bQ}$ and
$\tilde{\bQ}^{\dagger}$ given by (\ref{nq}). As the result, we
will find \bea a_1^f &=&1\, , ~~~~a_2^f=0\, , ~~~a_k^f\equiv
a_k(1,0;\tilde{\sfa},\tilde{\sfb},\tilde{\sfc},\tilde{\sfd})\, ,~~~k>2, \nonumber \\
a_1^s&=&0\, , ~~~~a_2^s=1\, , ~~~~a_k^s\equiv
a_s(0,1;\tilde{\sfa},\tilde{\sfb},\tilde{\sfc},\tilde{\sfd})\,
,~~~k>2, \nonumber
 \eea
where the $\sl(2)$-transformed representation coefficients are
given by eqs.(\ref{tQc}) and they depend on arbitrary parameters
$u_1,u_2$ and $v_1,v_2$ modulo the constraint $u_1v_1-u_2v_2=1$.
On the other hand, the coefficients $a_k^s$ and $a_k^f$ are
$\sl(2)$ invariant and they must be independent of $u$ and $v$. By
picking up various values of these parameters one can achieve
enormous simplification of $a_k$. In this way we find \bea\nonumber
q&=&\frac{y^+_1y^-_2(y_1^--y^+_2)(-1+y_1^-y_2^+)}{y_1^-y_2^+
(-1+y_1^-y_2^-)(y_2^--y_1^+)}\times\\
\nonumber &\times &\frac{-3y_1^+y_2^-+2y_2^+y_2^-+
y_1^-(y_2^-)^2y_2^++y_1^+y_2^++2y_1^-y_2^-y_1^+y_2^+-3y_1^-y_2^-(y_2^+)^2}
{y_1^+y_2^--
2y_2^-y_2^++y_1^+y_2^++y_1^+y_2^+(y_2^-)^2-2y_2^+y_2^-(y_1^+)^2
+y_1^+y_2^-(y_2^+)^2}\, . \eea The final form of the coefficients
$a_k$ is given in appendix \ref{SBBcoef}. Quite remarkably, we
find that $a_{45}=a_{46}=a_{47}=a_{48}=0$. Finally, one can check
that with these coefficients $a_k$ the YB equations (\ref{YB1})
and (\ref{YB2}) are satisfied. Thus, solving the invariance
condition together with the YB equations we found the unique, up
to an overall scale, S-matrix describing the scattering of
two-particle bound states.

\medskip

The found S-matrix $\S^{BB}$ enjoys all the properties discussed
in section 3.4. Below we outline the transformation rules for
invariants $\Lambda_k$ under the hermitian conjugation, graded
permutation and parity.

\smallskip

 First, by using the hermitian
conjugation rules (\ref{hermc}) we find the following relations
\bea\nonumber \left(\Lambda_i\right)^\dagger = \Lambda_i\,,\quad i=1,\ldots
,16\,;\qquad \left(\Lambda_{17+2k}\right)^\dagger =
\Lambda_{18+2k}\,,\quad k=0,\ldots ,15\,. \eea

\smallskip

Second, the S-matrix is a symmetric operator, c.f. (\ref{symc}),
and, as a consequence, the coefficients $a_i$ satisfy the
following relations \bea\la{relbb1} &&a_{17+2k}(z_1,z_2) =
a_{18+2k}(z_1,z_2)\,,\quad k=0,\ldots ,15\,.~~~~~~~~ \eea It is
worth emphasizing that these relations hold only for the choice of
$\eta_i$ as in eq.(\ref{etakbb}).

\smallskip

Third, the parity transformation ${\mathscr P}$ acts on the
invariants $\Lambda_k$ as follows \bea \nonumber
\Lambda_k^{\mathscr P}&=&\Lambda_k\,~~~~~{\rm for}~~~~k=1,\ldots,
18 ~~~{\rm and}~~~~~k=25,\ldots,
36;\\
\nonumber \Lambda_k^{\mathscr P}&=&-\Lambda_k\, ~~~{\rm for}
~~~~k=19,\ldots, 24 ~~~{\rm and}~~~k=37,\ldots, 48.
 \eea

\smallskip
Finally, the graded permutation acts on $\Lambda_i$ in the
following way
\bea
\nonumber &&\Lambda_i  \leftrightarrow \Lambda_i \,,\ i=1,\ldots
,8\,;\quad \Lambda_{9+2k}  \leftrightarrow \Lambda_{10+2k}\,,\
k=0,1,2,3\,;\quad \Lambda_i  \leftrightarrow \Lambda_i \,,\
i=17,18\,;\\\nonumber &&\Lambda_i  \leftrightarrow -\Lambda_i \,,\
i=19,\ldots ,24\,;\quad \Lambda_{25+2k}  \leftrightarrow
\Lambda_{26+2k}\,,\  k=0,1,2,3\,;\quad\\\nonumber&&
\Lambda_{33+k}  \leftrightarrow \Lambda_{35+k}\,,\  k=0,1\,; \quad
\Lambda_{37+k +4m}  \leftrightarrow -\Lambda_{39+k+4m}\,,\
k=0,1\,,\ m=0,1,2\,,~~~~~ \eea leading to the unitarity
condition $\bS(z_2,z_1)^\spg\cdot \bS (z_1,z_2)= \bI$, and to the
hermitian analyticity condition $\bS(z_1^*,z_2^*)^\dagger =
\bS(z_2,z_1)^\spg\,,$ which together with (\ref{relbb1})  implies
the following relations between the coefficients $a_i$ of the
S-matrix
\bea
\nonumber &&a_i(z_2,z_1)^* = a_i(z_1^*,z_2^*)\,,\ i=1,\ldots
,8\,;\quad a_{9+2k}(z_2,z_1)^* = a_{10+2k}(z_1^*,z_2^*)\,,\
k=0,1,2,3\,;\\\nonumber &&a_{17}(z_2,z_1)^* =
a_{18}(z_1^*,z_2^*)\,;\quad a_{19+k}(z_2,z_1)^* =
-a_{19+k}(z_1^*,z_2^*)\,,\  k=0,\ldots,5\,;\\\nonumber
&&a_{25+k}(z_2,z_1)^* = a_{25+k}(z_1^*,z_2^*)\,,\
k=0,\ldots,7\,;\quad a_{33+k}(z_2,z_1)^* =
a_{35+k}(z_1^*,z_2^*)\,,\  k=0,1\,; \\\nonumber&&
a_{37+k+4m}(z_2,z_1)^* =- a_{39+k+4m}(z_1^*,z_2^*) \,,\  k=0,1\,,\
m=0,1,2\,.~~~~~ \eea One can check that the relations do hold.

\medskip

Further, one can see that on solutions of the equation \bea
\la{boundSBB} y_1^-=y_2^+ \eea the coefficients $a_2,a_4,a_6,
a_7,a_8,a_{11},a_{12},a_{15},\ldots,a_{22},a_{27},\ldots,a_{30},a_{41},\ldots,a_{48}$
vanish. As a result, the S-matrix $\S^{BB}$ degenerates and has
rank 16. This precisely corresponds to forming a 16-dim
four-particle bound state multiplet. Under the condition
(\ref{boundSBB}) the 48-dim component $\W^4$ of $\V^B$ becomes
reducible but indecomposable. The invariant subspace is a 32-dim
short multiplet, while  the four-particle bound state
representation is the factor representation.

\section{Crossing symmetry}\la{seccross}
In this section we introduce the S-matrices which describe  the
bound state scattering in the light-cone string theory on $\AdS$,
and discuss the conditions imposed on them by crossing symmetry.

\subsection{String S-matrix}
The canonical $\su(2|2)$ invariant S-matrices  can be used to find
the  corresponding $\su(2|2)\oplus\su(2|2)$ invariant string
S-matrices which describe the scattering of bound states in the
light-cone string theory on $\AdS$. To this end, one should
multiply the tensor product of two copies of the canonical
$\su(2|2)$ S-matrix by a scalar factor so that the resulting
matrix would satisfy the crossing symmetry relations. Thus, the
string S-matrix describing the scattering of $M$-particle with
$N$-particle bound states is \bea\la{sfull} {\cal S}^{MN}
(z_1,z_2) = S_0^{MN}(z_1,z_2)\, \S^{MN}(z_1,z_2)\otimes
\S^{MN}(z_1,z_2)\,. \eea The scalar factor $S_0^{MN}$ can be
easily identified by noting that, since $\Lambda_1$ is the
diagonal invariant operator corresponding to the component
$V^{M+N\ov 2}_{(M,N)}\times V^{0}_{(0,0)}$ with the maximum spin
$j_B = (M+N)/2$,  the term
$$S_0^{MN}(z_1,z_2)a_1(z_1,z_2)^2\,  \Lambda_1\otimes\Lambda_1$$
describes the bound state scattering in the $\su(2)$ sector of the
theory, the latter contains bound states with the single charge
non-vanishing\footnote{Recall that in the light-cone gauge the
particles do not carry the charge $J_1\equiv J$ that is
taken to infinity in the decompactification limit, see \cite{AAF,
AFlc,FPZ} for detail.}. The S-matrix $S_{\su(2)}^{MN}$
corresponding to this sector  is just a scalar function equal to
the coefficient in front of $\Lambda_1\otimes\Lambda_1$. Since we
have set the coefficient $a_1$ equal to unity, we see that the
scalar factor $S_0$ in eq.(\ref{sfull}) must be equal to the
S-matrix of the $\su(2)$ sector \bea\nonumber S_0^{MN}(z_1,z_2) =
S_{\su(2)}^{MN}(z_1,z_2)\,. \eea In its turn, the S-matrix in the
$\su(2)$ sector can be determined either by using the fusion
procedure \cite{Dorey2,Roiban} or simply   by evoking  the
$\su(2)$ sector of the asymptotic Bethe equations \cite{AFS} for
fundamental particles.\footnote{The $\su(2)$-sector S-matrix
exhibits an intricate structure of poles and zeroes which has been
shown in \cite{DHM,CDO2} to have a natural interpretation in terms
of on-shell intermediate bound states.} The only subtlety which
escaped from considerations in \cite{Dorey2,Roiban} is that the
$\su(2)$-sector S-matrix depends on extra phases which come from
the parameters $\eta_k$ \cite{AFZ}.  Taking these phases into
account and using the results by  \cite{Dorey2,Roiban}, we get the
following $\su(2)$-sector S-matrix \bea\nonumber
&&S_{\su(2)}^{MN}(z_1,z_2) = e^{i
a(p_1  \epsilon_2-p_2 \epsilon_1)}\, \left({y_1^-\ov y_1^+} \right)^N\left({y_2^+\ov y_2^-} \right)^{M}\,  \s(y_1^\pm,y_2^\pm)^2\, \times\\
&&\hspace{4cm} \times\, G(N-M)G(N+M) \prod_{k=1}^{M-1}G(N-M+
2k)^2\,,\qquad\la{ssu2} \eea where we assumed $M\le N$ and
introduced the  function \bea\nonumber
G(\ell)=\frac{u_1-u_2+\frac{i\ell
}{g}}{u_1-u_2-\frac{i\ell}{g}}\,. \eea In eq.(\ref{ssu2}) the
parameters $y^\pm_k$ satisfy the following relations \bea\nonumber
y^+_k +{1\ov y_k^+} -y_k^- -{1\ov y_k^-} = M_k {2i\ov g}\,,\quad
M_1=M\,,\ M_2=N\,
 \eea and the spectral parameters $u_k$ are
expressed in terms of $y_k^\pm$ as follows \bea\nonumber u_k=
{1\ov 2}\left(y^+_k +{1\ov y_k^+} +y_k^- +{1\ov y_k^-} \right)\,.
\eea The first factor in eq.(\ref{ssu2}) depends on the parameter
$a$ which is the parameter of the generalized light-cone gauge
\cite{magnon}, and $\epsilon_i = H(p_i)$ is the energy of a bound
state. The gauge-dependent factor solves the homogeneous crossing
equation \cite{AF06}. In what follows we set $a=0$.

\smallskip
The gauge-independent dressing factor is $\s(y^\pm_1,y^\pm_2) =
e^{i\theta(y^\pm_1,y^\pm_2)}$. Here
 the dressing phase \cite{AFS}
\bea\theta(y_1^+,y_1^-,y_2^+,y_2^-) &=& \sum_{r=2}^\infty
\sum_{n=0}^\infty c_{r,r+1+2n}(g)\Big[
q_r(y_1^\pm)q_{r+1+2n}(y_2^\pm) - q_r(y_2^\pm)q_{r+1+2n}(y_1^\pm)
\Big]\,  \nonumber\la{phase} \eea is a two-form on the vector
space of conserved charges $q_r(y^\pm)$    \bea \nonumber
q_r(y_k^-,y_k^+) &=& {i\ov r - 1}\left[ \left({1\ov
y^+_k}\right)^{r - 1} - \left({1\ov y^-_k}\right)^{r -
1}\right]\,.\eea An important observation made in
\cite{Dorey2,Roiban} is that as a function of {\it four}
parameters $y_1^+,y_1^-,y_2^+,y_2^-$ the dressing factor is
universal for scattering of any bound states. We point out,
however, that as a function of the torus rapidity variables
$z_1,z_2$, the dressing factor depends on $M$ and $N$, and this
results in different crossing symmetry equations satisfied by each
of the dressing factors.

\medskip
The general crossing equations can be derived by combining the
fusion procedure with the known crossing equations for the
dressing factor of the fundamental S-matrix \cite{Janik}. In fact,
the simplest form of these equations is obtained for a variable
$\Sigma$ which differs from $\s$ by inclusion of the extra phases
in eq.(\ref{ssu2}) \bea\la{dressn} \Sigma^{MN}(z_1,z_2) =
\left({y_1^-\ov y_1^+} \right)^{N\ov 2}\left({y_2^+\ov y_2^-}
\right)^{M\ov 2}\, \s(y_1^\pm,y_2^\pm)\,, \eea where on the r.h.s.
the variables $y_i^\pm$ should be expressed through $z_i$ by using
eq.(\ref{xpxmz}).

\smallskip
Then, it is not difficult to show that $\Sigma^{MN}$ should
satisfy the following crossing symmetry equations\footnote{The
second equation in (\ref{crosseqg}) follows from the first one by
using the unitarity condition
$\Sigma^{MN}(z_1,z_2)\Sigma^{NM}(z_2,z_1)=1$.}
\bea\la{crosseqg}\begin{aligned}
\Sigma^{MN}(z_1,z_2)\,\Sigma^{MN}(z_1+\om_2^{(M)},z_2)  &=
h(y_1^\pm,y_2^\pm)\, \prod_{k=0}^{M-1} G(M-N-2k)\,,\\
\Sigma^{MN}(z_1,z_2)\,\Sigma^{MN}(z_1,z_2-\om_2^{(N)})  &=
h(y_1^\pm,y_2^\pm) \, G(M-N)\,\prod_{k=0}^{N-1} G(M-N+2k)\,,
\end{aligned}\eea where the function $h(y_1^\pm,y_2^\pm)$ is given
by \bea\nonumber
h(y_1^\pm,y_2^\pm)=\frac{(y_1^--y_2^+)\Big(1-\frac{1}{y_1^-y_2^-}\Big)}{(y_1^+-y_2^+)\Big(1-\frac{1}{y_1^+y_2^-}\Big)}\,.
\eea Quite interesting, for $M=1$ the right hand sides  of two
crossing relations in eq.(\ref{crosseqg}) are the same
\bea\la{crosseqg1}
\begin{aligned}
\Sigma^{1N}(z_1,z_2)\,\Sigma^{1N}(z_1+\om_2^{(1)},z_2)  &=
h(y_1^\pm,y_2^\pm)G(1-N)\,,\\
\Sigma^{1N}(z_1,z_2)\,\Sigma^{1N}(z_1,z_2-\om_2^{(N)}) &=
h(y_1^\pm,y_2^\pm)G(1-N) \,. \end{aligned}\eea Although the
function $\s(y_1^+,y_1^-,y_2^+,y_2^-)$ does not have an explicit
dependence on $M$ and $N$, it satisfies the crossing equations
which do depend on these numbers, the reason for this lies in the
fact that the crossing equations hold on the constraint surfaces
of $y_i^\pm$ only and the latter depend on $M$ and $N$.

\subsection{Crossing symmetry equations for the S-matrices}
In this section we show that the canonical $\su(2|2)$ invariant
S-matrices we discussed in the previous section are compatible
with the crossing equations (\ref{crosseqg}). To simplify the
notation, we consider only one copy of $\su(2|2)$.

\medskip

{\it  Fundamental S-matrix}

\medskip

We start with the canonical S-matrix $\S^{AA}\equiv \S$ and
multiply it with the square root of the $\su(2)$-sector S-matrix
(\ref{ssu2})
$$
{\cal S}(z_1,z_2)=\Sigma(z_1,z_2)\, G(2)^{1/2}\, \S(z_1,z_2),
$$
where $\Sigma(z_1,z_2)\equiv\Sigma^{11}(z_1,z_2)$ is the dressing phase (\ref{dressn}).

\smallskip
The simplest  way to formulate the crossing symmetry equations for
${\cal S}$ is to use its matrix form (\ref{Sm}). Then, as was
shown in \cite{AFZ,AFTBA}, requiring the dressing factor $\Sigma$
to satisfy eqs.(\ref{crosseqg}) \bea\nonumber\la{crosseqsaa}
\Sigma(z_1,z_2)\,\Sigma(z_1+\om_2,z_2)  =
h(x_1^\pm,x_2^\pm)\,,\quad \Sigma(z_1,z_2)\,\Sigma(z_1,z_2-\om_2)
= h(x_1^\pm,x_2^\pm) \,,~~~~~ \eea one finds that the string
S-matrix exhibits the following crossing symmetry relations \bea
\label{cross1}\nonumber
 {\mathscr C}_1^{-1}{\cal S}_{12}^{t_1}(z_1,z_2){\mathscr C}_1{\cal
S}_{12}(z_1+\om_2,z_2)= I \, ,\quad
 {\mathscr C}_1^{-1}{\cal
S}_{12}^{t_1}(z_1,z_2){\mathscr C}_1{\cal
S}_{12}(z_1,z_2-\om_2)= I \,  .~~~~~
 \eea
Here $t_1$ denotes transposition in the first matrix space and
${\mathscr C}$ is a constant charge conjugation
matrix \bea\la{cc} \mathscr{C}=\left(\begin{array}{cc} \sigma_2 &
0 \\ 0 & i\, \sigma_2
\end{array}\right)\, ,
\eea where $\sigma_2$ is the Pauli matrix.

\medskip

It is of interest to understand  how the crossing relations are
formulated in the operator language. To this end, we need to find
out how the operator corresponding to the charge-conjugated and
transposed in the first space S-matrix ${\mathscr C}_1^{-1}{\cal
S}_{12}^{t_1}{\mathscr C}_1$ can be obtained from the invariant
operators $\Lambda_k$. The corresponding transformation rule for
$\Lambda_k$ appears to be simple and natural. Denoting the
transformed operator as $\Lambda_k^{c_1}$, we find that the latter
can be obtained from $\Lambda_k$ by means of the following
substitution \bea\la{crtr} w_a^1 \to \eps_{ab}{\pa\ov\pa
w_b^1}\,,\quad {\pa\ov\pa w_a^1} \to \eps^{ab}w_b^1\,,\quad
\theta_\a^1 \to i \eps_{\a\b}{\pa\ov\pa \theta_\b^1}\,,\quad
{\pa\ov\pa \theta_\a^1} \to i\eps^{\a\b}\theta_\b^1\,,~~~ \eea and
in the product of several variables entering $\Lambda_k$ we should
also change their order according to the rule $({\mathbb{A
B}})^{c_1} = (-1)^{\eps_{\mathbb A}\eps_{\mathbb B}}
{\mathbb{BA}}$. With these rules we find that the invariant
operators transform under the crossing transformation as follows
\bea\nonumber &&\Lambda_1^{c_1} = {1\ov 2}\Lambda_1 +{3\ov
2}\Lambda_2\,,\ \Lambda_2^{c_1} = {1\ov 2}\Lambda_1 -{1\ov
2}\Lambda_2\,,\  \Lambda_3^{c_1} = {1\ov 2}\Lambda_3 +{3\ov
2}\Lambda_4\,,\ \Lambda_4^{c_1} = {1\ov 2}\Lambda_3 -{1\ov
2}\Lambda_4\,,\\\nonumber\la{crsaa} &&\Lambda_5^{c_1} =
\Lambda_5\,,\ \Lambda_6^{c_1} = \Lambda_6\,,\  \Lambda_7^{c_1} =
-i \Lambda_{10}\,,\  \Lambda_8^{c_1} =- i \Lambda_{9}\,,\
\Lambda_9^{c_1} =- i \Lambda_{8}\,,\  \Lambda_{10}^{c_1} = -i
\Lambda_{7}\,.~~~~~~~~ \eea Notice that the crossing
transformation does not square to the identity because the
off-diagonal operators change their sign.

\smallskip

Then the crossing symmetry relations take the following operator
form \bea\nonumber {\cal S}^{c_1}(z_1,z_2)\cdot {\cal
S}(z_1+\om_2, z_2)=\mathbb{I} \,, \eea or, by using the unitarity
condition (\ref{uuc}), we can write them as \bea\nonumber {\cal
S}^{c_1}(z_1,z_2)= {\cal S}^\spg(z_1+\om_2, z_2)\,. \eea The last
equation implies a number of non-trivial relations between the
S-matrix coefficients which can be shown to hold.

\bigskip

The fact that the operators $\Lambda_1,\ldots,\Lambda_4$ transform
non-trivially under crossing indicates that the basis of invariant
operators we use is not particularly suitable for exhibiting the
crossing symmetry. It is not difficult to see that a better basis
can be obtained by regarding a differential operator acting in the
space $\V^M\otimes\V^N$ not as an element of
$(\V^M\otimes\V^N)\otimes (\D_M\otimes\D_N)$, but  as an element
of $(\V^M\otimes\D_M)\otimes (\V^N\otimes\D_N)$. The relation
between two bases is a linear transformation arising due to
the reordering of the factors in the tensor product. The brackets in the tensor products above indicate the order of the tensor product decomposition. In the new basis the crossing transformation
would act in the first factor $\V^M\otimes\D_M$ only leading to a
simpler transformation of the invariant operators. A disadvantage
of this crossing-adjusted basis is that the corresponding
coefficients of the S-matrix will be given by some linear
combinations of our simple coefficients $a_k$, and, for this
reason, they will not be simple anymore. Moreover, it would be
difficult to find a common normalization for all S-matrices
similar  to the condition $a_1=1$ we imposed.

\newpage

{\it S-matrix $ \S^{AB}$}

\medskip

Multiplying the  canonical S-matrix $\S^{AB}$ with the square root
of the $\su(2)$-sector S-matrix (\ref{ssu2}) taken for $M=1,N=2$,
we get
$$
{\cal S}^{AB}(z_1,z_2)=\Sigma^{AB}(z_1,z_2)\, \left(G(1)G(3)\right)^{1/2}\, \S^{AB}(z_1,z_2),
$$
where $\Sigma^{AB}\equiv\Sigma^{12}$ is the dressing phase (\ref{dressn}).

\smallskip
We use again the matrix form (\ref{Sm}) of ${\cal S}^{AB}$  to
formulate the crossing symmetry equations. One can show that  if
the dressing factor $\Sigma^{AB}$ satisfies eq.(\ref{crosseqg})
\bea\la{crosseqsab1}\nonumber
\Sigma^{AB}(z_1,z_2)\,\Sigma^{AB}(z_1+\om_2,z_2)  =
h(x_1^\pm,y_2^\pm)G(-1) =
h(x_1^\pm,y_2^\pm)\frac{u_1-u_2-\frac{i}{g}}{u_1-u_2+\frac{i}{g}}\,,~~~~~
\eea then the crossing  equation corresponding to the shift of the
torus rapidity variable $z_1$ of the fundamental particle takes
the same form as eq(\ref{cross1}) \bea \nonumber\label{crossab1}
 {\mathscr C}_1^{-1}{\cal S}^{AB,t_1}_{12}(z_1,z_2){\mathscr C}_1{\cal
S}_{12}^{AB}(z_1+\om_2,z_2)= I  \,  ,~~~~~
 \eea
where $\om_2\equiv \om_2^{(1)}$ is the imaginary half-period of
the fundamental particle rapidity torus, and ${\mathscr C}$ is
defined in eq.(\ref{cc}).

\smallskip
A matrix equation corresponding to the shift of the  second
rapidity $z_2$ of the two-particle bound state takes its simplest
form not for the S-matrix ${\cal S}^{AB}_{12}(z_1,z_2)$ acting in
the space $\V^A\otimes \V^B$ but for the S-matrix ${\cal
S}^{BA}_{12}(z_2,z_1)$ acting in  $\V^B\otimes \V^A$. The dressing
factor $\Sigma^{BA}$ of  ${\cal S}^{BA}$ is related to
$\Sigma^{AB}$ through the unitarity condition
$\Sigma^{AB}(z_1,z_2)\Sigma^{BA}(z_2,z_1)=1$ and it satisfies the
crossing equation \bea\la{crosseqsba}
\Sigma^{BA}(z_2,z_1)\,\Sigma^{BA}(z_2+\om_2^{(2)},z_1)  =
h(y_2^\pm,x_1^\pm) \,.~~~~~ \eea With this equation for the
dressing factor, the crossing symmetry relation corresponding to
the shift of the two-particle bound state rapidity $z_2$ acquires
the form similar to eq.(\ref{crossab1}) \bea \label{crossab2}
 {\mathscr C}_{B,1}^{-1}{\cal S}^{BA,t_1}_{12}(z_2,z_1){\mathscr C}_{B,1}{\cal
S}_{12}^{BA}(z_2+\om_2^{(2)},z_1)= I  \,  ,~~~~~
 \eea
where
${\mathscr C_B}$ is an $8\times 8$ charge conjugation
matrix acting in the two-particle bound state representation
 {\small \bea\la{ccb} \mathscr{C}_B={\small \left(
\begin{array}{llllllll}
 0 & 0 & -1 & 0 & 0 & 0 & 0 & 0 \\
 0 & 1 & 0 & 0 & 0 & 0 & 0 & 0 \\
 -1 & 0 & 0 & 0 & 0 & 0 & 0 & 0 \\
 0 & 0 & 0 & -1 & 0 & 0 & 0 & 0 \\
 0 & 0 & 0 & 0 & 0 & 0 & 0 & -i \\
 0 & 0 & 0 & 0 & 0 & 0 & i & 0 \\
 0 & 0 & 0 & 0 & 0 & i & 0 & 0 \\
 0 & 0 & 0 & 0 & -i & 0 & 0 & 0
\end{array}
\right)}\, . \eea }

\smallskip
We can formulate the crossing transformations in the operator language too.
By using  (\ref{crtr}), we find that the invariant operators transform under
the crossing transformation of the fundamental particle space as follows
\bea\nonumber
&&\Lambda_1^{c_1} = {1\ov 3}\Lambda_1 +{4\ov 3}\Lambda_2\,,\quad
\Lambda_2^{c_1} = {2\ov 3}\Lambda_1 -{1\ov 3}\Lambda_2\,,
\quad \Lambda_3^{c_1} = {1\ov 2}\Lambda_3 +{3\ov 2}\Lambda_4\,,\quad
\Lambda_4^{c_1} = {1\ov 2}\Lambda_3 -{1\ov 2}\Lambda_4\,,\\\nonumber
&&\Lambda_5^{c_1} = \Lambda_5\,,\quad \Lambda_6^{c_1} =
\Lambda_6\,,\quad \Lambda_7^{c_1} = {1\ov 2}\Lambda_7 +{3\ov
2}\Lambda_8\,,\quad \Lambda_8^{c_1} = {1\ov 2}\Lambda_7 -{1\ov
2}\Lambda_8\,,\quad \Lambda_9^{c_1} = \Lambda_{9}\,,\\\nonumber &&
\Lambda_{10}^{c_1} = - \Lambda_{10}\,,\quad \Lambda_{11}^{c_1} = -
\Lambda_{11}\,,\quad  \Lambda_{12}^{c_1} =- i \Lambda_{19}\,,\quad
\Lambda_{13}^{c_1} = -i \Lambda_{18}\,,\quad \Lambda_{14}^{c_1} =
-i \Lambda_{16}\,,\\\nonumber && \Lambda_{15}^{c_1} =- i
\Lambda_{17}\,,\quad \Lambda_{16}^{c_1} =- i \Lambda_{14}\,,\quad
\Lambda_{17}^{c_1} =- i \Lambda_{15}\,,\quad \Lambda_{18}^{c_1} =-
i \Lambda_{13}\,,\quad \Lambda_{19}^{c_1} =- i \Lambda_{12}\,.~~~
\eea With these transformation rules, the first crossing symmetry
equation acquires the following operator form \bea\nonumber {\cal
S}^{AB,c_1}(z_1,z_2)\cdot {\cal S}^{AB}(z_1+\om_2, z_2)=\mathbb{I}
\,. \eea

\smallskip
The second crossing symmetry relation is found by transforming the
 invariant operators under the crossing transformation of the two-particle  bound space
 by means of the following substitution
\bea\la{crtr2} w_a^2 \to \eps_{ab}{\pa\ov\pa w_b^2}\,,\quad
{\pa\ov\pa w_a^2} \to \eps^{ab}w_b^2\,,\quad \theta_\a^2 \to -i
\eps_{\a\b}{\pa\ov\pa \theta_\b^2}\,,\quad {\pa\ov\pa \theta_\a^2}
\to -i\eps^{\a\b}\theta_\b^2\,.~~~ \eea Notice the opposite sign
in the transformation of fermions in comparison to
eq.(\ref{crtr}).
 We find
\bea\nonumber
&&\Lambda_1^{c_2} = {1\ov 3}\Lambda_1 +{4\ov 3}\Lambda_2\,,\quad
\Lambda_2^{c_2} = {2\ov 3}\Lambda_1 -{1\ov 3}\Lambda_2\,,\quad \Lambda_3^{c_2} = {1\ov 2}\Lambda_3 +{3\ov 2}\Lambda_4\,,\quad
\Lambda_4^{c_2} = {1\ov 2}\Lambda_3 -{1\ov 2}\Lambda_4\,,\\\nonumber
&&\Lambda_5^{c_2} = \Lambda_5\,,\quad \Lambda_6^{c_2} =
\Lambda_6\,,\quad \Lambda_7^{c_2} = {1\ov 2}\Lambda_7 +{3\ov
2}\Lambda_8\,,\quad \Lambda_8^{c_2} = {1\ov 2}\Lambda_7 -{1\ov
2}\Lambda_8\,,\quad \Lambda_9^{c_2} = \Lambda_{9}\,,\\\nonumber &&
\Lambda_{10}^{c_2} = - \Lambda_{11}\,,\quad \Lambda_{11}^{c_2} = -
\Lambda_{10}\,,\quad  \Lambda_{12}^{c_2} = -i \Lambda_{18}\,,\quad
\Lambda_{13}^{c_2} = -i \Lambda_{19}\,,\quad \Lambda_{14}^{c_2} =
-i \Lambda_{17}\,,\\\nonumber && \Lambda_{15}^{c_2} =- i
\Lambda_{16}\,,\quad \Lambda_{16}^{c_2} =- i \Lambda_{15}\,,\quad
\Lambda_{17}^{c_2} =- i \Lambda_{14}\,,\quad \Lambda_{18}^{c_2} =
-i \Lambda_{12}\,,\quad \Lambda_{19}^{c_2} =- i \Lambda_{13}\,.~~~
\eea Then the second crossing symmetry equation acquires the
following operator form \bea\nonumber {\cal
S}^{AB,c_2}(z_1,z_2)\cdot {\cal S}^{AB}(z_1,
z_2-\om_2^{(2)})=\mathbb{I} \, . \eea The last equation implies a
number of non-trivial relations between the S-matrix coefficients
which can be shown to hold.

\bigskip

{\it S-matrix $\S^{BB}$}

\medskip

The crossing equations for the string S-matrix ${\cal S}^{BB}$ are
basically the same as for ${\cal S}^{AA}$. All one needs  is to
replace ${\mathscr C}$ by ${\mathscr C}_{B}$ in
eqs.(\ref{cross1}).

\smallskip
We multiply the canonical S-matrix  $\S^{BB}\equiv \S^{22}$ with
the square root of the $\su(2)$-sector S-matrix (\ref{ssu2})
$$
{\cal S}^{BB}(z_1,z_2)=\Sigma^{BB}(z_1,z_2)\,\left( G(2)^2G(4)\right)^{1/2}\, \S^{BB}(z_1,z_2),
$$
where $\Sigma^{BB}(z_1,z_2)\equiv\Sigma^{22}(z_1,z_2)$ is the
dressing phase (\ref{dressn})  satisfying the equations
\bea\la{crosseqsbb}\begin{aligned}
&\Sigma^{BB}(z_1,z_2)\,\Sigma^{BB}(z_1+\om_2^{(2)},z_2)  =
h(y_1^\pm,y_2^\pm)G(-2)
=h(y_1^\pm,y_2^\pm)\frac{u_1-u_2-\frac{2i}{g}}{u_1-u_2+\frac{2i}{g}}\,,~~~~~~~~\\
&\Sigma^{BB}(z_1,z_2)\,\Sigma^{BB}(z_1,z_2-\om_2^{(2)}) =
h(y_1^\pm,y_2^\pm)G(2)
=h(y_1^\pm,y_2^\pm)\frac{u_1-u_2+\frac{2i}{g}}{u_1-u_2-\frac{2i}{g}}\,.~~~~~
\end{aligned}
\eea With these equations for the dressing phase, the string
S-matrix exhibits the following crossing symmetry relations
\bea\begin{aligned} \label{crossbb} & {\mathscr C}_{B,1}^{-1}{\cal
S}^{BB,t_1}_{12}(z_1,z_2){\mathscr
C}_{B,1}{\cal S}^{BB}_{12}(z_1+\om_2^{(2)},z_2)= I \, ,\\
& {\mathscr C}_{B,1}^{-1}{\cal S}^{BB,t_1}_{12}(z_1,z_2){\mathscr
C}_{B,1}{\cal S}^{BB}_{12}(z_1,z_2-\om_2^{(2)})= I \,  .~~~~~
\end{aligned} \eea

\smallskip
To formulate the crossing transformations  in the operator
language we use  (\ref{crtr}) to find how the invariant operators
transform under the crossing transformation \bea\nonumber
&&\Lambda_1^{c_1} = {1\ov 6}\Lambda_1 +{5\ov 3}\Lambda_2+{5\ov
6}\Lambda_3\,,\quad \Lambda_2^{c_1} =  {1\ov 3}\Lambda_1 +{1\ov
3}\Lambda_2-{1\ov 3}\Lambda_3 \,,\quad \Lambda_3^{c_1} =  {1\ov
2}\Lambda_1 -\Lambda_2+{1\ov 2}\Lambda_3\,,\\\nonumber &&
\Lambda_4^{c_1} =  {1\ov 4}\Lambda_4 +{3\ov 4}\Lambda_5+{3\ov
4}\Lambda_6+{9\ov 4}\Lambda_7\,,\quad\Lambda_5^{c_1} =  {1\ov
4}\Lambda_4 -{1\ov 4}\Lambda_5+{3\ov 4}\Lambda_6-{3\ov
4}\Lambda_7\,,\\\nonumber && \Lambda_6^{c_1} =  {1\ov 4}\Lambda_4
+{3\ov 4}\Lambda_5-{1\ov 4}\Lambda_6-{3\ov
4}\Lambda_7\,,\quad\Lambda_7^{c_1} =  {1\ov 4}\Lambda_4 -{1\ov
4}\Lambda_5-{1\ov 4}\Lambda_6+{1\ov 4}\Lambda_7\,, \eea \bea
\nonumber &&\Lambda_8^{c_1} = \Lambda_8\,,\quad
 \Lambda_9^{c_1} = {1\ov 3}\Lambda_9 +{4\ov 3}\Lambda_{11}\,,\quad
\Lambda_{11}^{c_1} = {2\ov 3}\Lambda_9 -{1\ov 3}\Lambda_{11}\,,
\\\nonumber
&&
 \Lambda_{10}^{c_1} = {1\ov 3}\Lambda_{10} +{4\ov 3}\Lambda_{12}\,,\quad
\Lambda_{12}^{c_1} = {2\ov 3}\Lambda_{10} -{1\ov 3}\Lambda_{12}\,,
\quad\Lambda_k^{c_1} = \Lambda_k\,,~~ k=13,14,15,16\,,~~~~~
\\\nonumber
&&
 \Lambda_{17}^{c_1} = \Lambda_{26}\,,\quad
\Lambda_{26}^{c_1} = \Lambda_{17}\,,\quad\Lambda_{18}^{c_1} =
\Lambda_{25}\,,\quad \Lambda_{25}^{c_1} = \Lambda_{18}\,, \eea
\bea \nonumber && \Lambda_{19}^{c_1} = i\Lambda_{28}
-i\Lambda_{32}\,,\ \Lambda_{23}^{c_1} = -2i\Lambda_{28}
-i\Lambda_{32}\,,\ \Lambda_{28}^{c_1} = {i\ov 3}\Lambda_{19}
-{i\ov 3}\Lambda_{23}\,,\nonumber
\\\nonumber
&& \Lambda_{20}^{c_1} = i\Lambda_{27} -i\Lambda_{31}\,,\
\Lambda_{24}^{c_1} = -2i\Lambda_{27} -i\Lambda_{31}\,,\
\Lambda_{27}^{c_1} = {i\ov 3}\Lambda_{20} -{i\ov
3}\Lambda_{24}\,,\
\\\nonumber
&& \Lambda_{31}^{c_1} = -{2i\ov 3}\Lambda_{20} -{i\ov
3}\Lambda_{24}\,,~~~~~~~\Lambda_{32}^{c_1} = -{2i\ov
3}\Lambda_{19} -{i\ov 3}\Lambda_{23}\,,
\\
&& \nonumber
 \Lambda_{21}^{c_1} = i\Lambda_{29}\,,\quad
\Lambda_{29}^{c_1} = i\Lambda_{21}\,,\quad\Lambda_{22}^{c_1} =
i\Lambda_{30}\,,\quad \Lambda_{30}^{c_1} = i\Lambda_{22}\,, \eea
\bea \nonumber &&  \Lambda_{33}^{c_1} = -i\Lambda_{42}\,,\quad
\Lambda_{42}^{c_1} = -i\Lambda_{33}\,,  \quad \Lambda_{34}^{c_1} =
-i\Lambda_{41}\,,\quad \Lambda_{41}^{c_1} = -i\Lambda_{34}\,,
\\
&& \nonumber \Lambda_{35}^{c_1} = -i\Lambda_{43}\,,\quad
\Lambda_{43}^{c_1} = -i\Lambda_{35}\,, \quad\Lambda_{36}^{c_1} =
-i\Lambda_{44}\,,\quad \Lambda_{44}^{c_1} = -i\Lambda_{36}\,,
\\
\nonumber && \Lambda_{37}^{c_1} = -\Lambda_{37}\,,\quad~
\Lambda_{38}^{c_1} = -\Lambda_{38}\,,\quad ~\Lambda_{39}^{c_1} =
-\Lambda_{40}\,,\quad ~\Lambda_{40}^{c_1} = -\Lambda_{39}\,.~~~
\eea With this formulae the first crossing symmetry equation
acquires the following operator form \bea\nonumber {\cal
S}^{BB,c_1}(z_1,z_2)\cdot {\cal S}^{BB}(z_1+\om_2^{(2)},
z_2)=\mathbb{I} \,, \eea and leads to non-trivial relations
between the S-matrix coefficients. For instance, one finds
$$
{1\ov 6} + {1\ov 3}  a_2 + {1\ov 2}  a_3 = {G(-2)G(-4)\ov h(y^\pm_1,y^\pm_2)}\,,
$$
where $a_i$ are the coefficients of the canonical S-matrix, see
Appendix \ref{SBBcoef}. We have checked that all the relations
between the S-matrix coefficients are satisfied. This concludes
our discussion of the crossing symmetry for the string S-matrices.

\section*{Acknowledgements}
We thank  M. de Leeuw  for  discussions.
The work of G.~A. was
supported in part by the RFBI grant N05-01-00758, by the grant
NSh-672.2006.1, by NWO grant 047017015 and by the INTAS contract
03-51-6346. The work of S.F. was supported in part by the Science
Foundation Ireland under Grant No. 07/RFP/PHYF104. The work of G.~A. and S.~F.~was
supported in part by the EU-RTN network {\it Constituents,
Fundamental Forces and Symmetries of the Universe}
(MRTN-CT-2004-512194).

\section{Appendices}

\subsection{The S-matrix $\S^{AB}$}
\subsubsection{Invariant differential operators for
$\S^{AB}$}\label{ASAB1} To construct a corresponding basis of
invariant differential operators $\Lambda_k$, we have to figure
out all $\su(2)\oplus \su(2)$ singlet components in the tensor
product decomposition of $\W^3\otimes \D_3$. The basis of $\W^3$:
$$
\W^3={\rm
Span}\Big\{w_a^1w_c^2w_d^2,w_a^1w_b^2\theta_{\a}^2,\theta_{\a}^1w_a^2w_b^2,\theta_{\a}^1
w_a^2\theta_{\b}^2,w_a^1\theta^2_{\a}\theta_{\b}^2,\theta_{\a}^1\theta_{\b}^2\theta_{\gamma}^2
\Big\}\, , $$ as well as the basis of the dual space $\D_3$, can
be adopted to the decomposition (\ref{decom12}). For instance,
\bea \nonumber
w_a^1w_c^2w_d^2=\frac{1}{3}(w_a^1w_c^2w_d^2+w_a^2w_c^1w_d^2+w_a^2w_c^2w_d^1)+
\frac{1}{3}(\eps_{ab}w_c^2+\eps_{ac}w_b^2)\eps^{kl}w_k^1w_l^2 \, ,
\eea where two terms in the r.h.s.  provide the bases for the
spaces $V^{3/2}_{(1,2)}\times V^0_{(0,0)}$ and
$V^{1/2}_{(1,2)}\times V^0_{(0,0)}$, respectively. Similarly,
 \bea
\nonumber w_a^1w_b^2\theta_{\a}^2 & = &
\frac{1}{2}(w_a^1w_b^2+w_a^2w_b^1)\theta_{\a}^2
+\frac{1}{2}\eps_{ab}\eps^{kl}w_k^1w_l^2\theta_{\a}^2\to
(V^{1}_{(1,1)}+V^{0}_{(1,1)} )\times V^{1/2}_{(0,1)} \\
\nonumber \theta_{\a}^1 w_a^2\theta_{\b}^2&=&
\frac{1}{2}w_a^2(\theta_{\a}^1\theta_{\b}^2+\theta_{\b}^1\theta_{\a}^2)
+\frac{1}{2}w_a^2(\theta_{\a}^1\theta_{\b}^2-\theta_{\b}^1\theta_{\a}^2)
\to V^{1/2}_{(0,1)}\times(V^{1}_{(1,1)}+V^0_{(1,1)})\\
\nonumber \theta_{\a}^1w_a^2w_b^2 &\to & V^{1}_{(0,2)}\times
V^{1/2}_{(1,0)}\, ,~~~ \nonumber w_a^1\theta^2_{\a}\theta_{\b}^2
\to V^{1/2}_{(1,0)}\times V^0_{(1,1)} \, , ~~~ \nonumber
\theta_{\a}^1\theta_{\b}^2\theta_{\gamma}^2 \to V^0_{(0,0)}\times
V^{1/2}_{(1,2)}
 \eea
Projecting on the singlet components of the tensor product
decomposition $\V^3\otimes \D_3$, one obtains 19 invariant
differential operators of the third order whose explicit form is
given below. \bea \nonumber~\qquad \qquad
\Lambda_1&=&\frac{1}{6}(w_a^1w_b^2w_c^2+w_a^2w_b^1w_c^2+w_a^2w_b^2w_c^1)
~\frac{\pa^3}{\pa w_c^2 \pa w_b^2\pa w_a^1 } \hspace{15cm} \eea

\vspace{-0.6cm}

\bea\nonumber~\qquad \qquad
\Lambda_2&=&\frac{1}{6}(\eps_{ab}w_c^2+\eps_{ac}w_b^2)\eps^{kl}w_k^1w_l^2
~\frac{\pa^3}{ \pa w_c^2 \pa w_b^2\pa w_a^1 } \hspace{15cm}
\eea

\vspace{-0.6cm}

\bea\nonumber~\qquad \qquad
\Lambda_3&=&\frac{1}{2}(w_a^1w_b^2+w_a^2w_b^1)\theta_{\a}^2
~\frac{\pa^3}{\pa \theta_{\a}^2\pa w_b^2\pa w_a^1  } \hspace{15cm}
\eea

\vspace{-0.6cm}

\bea\nonumber~\qquad \qquad
\Lambda_4&=&\frac{1}{2}(w_a^1w_b^2-w_a^2w_b^1)\theta_{\a}^2
~\frac{\pa^3}{\pa \theta_{\a}^2\pa w_b^2\pa w_a^1  } \hspace{15cm}
\eea

\vspace{-0.6cm}

\bea\nonumber~\qquad \qquad
\Lambda_5&=& \frac{1}{2}w_a^2w_b^2\theta_{\a}^1
~\frac{\pa^3}{\pa w_b^2\pa w_a^2 \pa \theta_{\a}^1} \hspace{15cm}
\eea

\vspace{-0.6cm}

\bea\nonumber~\qquad \qquad
\Lambda_6&=&\frac{1}{2}w_a^1\theta_{\a}^2\theta_{\b}^2
~\frac{\pa^3}{\pa \theta_{\b}^2\pa \theta_{\a}^2\pa w_a^1 } \hspace{15cm}
\eea

\vspace{-0.6cm}

\bea\nonumber~\qquad \qquad
\Lambda_7&=&\frac{1}{2}w_a^2(\theta_{\a}^1\theta_{\b}^2+\theta_{\b}^1\theta_{\a}^2)
~\frac{\pa^3}{\pa w_a^2\pa \theta_{\b}^2\pa \theta_{\a}^1 } \hspace{15cm}
\eea

\vspace{-0.6cm}

\bea\nonumber~\qquad \qquad
\Lambda_8&=&\frac{1}{2}w_a^2(\theta_{\a}^1\theta_{\b}^2-\theta_{\b}^1\theta_{\a}^2)
~\frac{\pa^3}{\pa w_a^2\pa \theta_{\b}^2 \pa \theta_{\a}^1} \hspace{15cm}
\eea

\vspace{-0.6cm}

\bea\nonumber~\qquad \qquad
\Lambda_9&=&\frac{1}{2}\theta_{\a}^1\theta_{\beta}^2\theta_{\gamma}^2
~\frac{\pa^3}{\pa \theta_{\g}^2\pa \theta_{\b}^2\pa \theta_{\a}^1 } \hspace{15cm}
\eea

\vspace{-0.6cm}

\bea\nonumber~\qquad \qquad
\Lambda_{10}&=&\frac{1}{2}\eps^{kl}w_k^1w_l^2
w_a^2\eps_{\a\b} ~\frac{\pa^3}{\pa \theta_{\b}^2\pa
\theta_{\a}^2\pa w_a^1 }  \hspace{15cm}
\eea

\vspace{-0.6cm}

\bea\nonumber~\qquad \qquad
 \Lambda_{11}&=&\frac{1}{2}\eps^{\a\b}\theta_{\a}^2\theta_{\b}^2\eps_{ab}w_c^1
 ~\frac{\pa^3}{\pa w_c^2 \pa w_b^2\pa w_a^1}
 \hspace{15cm}
\eea

\vspace{-0.6cm}

\bea\nonumber~\qquad \qquad
 \Lambda_{12}&=&\eps^{kl}w_k^1w_l^2 \eps_{\a\b}w_a^2
~\frac{\pa^3}{\pa w_a^2 \pa \theta_{\b}^2\pa \theta_{\a}^1}
 \hspace{15cm}
\eea

\vspace{-0.6cm}

\bea\nonumber~\qquad \qquad
\Lambda_{13}&=&\eps^{\a\b}\theta_{\a}^1\theta_{\b}^2\eps_{ab}w_c^2
 ~\frac{\pa^3}{\pa w_c^2 \pa w_b^2\pa w_a^1} \hspace{15cm}
\eea

\vspace{-0.6cm}

\bea\nonumber~\qquad \qquad
\Lambda_{14}&=&\frac{1}{2}\eps^{\a\b}\theta_{\a}^2\theta_{\b}^2\eps_{\gamma\delta}w_a^1
~\frac{\pa^3}{\pa w_a^2\pa \theta_{\delta}^2\pa \theta_{\gamma}^1
} \hspace{15cm}
\eea

\vspace{-0.6cm}

\bea\nonumber~\qquad \qquad
 \Lambda_{15}&=&
\frac{1}{2}\eps^{\a\b}\theta_{\a}^1\theta_{\b}^2\eps_{\gamma\delta}w_a^2
~\frac{\pa^3}{\pa \theta_{\delta}^2\pa \theta_{\g}^2\pa w_a^1
} \hspace{15cm}
\eea

\vspace{-0.6cm}

\bea\nonumber~\qquad \qquad
\Lambda_{16}&=&\eps_{ab}\eps^{\a\b}\theta_{\a}^1\theta_{\b}^2\theta_{\gamma}^2
~\frac{\pa^3}{\pa \theta_{\gamma}^2\pa w_b^2 \pa w_a^1} \hspace{15cm}
\eea

\vspace{-0.6cm}

\bea\nonumber~\qquad \qquad
\Lambda_{17}&=&\eps^{kl}w_k^1w_l^2
\eps_{\a\b}\theta_{\gamma}^2 ~\frac{\pa^3}{ \pa
\theta_{\g}^2\pa \theta_{\b}^2\pa \theta_{\a}^1 } \hspace{15cm}
\eea

\vspace{-0.6cm}

\bea\nonumber~\qquad \qquad
\Lambda_{18}&=&w_a^1w_b^2\theta_{\a}^2
~\frac{\pa^3}{\pa w_b^2\pa w_a^2\pa \theta_{\a}^1 } \hspace{15cm}
\eea

\vspace{-0.6cm}

\bea\nonumber~\qquad \qquad
\Lambda_{19}&=&w_a^2w_b^2\theta_{\a}^1 ~\frac{\pa^3}{\pa
\theta_{\a}^2\pa w_b^2 \pa w_a^1}\hspace{15cm}
 \eea

\subsubsection{Coefficients of $\S^{AB}$}\label{ASAB2} Here
$x^\pm_1$ are the parameters of the fundamental  particle with
momentum \mbox{$p_1=2\,{\rm am}z_1$}, and $y^\pm_2$ are the
parameters of the two-particle bound state with momentum
\mbox{$p_2=2\,{\rm am}z_2$}. The parameters $\eta_k$ are given by
\bea\la{etak2} \eta_1 = e^{ip_2/2}\eta(z_1,1)\,,\quad \eta_2
=\eta(z_2,2)\,,\quad \tilde{\eta}_1 = \eta(z_1,1)\,,\quad
\tilde{\eta}_2 = e^{ip_1/2}\eta(z_2,2)\,, \eea where $\eta(z,M)$
is defined by (\ref{etaz}). The coefficients are meromorphic
functions of $z_1$ and $z_2$ defined on the product of two tori.
Note, however, that the tori have different moduli.
\bea\nonumber~\qquad \qquad a_1&=&1 \hspace{15cm} \eea

\vspace{-0.6cm}

\bea\nonumber~\qquad \qquad
a_2&=&
-\frac{1}{2}-\frac{3}{2}\frac{(1-\xmao \xpbt
)(\xpao-\xpbt)\xmbt}{(1-\xmao \xmbt)(\xmbt-\xpao)\xpbt}\hspace{15cm}
\eea

\vspace{-0.6cm}

\bea\nonumber~\qquad \qquad
a_3&=& -\frac{\xmao-\xmbt}{ \xmbt-\xpao  }
\frac{\tilde{\eta}_2}{\eta_2}\hspace{15cm}\eea

\vspace{-0.6cm}

\bea\nonumber~\qquad \qquad
a_4&=&
-\frac{(1-\xmao\xpbt)(\xmao-\xpbt)\xmbt}
{(1-\xmao\xmbt)(\xmbt-\xpao)\xpbt}
\frac{\tilde{\eta}_2}{\eta_2}\hspace{15cm}
\eea

\vspace{-0.6cm}

\bea\nonumber~\qquad \qquad
a_5&=&-\frac{\xpao-\xpbt}{\xmbt-\xpao}\frac{\tilde{\eta}_1}{\eta_1}\hspace{15cm}
\eea

\vspace{-0.6cm}

\bea\nonumber~\qquad \qquad
a_6&=&\frac{x^-_1 \left(2
   y^+_2 y^-_2 - x^-_1
   y^-_2 - x^-_1 y^+_2-x^+_1 y^+_2
   y^-_2{}^2-x^+_1 y^+_2{}^2 y^-_2+2 x^-_1 x^+_1 y^+_2 y^-_2{}\right)}{2 x^+_1 (x^+_1-y^-_2) (x^-_1
   y^-_2-1) y^+_2 }\frac{ \tilde{\eta}_2^2}{\eta _2^2}
\hspace{15cm}\eea

\vspace{-0.6cm}

\bea\nonumber~\qquad \qquad
a_7&=&-\frac{\xmao-\xpbt}{\xmbt-\xpao}\frac{\tilde{\eta}_1\tilde{\eta}_2}{\eta_1
\eta_2}\hspace{15cm}\eea

\vspace{-0.6cm}

\bea\nonumber~\qquad \qquad
a_8&=&
\frac{\xmao-\xpbt}{\xmbt-\xpao}\left(1-2\frac{(1-\xmbt\xpao)(\xpao-\xpbt)\xmao
}{ (1-\xmao\xmbt)(\xmao-\xpbt)\xpao
}\right)\frac{\tilde{\eta}_1\tilde{\eta}_2}{\eta_1\eta_2}\hspace{15cm}\eea

\vspace{-0.6cm}

\bea\nonumber~\qquad \qquad
a_9&=&-\frac{(1-\xmbt\xpao)(\xmao-\xpbt)}{(1-\xmao\xmbt)(\xmbt-\xpao)}\frac{\tilde{\eta}_1}{\eta_1}\hspace{15cm}\eea

\vspace{-0.6cm}

\bea\nonumber~\qquad \qquad
a_{10}&=&-\frac{i}{2}\frac{(\xmao-\xpao)(\xmbt-\xpbt)^2}{(1-\xmao\xmbt)(\xmbt-\xpao)}\frac{1}{\eta_2^2}\hspace{15cm}
\eea

\vspace{-0.6cm}

\bea\nonumber~\qquad \qquad
a_{11}&=&\frac{i}{2}\frac{\xmao\xmbt}{\xpao\xpbt}\frac{(\xmao-\xpao)\,
\tilde{\eta}_2^2}{(1-\xmao\xmbt)(\xmbt-\xpao)}\hspace{15cm}\eea

\vspace{-0.6cm}

\bea\nonumber~\qquad \qquad
a_{12}&=&\frac{i}{\sqrt{2}}\frac{(\xmao-\xpao)(\xpao-\xpbt)(\xmbt-\xpbt)}{(1-\xmao\xmbt)(\xmbt-\xpao)
}\frac{1}{\eta_1\eta_2} \hspace{15cm}\eea

\vspace{-0.6cm}

\bea\nonumber~\qquad \qquad
a_{13}&=&-\frac{i}{\sqrt{2}}\frac{\xmao\xmbt}{\xpao\xpbt}\frac{(\xmao-\xpbt
)\,
\tilde{\eta}_1\tilde{\eta}_2}{(1-\xmao\xmbt)(\xmbt-\xpao)}\hspace{15cm}\eea

\vspace{-0.6cm}

\bea\nonumber~\qquad \qquad
a_{14}&=&\frac{1}{\sqrt{2}}\frac{\xmao}{\xpao}\frac{(1-\xmbt\xmao)(\xmao-\xpao)}{(1-\xmao\xmbt)(\xmbt-\xpao)}
\frac{\tilde{\eta}_2^2}{\eta_1\eta_2}\hspace{15cm}\eea

\vspace{-0.6cm}

\bea\nonumber~\qquad \qquad
a_{15}&=&\frac{1}{\sqrt{2}}\frac{\xmao}{\xpao}\frac{(1-\xmbt\xpao)(\xmbt-\xpbt)}{(1-\xmao\xmbt)(\xmbt-\xpao)}
\frac{\tilde{\eta}_1\tilde{\eta}_2}{\eta_2^2}\hspace{15cm} \eea

\vspace{-0.6cm}

\bea\nonumber~\qquad \qquad
a_{16}&=&-\frac{i}{\sqrt{2}}\frac{\xmao\xmbt}{\xpao\xpbt}\frac{(\xmao-\xpbt)}{(1-\xmao\xmbt)(\xmbt-\xpao)}
\frac{\tilde{\eta}_1\tilde{\eta}_2^2}{\eta_2}\hspace{15cm}
\eea

\vspace{-0.6cm}

\bea\nonumber~\qquad \qquad
a_{17}&=&\frac{i}{\sqrt{2}}
\frac{(\xmao-\xpao)(\xmao-\xpbt)(\xmbt-\xpbt)}{(1-\xmao\xmbt)(\xmbt-\xpao)}\frac{\tilde{\eta}_2}{\eta_1\eta_2^2}\hspace{15cm}\eea

\vspace{-0.6cm}

\bea\nonumber~\qquad \qquad
a_{18}&=&\frac{1}{\sqrt{2}}\frac{\xmao-\xpao}{\xmbt-\xpao}\frac{\tilde{\eta}_2}{\eta_1
}\hspace{15cm}\eea

\vspace{-0.6cm}

\bea\nonumber~\qquad \qquad
a_{19}&=&\frac{1}{\sqrt{2}}\frac{\xmbt-\xpbt}{\xmbt-\xpao}\frac{\tilde{\eta}_1}{\eta_2
} \hspace{15cm}\eea

\subsubsection{Matrix form of invariant differential operators
$\Lambda_k$}

We use the basis of monomials (\ref{basismon}). The basis vectors
$|e_{i}\rangle$  of the fundamental representation are $
|e_{a}\rangle = w_a^1\,,\ |e_{\a}\rangle = \theta_\a^1$, and the
basis vectors $|e_{J}\rangle$  of the two-particle bound state
representation are \bea \begin{aligned} |e_{1}\rangle =
&{w_1^2w_1^2\ov\sqrt{2}}\,,&\  |e_{2}\rangle &= w_1^2w_2^2\,, &\
|e_{3}\rangle &= {w_2^2w_2^2\ov\sqrt{2}}\,, &\ |e_{4}\rangle &= \theta_3^2\theta_4^2\,,\\
\la{basis2p}  |e_{5}\rangle =& w_1^2\theta_3^2\,,&\ |e_{6}\rangle
&= w_1^2\theta_4^2\,,&\  |e_{7}\rangle &= w_2^2\theta_3^2\,,&\
|e_{8}\rangle &= w_2^2\theta_4^2\,. \end{aligned}\eea A
$\su(2)\oplus\su(2)$ invariant differential operator is
represented in the matrix form as a sum over symbols $E_{kiLJ}$
which can be equal to either $E_k{}^i \otimes E_L{}^J$ or to
\mbox{$(-1)^{\epsilon_k\epsilon_L}\, E_k{}^i \otimes E_L{}^J$} (or
anything else one wants).

\bea\nonumber \Lambda_1 &=& E_{1111}+\frac{2}{3}
   E_{1122}+\frac{1}{3}
   E_{1133}+\frac{1}{3} \sqrt{2}
   E_{1221}+\frac{1}{3} \sqrt{2}
   E_{1232}+\frac{1}{3} \sqrt{2}
   E_{2112}+\frac{1}{3} \sqrt{2}
   E_{2123}\\\nonumber &+& \frac{1}{3}
   E_{2211}+\frac{2}{3}
   E_{2222}+E_{2233}
\hspace{15cm}
\eea

\vspace{-0.6cm}

\bea\nonumber
\Lambda_2 &=& \frac{1}{3} \left(E_{1122}+2
   E_{1133}-\sqrt{2}
   E_{1221}-\sqrt{2}
   E_{1232}-\sqrt{2}
   E_{2112}-\sqrt{2}
   E_{2123}+2
   E_{2211}+E_{2222}\right)\hspace{15cm}
\eea

\vspace{-0.6cm}

\bea\nonumber
\Lambda_3 &=& E_{1155}+E_{1166}+\frac
   {1}{2} E_{1177}+\frac{1}{2}
   E_{1188}+\frac{1}{2}
   E_{1275}+\frac{1}{2}
   E_{1286}+\frac{1}{2}
   E_{2157}+\frac{1}{2}
   E_{2168}+\frac{1}{2}
   E_{2255}\\\nonumber &+&\frac{1}{2}
   E_{2266}+E_{2277}+E_
   {2288}
\hspace{15cm}
\eea

\vspace{-0.6cm}

\bea\nonumber
\Lambda_4 &=& \frac{1}{2}
   \left(E_{1177}+E_{1188}-E_{1275}-E_{1286
   }-E_{2157}-E_{2168}+
   E_{2255}+E_{2266}\right)
\hspace{15cm}
\eea

\vspace{-0.6cm}

\bea\nonumber
\Lambda_5 &=&
E_{3311}+E_{3322}+E_{3333}+E_{4411}+E_{4422}+E_{4433}
\hspace{15cm}
\eea

\vspace{-0.6cm}

\bea\nonumber
\Lambda_6 &=& E_{1144}+E_{2244}
\hspace{15cm}
\eea

\vspace{-0.6cm}

\bea\nonumber
\Lambda_7 &=& E_{3355}+\frac{1}{2}
   E_{3366}+E_{3377}+\frac{1}{2} E_{3388}+\frac{1}{2}
   E_{3465}+\frac{1}{2}
   E_{3487}+\frac{1}{2}
   E_{4356}+\frac{1}{2}
   E_{4378}\\\nonumber &+&\frac{1}{2}
   E_{4455}+E_{4466}+\frac{1}{2}
   E_{4477}+E_{4488}
\hspace{15cm}
\eea

\vspace{-0.6cm}

\bea\nonumber
\Lambda_8 &=& \frac{1}{2}
   \left(E_{3366}+E_{3388}-E_{3465}-E_{3487
   }-E_{4356}-E_{4378}+
   E_{4455}+E_{4477}\right)
\hspace{15cm}
\eea

\vspace{-0.6cm}

\bea\nonumber
\Lambda_9 &=& E_{3344}+E_{4444}
\hspace{15cm}
\eea

\vspace{-0.6cm}

\bea\nonumber
\Lambda_{10} &=& E_{1124}+\sqrt{2}
   E_{1234}-\sqrt{2}
   E_{2114}-E_{2224}
\hspace{15cm}
\eea

\vspace{-0.6cm}

\bea\nonumber
\Lambda_{11} &=& E_{1142}-\sqrt{2}
   E_{1241}+\sqrt{2}
   E_{2143}-E_{2242}
\hspace{15cm}
\eea

\vspace{-0.6cm}

\bea\nonumber
\Lambda_{12} &=& E_{1326}+\sqrt{2}
   E_{1338}-E_{1425}-\sqrt{2} E_{1437}-\sqrt{2}
   E_{2316}-E_{2328}+\sqrt{2}
   E_{2415}+E_{2427}
\hspace{15cm}
\eea

\vspace{-0.6cm}

\bea\nonumber
\Lambda_{13} &=& E_{3162}+\sqrt{2}
   E_{3183}-\sqrt{2}
   E_{3261}-E_{3282}-E_
   {4152}-\sqrt{2}
   E_{4173}+\sqrt{2}
   E_{4251}+E_{4272}
\hspace{15cm}
\eea

\vspace{-0.6cm}

\bea\nonumber
\Lambda_{14} &=&
E_{1346}-E_{1445}+E_{2348}-E_{2447}
\hspace{15cm}
\eea

\vspace{-0.6cm}

\bea\nonumber
\Lambda_{15} &=& E_{3164}+E_{3284}-E_{4154}-E_{4274}
\hspace{15cm}
\eea

\vspace{-0.6cm}

\bea\nonumber
\Lambda_{16} &=& -E_{3147}+E_{3245}-E_{4
   148}+E_{4246}
\hspace{15cm}
\eea

\vspace{-0.6cm}

\bea\nonumber
\Lambda_{17} &=& -E_{1374}-E_{1484}+E_{2
   354}+E_{2464}
\hspace{15cm}
\eea

\vspace{-0.6cm}

\bea\nonumber
\Lambda_{18} &=& \sqrt{2}
   E_{1351}+E_{1372}+\sqrt{2}
   E_{1461}+E_{1482}+E_
   {2352}+\sqrt{2}
   E_{2373}+E_{2462}+\sqrt{2} E_{2483}
\hspace{15cm}
\eea

\vspace{-0.6cm}

\bea\nonumber
\Lambda_{19} &=& \sqrt{2}
   E_{3115}+E_{3127}+E_
   {3225}+\sqrt{2}
   E_{3237}+\sqrt{2}
   E_{4116}+E_{4128}+E_
   {4226}+\sqrt{2}
   E_{4238}\hspace{15cm}
\eea


\subsection{The S-matrix $\S^{BB}$}\la{ASBB}

\subsubsection{Invariant differential operators for $\S^{BB}$}
Recall that $\V^B\otimes \V^B=\W^2+\W^4$, where $\W^2$ and $\W^4$
are long supermultiplets of dimension $16$ and $48$, respectively.
The branching rule for $\W^2$ under the action of $\su(2)\oplus
\su(2)$ is given by eq.(\ref{decW2}). As to $\W^4$, it branches as
follows \bea\nonumber  \W^4&=&V^0\times V^0+3V^1\times
V^0+V^2\times V^0+V^1\times V^1 +2V^{3/2}\times
V^{1/2}+2V^{1/2}\times V^{1/2}\, , \eea where the integers in the
r.h.s. denote the multiplicities of the corresponding
representations. Thus, the sum $\W^2+\W^4$ contains $6+10=16$
$\su(2)\oplus \su(2)$ multiplets. The basis vectors of
$\V^B\otimes \V^B$ give rise to the corresponding basis vectors
adopted to the $\su(2)\oplus \su(2)$ decomposition. Schematically,

\bea
\begin{array}{ll}
\nonumber w_a^1w_b^1w_c^2w_d^2 \to  (V^0+V^1+V^2)\times V^0
~~~~&~~~~ w_a^1\theta_{\a}^1\theta_{\b}^2\theta_{\gamma}^2 \to
V^{1/2}\times V^{1/2} \\
w_a^1w_b^1w_c^2\theta_{\a}^2 \to (V^{3/2}+V^{1/2})\times V^{1/2}
~~~~&~~~~ w_a^2w_b^2\theta_{\a}^1\theta_{\b}^1 \to V^{1}\times
V^0\\
w_a^1w_b^1\theta_{\a}^2\theta_{\beta}^2 \to  V^1\times
V^0~~~~&~~~~w_a^2\theta_{\a}^1\theta_{\b}^1\theta_{\gamma}^2 \to
V^{1/2}\times V^{1/2}\\
w_a^1w_b^2w_c^2\theta_{\a}^1 \to  (V^{3/2}+V^{1/2})\times
V^{1/2}~~~~&~~~~\theta_{\a}^1\theta_{\b}^1\theta_{\gamma}^2\theta_{\delta}^2
\to V^0\times V^0\\
w_a^1w_b^2\theta_{\a}^1\theta_{\b}^2 \to  (V^0+V^1)\times
(V^0+V^1) \\
\end{array}
\eea Here the basis elements of the spaces $V^k\times V^m$ are
obtained from those of $\V^B\otimes \V^B$ by proper
(anti)symmetrization of the $\su(2)$ indices. Analogously, one can
construct the $\su(2)\oplus \su(2)$ basis of the dual space
$\D_2+\D_4$. Finally, by analyzing the tensor product
$(\W^2+\W^4)\otimes (\D_2+\D_4)$, one finds that it has 48 singlet
components. Thus, there are 48 independent invariant differential
operators $\Lambda_k$ which we present below.
 \bea
\nonumber \Lambda_1&=&\frac{1}{24}\Big[w_a^2 w_b^2 w_c^1
w_d^1 +4 w_a^1 w_b^2
w_c^1 w_d^2+w_a^1 w_b^1 w_c^2 w_d^2 \Big]~\frac{\pa^4}{\pa
w_a^{1}\pa w_b^{1}\pa w_c^2\pa w_d^2} \hspace{15cm}
\eea

\vspace{-0.6cm}

\bea\nonumber
\Lambda_2&=&\frac{1}{12}\eps_{ac}\eps_{bd}
 \eps^{kl}\eps^{mn}w_k^1w_l^2w_m^1w_n^2~\frac{\pa^4}{\pa
w_a^{1}\pa w_b^{1}\pa w_c^2\pa w_d^2} \hspace{15cm}
\eea

\vspace{-0.6cm}

\bea\nonumber
\Lambda_3&=&\frac{1}{8}\eps_{ad}(w_b^1w_c^2+w_b^2w_c^1)\eps^{kl}w_k^1w_l^2
~\frac{\pa^4}{\pa
w_a^{1}\pa w_b^{1}\pa w_c^2\pa w_d^2} \hspace{15cm}
\eea

\vspace{-0.6cm}

\bea\nonumber
\Lambda_{4}&=&\frac{1}{4}(w_a^1w_b^2+w_a^2w_b^1)(\theta_{\a}^1\theta_{\b}^2+\theta_{\b}^1\theta_{\a}^2)
~\frac{\pa^4}{\pa w_a^{1}\pa w_b^{2}\pa
\theta_{\beta}^2\pa \theta_{\alpha}^1} \hspace{15cm}
\eea

\vspace{-0.6cm}

\bea\nonumber
\Lambda_{5}&=&\frac{1}{4}(w_a^1w_b^2+w_a^2w_b^1)(\theta_{\a}^1\theta_{\b}^2-\theta_{\b}^1\theta_{\a}^2)
~\frac{\pa^4}{\pa w_a^{1}\pa w_b^{2}\pa
\theta_{\beta}^2\pa \theta_{\alpha}^1} \hspace{15cm}
\eea

\vspace{-0.6cm}

\bea\nonumber
\Lambda_{6}&=&\frac{1}{4}\eps_{ab}\eps^{kl}w_k^1w_l^2(\theta_{\a}^1\theta_{\b}^2+\theta_{\b}^1\theta_{\a}^2)
~\frac{\pa^4}{\pa w_a^{1}\pa w_b^{2}\pa
\theta_{\beta}^2\pa \theta_{\alpha}^1}\hspace{15cm}
\eea

\vspace{-0.6cm}

\bea\nonumber
\Lambda_{7}&=&\frac{1}{4}\eps_{ab}\eps^{kl}w_k^1w_l^2(\theta_{\a}^1\theta_{\b}^2-\theta_{\b}^1\theta_{\a}^2)
~\frac{\pa^4}{\pa w_a^{1}\pa w_b^{2}\pa
\theta_{\beta}^2\pa \theta_{\alpha}^1}\hspace{15cm}
\eea

\vspace{-0.6cm}

\bea\nonumber
\Lambda_{8}&=&\frac{1}{4}\theta_{\a}^1\theta_{\b}^1\theta_{\gamma}^2\theta_{\delta}^2
~\frac{\pa^4}{\pa
\theta_{\delta}^2\pa
\theta_{\g}^2\pa \theta_{\b}^1\pa \theta_{\alpha}^1}\hspace{15cm}
\eea

\vspace{-0.6cm}

\bea\nonumber
 \Lambda_9&=&
\frac{1}{6}(w_a^1w_b^1w_c^2+2w_a^1w_b^2w_c^1)\theta_{\alpha}^2~\frac{\pa^4}{\pa
w_a^{1}\pa w_b^{1}\pa w_c^2\pa \theta_{\alpha}^2}
\hspace{15cm}
\eea

\vspace{-0.6cm}

\bea\nonumber
\Lambda_{10}&=&\frac{1}{6}(2w_a^2w_b^2w_c^1+w_a^1w_b^2w_c^2)
\theta_{\alpha}^1~\frac{\pa^4}{\pa w_a^{1}\pa w_b^{2}\pa w_c^2\pa
\theta_{\alpha}^1}\hspace{15cm}
\eea

\vspace{-0.6cm}

\bea\nonumber
\Lambda_{11}&=&\frac{1}{3}\eps_{bc}w_a^1\eps^{kl}w_k^1w_l^2
\theta_{\alpha}^2~\frac{\pa^4}{\pa w_a^{1}\pa w_b^{1}\pa w_c^2\pa
\theta_{\alpha}^2} \hspace{15cm}
\eea

\vspace{-0.6cm}

\bea\nonumber
\Lambda_{12}&=&\frac{1}{3}\eps_{ab}w_c^2\eps^{kl}w_k^1w_l^2
\theta_{\alpha}^1~\frac{\pa^4}{\pa w_a^{1}\pa w_b^{2}\pa w_c^2\pa
\theta_{\alpha}^1}
\hspace{15cm}
\eea

\vspace{-0.6cm}

\bea\nonumber
\Lambda_{13}&=&\frac{1}{4}w_a^1w_b^1\theta_{\alpha}^2\theta_{\beta}^2
~\frac{\pa^4}{\pa w_a^{1}\pa w_b^{1}\pa \theta_{\b}^2\pa
\theta_{\alpha}^2} \hspace{15cm}
\eea

\vspace{-0.6cm}

\bea\nonumber
\Lambda_{14}&=&\frac{1}{4}w_a^2w_b^2\theta_{\alpha}^1\theta_{\beta}^1
~\frac{\pa^4}{\pa w_a^{2}\pa w_b^{2}\pa \theta_{\b}^1\pa
\theta_{\alpha}^1} \hspace{15cm}
\eea

\vspace{-0.6cm}

\bea\nonumber
\Lambda_{15}&=&w_a^2\eps_{\beta\gamma}\theta_{\a}^1\eps^{\rho\delta}\theta_{\rho}^1\theta_{\delta}^2~\frac{\pa^4}{\pa
w_a^{2}\pa
\theta_{\gamma}^2\pa \theta_{\b}^1\pa \theta_{\a}^1}\hspace{15cm}
\eea

\vspace{-0.6cm}

\bea\nonumber
\Lambda_{16}&=&w_a^1\eps_{\a\b}\theta_{\g}^2\eps^{\rho\delta}\theta_{\rho}^1\theta_{\delta}^2~\frac{\pa^4}{\pa
w_a^{1}\pa \theta_{\g}^2\pa
\theta_{\b}^2\pa \theta_{\a}^1}\hspace{15cm}
\eea

\vspace{-0.6cm}

\bea\nonumber
\Lambda_{17}&=&\frac{1}{4}\eps_{\a\gamma}\eps_{\beta\delta}\eps^{ab}w_a^1w_b^2\eps^{cd}w_c^1w_d^2
~\frac{\pa^4}{\pa \theta_{\delta}^2\pa
\theta_{\gamma}^2\pa \theta_{\b}^{1}\pa \theta_{\a}^1}
\hspace{15cm}
\eea

\vspace{-0.6cm}

\bea\nonumber
 \Lambda_{18}&=&\frac{1}{4}\eps^{\a\gamma}\eps^{\beta\delta}
 \theta_{\a}^1\theta_{\beta}^1\theta_{\gamma}^2\theta_{\delta}^2~\eps_{ac}\eps_{bd}
 ~\frac{\pa^4}{\pa w_a^{1}\pa w_{b}^1\pa
w_c^2 \pa w_d^2} \hspace{15cm}
\eea

\vspace{-0.6cm}

\bea\nonumber
\Lambda_{19}&=&\frac{1}{2}\eps^{kl}w_k^1w_l^2\eps^{cd}w_c^1w_d^2\eps_{ab}\eps_{\a\beta}~\frac{\pa^4}{\pa
w_a^{1}\pa w_{b}^2\pa \theta_{\beta}^2\pa \theta_{\alpha}^1 }
\hspace{15cm}
\eea

\vspace{-0.6cm}

\bea\nonumber
\Lambda_{20}&=&{1\ov 2}\eps^{kl}w_k^1w_{l}^2\eps^{\a\b}\theta_{\a}^1\theta_{\b}^2
\eps_{ac}\eps_{bd} ~\frac{\pa^4}{\pa w_a^{1}\pa w_b^{1}\pa
w_c^2\pa w_d^2}
\hspace{15cm}
\eea

\vspace{-0.6cm}

\bea\nonumber
\Lambda_{21}&=&\frac{1}{2}\eps_{\a\b}\eps_{ab}\eps^{\gamma\delta}\eps^{\rho\lambda}
\theta^1_{\gamma}\theta^1_{\rho}\theta^2_{\delta}\theta^2_{\lambda}
 ~\frac{\pa^4}{\pa w_a^{1}\pa w_b^{2}\pa \theta_{\b}^2\pa
\theta_{\a}^1}\hspace{15cm}
\eea

\vspace{-0.6cm}

\bea\nonumber
\Lambda_{22}&=&\frac{1}{2}\eps^{kl}w_k^1w_{l}^2\eps^{\mu\rho}\theta_{\mu}^1\theta_{\rho}^2
\eps_{\a\gamma}\eps_{\b\delta}
 ~\frac{\pa^4}{\pa \theta_{\delta}^2\pa
\theta_{\gamma}^2\pa \theta_{\b}^{1}\pa \theta_{\a}^{1}}\hspace{15cm}
\eea

\vspace{-0.6cm}

\bea\nonumber
\Lambda_{23}
 &=&\frac{1}{2}(w_b^1w_c^2+w_b^2w_c^1)\eps^{kl}w_k^1w_l^2\eps_{\a\b}
 ~\frac{\pa^4}{\pa w_b^{1}\pa w_c^{2}\pa
 \theta_{\b}^2 \pa \theta_{\a}^1}\hspace{15cm}
\eea

\vspace{-0.6cm}

\bea\nonumber
\Lambda_{24}
 &=&{1\ov 2} \eps^{\a\b}\theta_{\a}^1\theta_{\b}^2\, \eps_{ad}(w_b^1w_c^2+w_c^1w_b^2)
~\frac{\pa^4}{\pa w_a^1 \pa w_b^1 \pa w_c^2 \pa w_d^2}\hspace{15cm}
\eea

\vspace{-0.6cm}

\bea\nonumber
\Lambda_{25}
 &=&\frac{1}{4}w_a^1w_b^1\theta_{\a}^2\theta_{\b}^2
 ~\frac{\pa^4}{\pa w_a^2 \pa w_b^2 \pa \theta_{\b}^1\pa
 \theta_{\a}^1}\hspace{15cm}
\eea

\vspace{-0.6cm}

\bea\nonumber
 \Lambda_{26}
 &=&\frac{1}{4}w_a^2w_b^2\theta_{\a}^1\theta_{\b}^1
 ~\frac{\pa^4}{\pa w_a^1 \pa w_b^1 \pa \theta_{\b}^2\pa
 \theta_{\a}^2}\hspace{15cm}
\eea

\vspace{-0.6cm}

\bea\nonumber
 \Lambda_{27}
 &=&\frac{1}{3}\eps_{ac}w_b^1\theta_{\a}^2\eps^{kl}w_k^1w_l^2
 ~\frac{\pa^4}{\pa w_a^1 \pa w_b^2 \pa w_c^2\pa
 \theta_{\a}^1}\hspace{15cm}
\eea

\vspace{-0.6cm}

\bea\nonumber
 \Lambda_{28}
 &=&\frac{1}{3}\eps_{ac}w_b^2\theta_{\a}^1\eps^{kl}w_k^1w_l^2
 ~\frac{\pa^4}{\pa w_a^1 \pa w_b^1 \pa w_c^2\pa
 \theta_{\a}^2}\hspace{15cm}
\eea

\vspace{-0.6cm}

\bea\nonumber
\Lambda_{29}
 &=&{1\ov 2}\eps_{\a\b}\theta_{\gamma}^2
 \eps^{\rho\delta}\theta_{\rho}^1\theta_{\delta}^2
w_a^1 ~\frac{\pa^4}{\pa w_a^2
\pa
\theta_{\gamma}^2\pa \theta_{\b}^1 \pa \theta_{\a}^1}\hspace{15cm}
\eea

\vspace{-0.6cm}

\bea\nonumber
\Lambda_{30}
 &=&{1\ov 2}\eps_{\beta\gamma}\theta_{\a}^1
 \eps^{\rho\delta}\theta_{\rho}^1\theta_{\delta}^2
w_a^2 ~\frac{\pa^4}{\pa w_a^1 \pa \theta_{\g}^2 \pa
\theta_{\b}^2 \pa \theta_\a^1}\hspace{15cm}
\eea

\vspace{-0.6cm}

\bea\nonumber
 \Lambda_{31}
 &=&\frac{1}{6}(2w_a^1w_b^1w_c^2+w_a^2w_b^1w_c^1)\theta_{\a}^2
~\frac{\pa^4}{\pa w_a^1\pa w_b^2\pa w_c^2 \pa \theta_{\a}^1}\hspace{15cm}
\eea

\vspace{-0.6cm}

\bea\nonumber
 \Lambda_{32}
 &=&\frac{1}{6}(2w_a^1w_b^2w_c^2+w_a^2w_b^2w_c^1)\theta_{\a}^1
~\frac{\pa^4}{\pa w_a^1\pa w_b^1\pa w_c^2 \pa \theta_{\a}^2}\hspace{15cm}
 \eea

 \vspace{-0.6cm}

\bea\nonumber
\Lambda_{33}
 &=&w_a^1w_b^1\theta_{\a}^2\theta_{\b}^2
 ~\frac{\pa^4}{\pa w_a^1 \pa w_b^2\pa
 \theta_{\b}^2\pa \theta_{\a}^1}\hspace{15cm}
\eea

\vspace{-0.6cm}

\bea\nonumber
\Lambda_{34}
 &=&w_a^1w_b^2\theta_{\a}^1\theta_{\b}^2
 ~\frac{\pa^4}{\pa w_a^1 \pa w_b^1 \pa \theta_{\b}^2\pa
 \theta_{\a}^2}\hspace{15cm}
\eea

\vspace{-0.6cm}

\bea\nonumber
\Lambda_{35}
 &=&w_a^2w_b^2\theta_{\a}^1\theta_{\b}^1
 ~\frac{\pa^4}{\pa w_a^1 \pa w_b^2\pa
 \theta_{\b}^2 \pa \theta_{\a}^1}\hspace{15cm}
\eea

\vspace{-0.6cm}

\bea\nonumber
\Lambda_{36}
 &=&w_a^1w_b^2\theta_{\a}^1\theta_{\b}^2
 ~\frac{\pa^4}{\pa w_a^2 \pa w_b^2 \pa
 \theta_{\b}^1\pa \theta_{\a}^1}
\hspace{15cm}
\eea

\vspace{-0.6cm}

\bea\nonumber
 \Lambda_{37}&=&\frac{1}{4}w_b^1w_c^2
\eps^{kl}w_k^1w_l^2\eps_{\a\b}
 ~\frac{\pa^4}{\pa w_b^{1}\pa w_c^{1} \pa \theta_{\b}^2\pa
 \theta_{\a}^2}\hspace{15cm}
\eea

\vspace{-0.6cm}

\bea\nonumber
\Lambda_{38}&=&\frac{1}{4}\eps_{ad}w_b^1w_c^1\eps^{\a\b}\theta_{\a}^2\theta_{\beta}^2
 ~\frac{\pa^4}{\pa w_a^{1}\pa w_b^{1} \pa w_c^2\pa w_d^2}\hspace{15cm}
\eea

\vspace{-0.6cm}

\bea\nonumber
\Lambda_{39}&=&\frac{1}{4}w_b^1w_c^2
\eps^{kl}w_k^1w_l^2\eps_{\a\b}
 ~\frac{\pa^4}{\pa w_b^{2}\pa w_c^{2} \pa \theta_{\b}^1\pa
 \theta_{\a}^1}\hspace{15cm}
\eea

\vspace{-0.6cm}

\bea\nonumber
\Lambda_{40}&=&
\frac{1}{4}\eps_{ad}w_b^2w_c^2\eps^{\a\b}\theta_{\a}^1\theta_{\beta}^1
 ~\frac{\pa^4}{\pa w_a^{1}\pa w_b^{1} \pa w_c^2\pa w_d^2}\hspace{15cm}
\eea

\vspace{-0.6cm}

\bea\nonumber
\Lambda_{41}
 &=&w_a^1 \eps^{kl}w_k^1w_l^2
 \eps_{\a\b}\theta_{\gamma}^2
 ~\frac{\pa^4}{\pa w_a^1  \pa \theta_{\g}^2\pa
 \theta_{\b}^2\pa \theta_{\a}^1}\hspace{15cm}
\eea

\vspace{-0.6cm}

\bea\nonumber
 \Lambda_{42}
 &=&\theta_{\a}^2
 \eps^{\rho\delta}\theta_{\rho}^1\theta_{\delta}^2
 \eps_{bc}w_a^1
 ~\frac{\pa^4}{\pa w_a^1 \pa w_b^1 \pa w_c^2
\pa \theta_{\a}^2}\hspace{15cm}
\eea

\vspace{-0.6cm}

\bea\nonumber
\Lambda_{43}
 &=&w_c^2
 \eps^{kl}w_k^1w_l^2
\eps_{\b\gamma}\theta_{\a}^1 ~\frac{\pa^4}{\pa w_c^2\pa \theta_{\gamma}^2 \pa
\theta_{\b}^1 \pa \theta_{\a}^1 }\hspace{15cm}
\eea

\vspace{-0.6cm}

\bea\nonumber
 \Lambda_{44}
 &=&\theta_{\a}^1
 \eps^{\rho\delta}\theta_{\rho}^1\theta_{\delta}^2
\eps_{ab}w_c^2 ~\frac{\pa^4}{\pa w_a^1 \pa w_b^2 \pa w_c^2 \pa
\theta_{\a}^1}\hspace{15cm}
\eea

\vspace{-0.6cm}

\bea\nonumber
 \Lambda_{45}
 &=&w_a^1\eps^{kl}w_k^1w_l^2
 \eps_{\beta\gamma}\theta_{\a}^2
~\frac{\pa^4}{\pa w_a^2 \pa\theta_{\gamma}^2 \pa
\theta_{\b}^1\pa \theta_{\a}^1} \hspace{15cm}
\eea

\vspace{-0.6cm}

\bea\nonumber
\Lambda_{46}
 &=&\theta_{\a}^1
 \eps^{\rho\delta}\theta_{\rho}^1\theta_{\delta}^2
\eps_{bc}w_a^2 ~\frac{\pa^4}{\pa w_a^1 \pa w_b^1 \pa w_c^2 \pa
\theta_{\a}^2}\hspace{15cm} \eea

\vspace{-0.6cm}

\bea\nonumber
 \Lambda_{47}
 &=&w_c^2
 \eps^{kl}w_k^1w_l^2
\eps_{\a\b}\theta_{\gamma}^1 ~\frac{\pa^4}{\pa w_c^1 \pa
\theta_{\g}^2 \pa \theta_{\b}^2\pa \theta_{\a}^1}\hspace{15cm}
\eea

\vspace{-0.6cm}

\bea\nonumber
\Lambda_{48}
 &=&\theta_{\a}^2
 \eps^{\rho\delta}\theta_{\rho}^1\theta_{\delta}^2
\eps_{ab}w_c^1 ~\frac{\pa^4}{\pa w_a^1 \pa w_b^2 \pa w_c^2 \pa
\theta_{\a}^1}\hspace{15cm} \eea

\vskip 1cm

\subsubsection{Coefficients of the S-matrix $\S^{BB}$}\la{SBBcoef}
Below $y^\pm_1$ and $y^{\pm}_2$ parameterize the two-particle
bound state with momenta \mbox{$p_1=2\, {\rm am}z_1$} and $p_2=2\,
{\rm am}z_2$, respectively, and they are constrained by
eqs.(\ref{sc}). The parameters $\eta_k$ are given by
\bea\la{etakbb} \eta_1 = e^{ip_2/2}\eta(z_1,2)\,,\quad \eta_2
=\eta(z_2,2)\,,\quad \tilde{\eta}_1 = \eta(z_1,2)\,,\quad
\tilde{\eta}_2 = e^{ip_1/2}\eta(z_2,2)\,, \eea where $\eta(z,M)$
is defined by eq.(\ref{etaz}). The coefficients $a_k$ are
meromorphic functions of $z_1$ and $z_2$ defined on the product of
two equivalent tori. We also use the variables $u_1$ and $u_2$
$$
u_1=\frac{1}{2}\Big(y^+_1+\frac{1}{y^+_1}+y^-_1+\frac{1}{y^-_1}\Big)\,
, ~~~~~
u_2=\frac{1}{2}\Big(y^+_2+\frac{1}{y^+_2}+y^-_2+\frac{1}{y^-_2}\Big)\,
.
$$
With this notation the coefficients $a_k$ have the form

 {\small
\bea\nonumber a_1&=&1 \hspace{15cm}
\eea

\vspace{-0.8cm}

\bea\nonumber
a_2&=&-\frac{(u_1-u_2-\frac{2i}{g})
(y_{1}^{-}-y_{2}^{+}) (-1+y_{1}^{-} y_{2}^{+})y_{2}^{-} y_{1}^{+}
}{(u_1-u_2+\frac{2i}{g})(y_{2}^{-}-y_{1}^{+}) (-1+y_{1}^{-}
y_{2}^{-}) y_{1}^{-} y_{2}^{+}}-\frac{
(y_{1}^{-}-y_{2}^{+})(y_{1}^{-}-y_{1}^{+}) (-1+y_{1}^{-}
y_{2}^{+}) y_{2}^{-}}{2 (u_1-u_2+\frac{2i}{g}) (-1+y_{1}^{-}
y_{2}^{-}) y_{1}^{-} y_{2}^{+}}
\\ \nonumber
&~&- \frac{3 (y_{1}^{-}-y_{2}^{+}) (y_{1}^{-}-y_{1}^{+})
(y_{1}^{+}- y_{2}^{+})(-1+y_{1}^{-}
y_{2}^{+})y_{2}^{-}}{2(u_1-u_2+\frac{2i}{g})
(y_{2}^{-}-y_{1}^{+})(-1+y_{1}^{-} y_{2}^{-}) y_{1}^{-}
y_{2}^{+} }\hspace{15cm}
\eea

\vspace{-0.6cm}

\bea\nonumber
a_3&=&-\frac{(y_{1}^{-}-y_{1}^{+}) (y_{2}^{-}-y_{2}^{+})
(y_{1}^{+}-y_{2}^{+})}{(u_1-u_2+\frac{2i}{g}) (-1+y_{1}^{-}
y_{2}^{-})y_{1}^{+} y_{2}^{+}
 }\\
\nonumber &~&+ \frac{(y_{2}^{-}-y_{1}^{+})
(-y_{2}^{-}+y_{2}^{+})+y_{2}^{-} (y_{1}^{+}-y_{2}^{+})
(1-y_{1}^{-} y_{2}^{+})-y_{2}^{-} (-y_{2}^{-}+y_{2}^{+})
(1-y_{1}^{-} y_{2}^{+})}{(-1+y_{1}^{-} y_{2}^{-}) (y_{2}^{-}-y_{1}^{+}) y_{2}^{+} }\hspace{15cm}
\eea

\vspace{-0.6cm}

\bea\nonumber
a_{4}&=&-\frac{y_{1}^{-}-y_{2}^{+}
}{y_{2}^{-}-y_{1}^{+} }\frac{\tilde{\eta}_1 \tilde{\eta}_2} {\eta_1\eta_2}\hspace{15cm}
\eea

\vspace{-0.6cm}

\bea\nonumber
a_{5}&=&-\frac{2 (y_{1}^{-}-y_{2}^{-})
(y_{1}^{+}-y_{2}^{+})(-1+y_{2}^{-}
y_{1}^{+})}{(u_1-u_2+\frac{2i}{g})(y_{2}^{-}-y_{1}^{+}) y_{1}^{+}
y_{2}^{-}\text{ }}
\frac{\tilde{\eta}_1\tilde{\eta}_2}{\eta_1\eta_2}+\frac{
y_{1}^{-}-y_{2}^{+}}{
y_{2}^{-}-y_{1}^{+}}\frac{\tilde{\eta}_1\tilde{\eta}_2}{\eta_1\eta_2}\hspace{15cm}
\eea

\vspace{-0.6cm}

\bea\nonumber
a_{6}&=&-\frac{(u_1-u_2-\frac{2i}{g})(y_{1}^{-}-y_{2}^{+})}{(u_1-u_2+\frac{2i}{g})
(y_{2}^{-}-y_{1}^{+}) }\frac{\tilde{\eta}_1\tilde{\eta}_2}{\eta_1\eta_2}\hspace{15cm}
\eea

\vspace{-0.6cm}

\bea\nonumber
a_{7}&=& -\frac{(y_{1}^{-}-y_{2}^{+})}{2
(u_1-u_2+\frac{2i}{g})  (-1+y_{1}^{-} y_{2}^{-})
(y_{2}^{-}-y_{1}^{+}) y_{1}^{+} y_{2}^{+}y_{2}^{-}}
\left[y_{2}^{-} y_{1}^{+}+3 y_{1}^{-} (y_{2}^-)^2
y_{1}^{+}-4 (y_{2}^-)^2 (y_{1}^+)^2 \right. \\
\nonumber &~& -2 y_{2}^{-} y_{2}^{+}-2 y_{1}^{-} (y_{2}^-)^2
y_{2}^{+}+y_{1}^{+} y_{2}^{+}-5 y_{1}^{-} y_{2}^{-} y_{1}^{+}
y_{2}^{+}+5 (y_{2}^-)^2 y_{1}^{+} y_{2}^{+}-y_{1}^{-}
(y_{2}^{-})^3 y_{1}^{+}
y_{2}^{+}+2 y_{2}^{-} (y_{1}^+)^2 y_{2}^{+}\\
\nonumber &~& \left.+2 y_{1}^{-} (y_{2}^-)^2 (y_{1}^+)^2
y_{2}^{+}+4 y_{1}^{-} y_{2}^{-} (y_{2}^+)^2-3 y_{2}^{-} y_{1}^{+}
(y_{2}^+)^2-y_{1}^{-}
(y_{2}^-)^2 y_{1}^{+} (y_{2}^+)^2 \right]  \frac{\tilde{\eta}_1 \tilde{\eta}_2}{\eta_1 \eta_2}\hspace{15cm}
\eea

\vspace{-0.6cm}

\bea\nonumber
a_{8}&=&-\frac{
(y_{1}^{-}-y_{2}^{+}) (y_{1}^{+}-y_{2}^{+})(-1+y_{2}^{-}
y_{1}^{+})^2y_{1}^{-} }{2 (u_1-u_2+\frac{2i}{g})  (-1+y_{1}^{-}
y_{2}^{-}) (y_{2}^{-}-y_{1}^{+}) (y_{1}^+)^2y_{2}^{-}
}\frac{\tilde{\eta}_1^2\tilde{\eta}_2^2}{\eta_1^2\eta_2^2}-\frac{(y_{1}^{-}-y_{2}^{+})
(y_{1}^{-}-y_{2}^{+})(-1+y_{2}^{-} y_{1}^{+}) }{2
(u_1-u_2+\frac{2i}{g})
 (y_{2}^{-}-y_{1}^{+}) y_{1}^{+}y_{2}^{-}} \frac{\tilde{\eta}_1^2\tilde{\eta}_2^2}{\eta_1^2\eta_2^2}
\hspace{15cm}
\eea

\vspace{-0.6cm}

\bea\nonumber
a_9&=&-\frac{y_{1}^{-}-y_{2}^{-}}{y_{2}^{-}-y_{1}^{+}} \frac{\tilde{\eta}_2}{\eta_2}\hspace{15cm}
\eea

\vspace{-0.6cm}

\bea\nonumber
a_{10}&=&-\frac{y_{1}^{+}-y_{2}^{+}}{y_{2}^{-}-y_{1}^{+}}\frac{\tilde{\eta}_1}{\eta_1}\hspace{15cm}
\eea

\vspace{-0.6cm}

\bea\nonumber
a_{11}&=&-\frac{(y_{1}^{-}-y_{2}^{-})(y_{1}^{-}-y_{2}^{+})
(-1+y_{1}^{-} y_{2}^{+}) }{(u_1-u_2+\frac{2i}{g})
(y_{2}^{-}-y_{1}^{+})y_{1}^{-}y_{2}^{+}}\frac{\tilde{\eta}_2}{\eta_2}\hspace{15cm}
\eea

\vspace{-0.6cm}

\bea\nonumber
a_{12}&=&\frac{ (y_{1}^{-}-y_{2}^{+})
(-y_{1}^{+}+y_{2}^{+}) (-1+y_{1}^{-}
y_{2}^{+})}{(u_1-u_2+\frac{2i}{g}) (y_{2}^{-}-y_{1}^{+})y_{1}^{-}
y_{2}^{+}
}\frac{\tilde{\eta}_1}{\eta_1}\hspace{15cm}
\eea

\vspace{-0.6cm}

\bea\nonumber
a_{13}&=&\frac{y_{2}^{-}-y_{1}^{-}}{y_{2}^{-}-y_{1}^{+}}\frac{\tilde{\eta}_2^2}{\eta_2^2}-\frac{(y_{1}^{-}-y_{2}^{-})
(y_{1}^{-}-y_{1}^{+}) (y_{2}^{+}(-1+y_{1}^{+} y_{2}^{-})+y_{2}^{-}
(-1+y_{1}^{+} y_{2}^{+}))}{2 (u_1-u_2+\frac{2i}{g})(y_{2}^{-}-y_{1}^{+}) y_{1}^{+}y_{2}^{+}y_{2}^{-}  }\frac{\tilde{\eta}_2^2}{\eta_2^2}\hspace{15cm}
\eea

\vspace{-0.6cm}

\bea\nonumber
a_{14}&=&-\frac{(u_1-u_2-\frac{2i}{g})
(y_{1}^{+}-y_{2}^{+}) y_{2}^{+} }{(u_1-u_2+\frac{2i}{g})
(y_{2}^{-}-y_{1}^{+})y_{2}^{-}
}\frac{\tilde{\eta}_1^2}{\eta_1^2}-\frac{(y_{1}^{-}+y_{1}^{+}-2
y_{2}^{+}) (y_{2}^{-}-y_{2}^{+}) (y_{1}^{+}-y_{2}^{+}) }{2
(u_1-u_2+\frac{2i}{g})  (y_{2}^{-}-y_{1}^{+})y_{2}^{-}
}\frac{\tilde{\eta}_1^2}{\eta_1^2}\hspace{15cm}
\eea

\vspace{-0.6cm}

\bea\nonumber
 a_{15}&=&\frac{
(y_{1}^{-}-y_{2}^{+}) (y_{1}^{+}-y_{2}^{+})(1-y_{2}^{-} y_{1}^{+})
}{(u_1-u_2+\frac{2i}{g})  (y_{2}^{-}-y_{1}^{+}) y_{1}^{+}y_{2}^{-}
}
\frac{\tilde{\eta}_1^2\tilde{\eta}_2}{\eta_1^2 \eta_2}\hspace{15cm}
\eea

\vspace{-0.6cm}

\bea\nonumber
a_{16}&=&-\frac{(y_{1}^{-}-y_{2}^{-})(y_{1}^{-}-y_{2}^{+})
(-1+y_{2}^{-} y_{1}^{+}) }{(u_1-u_2+\frac{2i}{g})
(y_{2}^{-}-y_{1}^{+}) y_{1}^{+}y_{2}^{-}} \frac{\tilde{\eta}_1 \tilde{\eta}_2^2}{\eta_1\eta_2^2}\hspace{15cm}
\eea

\vspace{-0.6cm}

\bea\nonumber
a_{17}&=&-\frac{(y_{1}^{-}-y_{1}^{+})^2
(y_{1}^{-}-y_{2}^{+}) (y_{2}^{-}-y_{2}^{+})^2
(y_{1}^{+}-y_{2}^{+})}{2 (u_1-u_2+\frac{2i}{g})
 (-1+y_{1}^{-} y_{2}^{-}) (y_{2}^{-}-y_{1}^{+}) y_{1}^{-} y_{2}^{-} }\frac{1}{\eta_1^2 \eta_2^2}\hspace{15cm}
\eea

\vspace{-0.6cm}

\bea\nonumber
a_{18}&=&-\frac{y_{1}^{-} y_{2}^{-}
(y_{1}^{-}-y_{2}^{+}) (y_{1}^{+}-y_{2}^{+}) }{2 (u_1-u_2+\frac{2i}{g})  (y_{2}^{-}-y_{1}^{+}) (-1+y_{1}^{-} y_{2}^{-})(y_{1}^+)^2 (y_{2}^+)^2}\tilde{\eta}_1^2\tilde{\eta}_2^2\hspace{15cm}
\eea

\vspace{-0.6cm}

\bea\nonumber
a_{19}&=&-\frac{i (y_{1}^{-}-y_{1}^{+})
(y_{1}^{-}-y_{2}^{+}) (y_{2}^{-}-y_{2}^{+}) (y_{1}^{+}-y_{2}^{+})
(-1+y_{1}^{-} y_{2}^{+})}{2 (u_1-u_2+\frac{2i}{g})
(y_{2}^{-}-y_{1}^{+})(-1+y_{1}^{-} y_{2}^{-}) y_{1}^{-}y_{2}^{+}}
\frac{1}{\eta_1\eta_2}\hspace{15cm}
\eea

\vspace{-0.6cm}

\bea\nonumber
a_{20}&=&\frac{i
(y_{1}^{-}-y_{2}^{+}) (y_{1}^{+}-y_{2}^{+})
(-1+y_{1}^{-} y_{2}^{+})y_2^-}{2 (u_1-u_2+\frac{2i}{g})
(y_{2}^{-}-y_{1}^{+})(-1+y_{1}^{-} y_{2}^{-}) y_{1}^{+}y_{2}^{+}{}^2} \tilde{\eta}_1\tilde{\eta}_2
\hspace{15cm}
\eea

\vspace{-0.6cm}

\bea\nonumber
a_{21}&=&-\frac{i (y_{1}^{-}-y_{2}^{+})
(y_{1}^{+}-y_{2}^{+}) (-1+y_{2}^{-} y_{1}^{+})y_{1}^{-} }{2
(u_1-u_2+\frac{2i}{g})
(y_{2}^{-}-y_{1}^{+})(-1+y_{1}^{-} y_{2}^{-}) (y_{1}^+)^2 y_{2}^{+} }  \frac{\tilde{\eta}_1^2\tilde{\eta}_2^2}{\eta_1\eta_2}\hspace{15cm}
\eea

\vspace{-0.6cm}

\bea\nonumber
a_{22}&=& \frac{ i (y_{1}^{-}-y_{1}^{+})
 (y_{1}^{-}-y_{2}^{+})
(y_{2}^{-}-y_{2}^{+}) (y_{1}^{+}-y_{2}^{+})
(-1+y_{2}^{-}y_{1}^{+})}{ 2(u_1-u_2+\frac{2i}{g})
(y_{2}^{-}-y_{1}^{+})(-1+y_{1}^{-} y_{2}^{-})  y_{1}^{+} y_{2}^{-}
}\frac{\tilde{\eta}_1\tilde{\eta}_2}{\eta_1^2\eta_2^2}\hspace{15cm}
\eea

\vspace{-0.6cm}

\bea\nonumber
a_{23}&=&-\frac{i (y_{1}^{-}-y_{2}^{-})
(y_{1}^{-}-y_{1}^{+}) (y_{2}^{-}-y_{2}^{+})
(y_{1}^{+}-y_{2}^{+})}{2 (u_1-u_2+\frac{2i}{g})
 (y_{2}^{-}-y_{1}^{+})y_{1}^{-} y_{2}^{-} }\frac{1}{\eta_1\eta_2}\hspace{15cm}
\eea

\vspace{-0.6cm}

\bea\nonumber
a_{24}&=&\frac{i (y_{1}^{-}-y_{2}^{-})
(y_{1}^{+}-y_{2}^{+}) }{2 (u_1-u_2+\frac{2i}{g})
(y_{2}^{-}-y_{1}^{+}) y_{1}^{+} y_{2}^{+}} \tilde{\eta}_1 \tilde{\eta}_2\hspace{15cm}
\eea

\vspace{-0.6cm}

\bea\nonumber
a_{25}&=&-\frac{(y_{1}^{-}-y_{1}^{+})^2 (-1+y_{2}^{-}
y_{1}^{+}) }{2 (u_1-u_2+\frac{2i}{g}) (y_{2}^{-}-y_{1}^{+})
y_{1}^{+}y_{2}^{-}}
\frac{\tilde{\eta}_2^2}{\eta_1^2}    \hspace{15cm}
\eea

\vspace{-0.6cm}

\bea\nonumber
 a_{26}&=&-\frac{
(y_{2}^{-}-y_{2}^{+})^2(-1+y_{2}^{-} y_{1}^{+})}{2
(u_1-u_2+\frac{2i}{g})
 (y_{2}^{-}-y_{1}^{+}) y_{1}^{+}y_{2}^{-} }  \frac{\tilde{\eta}_1^2}{\eta_2^2}\hspace{15cm}
\eea

\vspace{-0.6cm}

\bea\nonumber
a_{27}&=&\frac{(y_{1}^{-}-y_{1}^{+})
(y_{1}^{-}-y_{2}^{+}) (-1+y_{1}^{-} y_{2}^{+}) }{2
(u_1-u_2+\frac{2i}{g})
 (y_{2}^{-}-y_{1}^{+}) y_{1}^{-}y_{2}^{+} } \frac{\tilde{\eta}_2}{\eta_1}
\hspace{15cm}
\eea

\vspace{-0.6cm}

\bea\nonumber
a_{28}&=&\frac{(u_1-u_2-\frac{2i}{g})
(y_{2}^{-}-y_{2}^{+}) }{(u_1-u_2+\frac{2i}{g})
(y_{2}^{-}-y_{1}^{+})} \frac{\tilde{\eta}_1}{\eta_2}
-\frac{(y_{2}^{-}-y_{2}^{+})
(y_{1}^{+}-y_{2}^{+}) (-1+y_{1}^{+} y_{2}^{+})}{2 (u_1-u_2+\frac{2i}{g}) (y_{2}^{-}-y_{1}^{+}) y_{1}^{+} y_{2}^{+} } \frac{\tilde{\eta}_1}{\eta_2}\hspace{15cm}
\eea

\vspace{-0.6cm}

\bea\nonumber
a_{29}&=&\frac{(y_{1}^{-}-y_{1}^{+})
(y_{1}^{-}-y_{2}^{+}) (-1+y_{2}^{-} y_{1}^{+})}{2
(u_1-u_2+\frac{2i}{g})
 (y_{2}^{-}-y_{1}^{+}) y_{1}^{+} y_{2}^{-}}
\frac{\tilde{\eta}_1\tilde{\eta}_2^2}{\eta_1^2\eta_2}
\hspace{15cm}
\eea

\vspace{-0.6cm}

\bea\nonumber
a_{30}&=&\frac{ (y_{1}^{-}-y_{2}^{+})
(y_{2}^{-}-y_{2}^{+}) (-1+y_{2}^{-} y_{1}^{+})}{2
(u_1-u_2+\frac{2i}{g})  (y_{2}^{-}-y_{1}^{+}) y_{1}^{+} y_{2}^{-}}
\frac{\tilde{\eta}_1^2\tilde{\eta}_2}{\eta_1\eta_2^2}\hspace{15cm}
\eea

\vspace{-0.6cm}

\bea\nonumber
a_{31}&=&\frac{y_{1}^{-}-y_{1}^{+}}{y_{2}^{-}-y_{1}^{+} } \frac{\tilde{\eta}_2}{\eta_1}\hspace{15cm}
\eea

\vspace{-0.6cm}

\bea\nonumber
a_{32}&=&\frac{y_{2}^{-}-y_{2}^{+} }{y_{2}^{-}-y_{1}^{+}
} \frac{\tilde{\eta}_1}{\eta_2}\hspace{15cm}\eea

\vspace{-0.6cm}

\bea\nonumber
a_{33}&=&\frac{(y_{1}^{-}-y_{2}^{-})
(y_{1}^{-}-y_{1}^{+}) (-1+y_{2}^{-} y_{1}^{+}) }{2
(u_1-u_2+\frac{2i}{g})
 (y_{2}^{-}-y_{1}^{+}) y_{1}^{+}y_{2}^{-}} \frac{\tilde{\eta}_2^2}{\eta_1\eta_2}\hspace{15cm}
\eea

\vspace{-0.6cm}

\bea\nonumber
a_{34}&=&\frac{(y_{1}^{-}-y_{2}^{-})
(y_{2}^{-}-y_{2}^{+})(-1+y_{2}^{-} y_{1}^{+})  }{2
(u_1-u_2+\frac{2i}{g})
 (y_{2}^{-}-y_{1}^{+}) y_{1}^{+}y_{2}^{-} }
\frac{\tilde{\eta}_1\tilde{\eta}_2}{\eta_2^2}\hspace{15cm}
\eea

\vspace{-0.6cm}

\bea\nonumber
a_{35}&=&\frac{(y_{2}^{-}-y_{2}^{+})
(y_{1}^{+}-y_{2}^{+})(-1+y_{2}^{-} y_{1}^{+})}{2
(u_1-u_2+\frac{2i}{g})
 (y_{2}^{-}-y_{1}^{+}) y_{1}^{+} y_{2}^{-}} \frac{\tilde{\eta}_1^2}{\eta_1\eta_2}\hspace{15cm}
\eea

\vspace{-0.6cm}

\bea\nonumber
a_{36}&=&\frac{(y_{1}^{-}-y_{1}^{+})
(y_{1}^{+}-y_{2}^{+})(-1+y_{2}^{-}
y_{1}^{+}) }{2 (u_1-u_2+\frac{2i}{g})  (y_{2}^{-}-y_{1}^{+}) y_{1}^{+} y_{2}^{-}} \frac{\tilde{\eta}_1\tilde{\eta}_2}{\eta_1^2}\hspace{15cm}
\eea

\vspace{-0.6cm}

\bea\nonumber
a_{37}&=&\frac{i (y_{1}^{-}-y_{2}^{-})
(y_{1}^{-}-y_{1}^{+}) (y_{2}^{-}-y_{2}^{+})^2}{2
(u_1-u_2+\frac{2i}{g})
(y_{2}^{-}-y_{1}^{+})y_{1}^{-} y_{2}^{-}  } \frac{1}{\eta_2^2 }\hspace{15cm}
\eea

\vspace{-0.6cm}

\bea\nonumber
a_{38}&=&-\frac{i (y_{1}^{-}-y_{2}^{-})
(y_{1}^{-}-y_{1}^{+}) }{2 (u_1-u_2+\frac{2i}{g})
(y_{2}^{-}-y_{1}^{+})
y_{1}^{+} y_{2}^{+}} \tilde{\eta}_2^2\hspace{15cm}
\eea

\vspace{-0.6cm}

\bea\nonumber
a_{39}&=&\frac{i (y_{1}^{-}-y_{1}^{+})
(y_{1}^{-}-y_{1}^{+}) (y_{2}^{-}-y_{2}^{+})
(y_{1}^{+}-y_{2}^{+})}{2 (u_1-u_2+\frac{2i}{g})
(y_{2}^{-}-y_{1}^{+}) y_{1}^{-} y_{2}^{-} }\frac{1}{\eta_1^2}\hspace{15cm}
\eea

\vspace{-0.6cm}

\bea\nonumber
a_{40}&=&-\frac{i (y_{1}^{+}-y_{2}^{+})
(y_{2}^{-}-y_{2}^{+})  }{2 (u_1-u_2+\frac{2i}{g})
(y_{2}^{-}-y_{1}^{+})
y_{1}^{+} y_{2}^{+}} \tilde{\eta}_1^2\hspace{15cm}
\eea

\vspace{-0.6cm}

\bea\nonumber
a_{41}&=&-\frac{i (y_{1}^{-}-y_{2}^{-})
(y_{1}^{-}-y_{1}^{+}) (y_{1}^{-}-y_{2}^{+}) (y_{2}^{-}-y_{2}^{+})
}{2 (u_1-u_2+\frac{2i}{g}) (y_{2}^{-}-y_{1}^{+}) y_{1}^{-}
y_{2}^{-}}
\frac{\tilde{\eta}_2^2}{\eta_1\eta_2^2}\hspace{15cm}
\eea

\vspace{-0.6cm}

\bea\nonumber
a_{42}&=&\frac{i (y_{1}^{-}-y_{2}^{-})
(y_{1}^{-}-y_{2}^{+}) }{2 (u_1-u_2+\frac{2i}{g})
(y_{2}^{-}-y_{1}^{+}) y_{1}^{+} y_{2}^{+} } \frac{\tilde{\eta}_1\tilde{\eta}_2^2}{\eta_2}\hspace{15cm}
\eea

\vspace{-0.6cm}

\bea\nonumber
a_{43}&=&-\frac{i (y_{1}^{-}-y_{1}^{+})
(y_{1}^{-}-y_{2}^{+}) (y_{2}^{-}-y_{2}^{+}) (y_{1}^{+}-y_{2}^{+})
}{2 (u_1-u_2+\frac{2i}{g}) y_{1}^{-} y_{2}^{-}
(y_{2}^{-}-y_{1}^{+})}
\frac{\tilde{\eta}_1}{\eta_1^2\eta_2}\hspace{15cm}
\eea

\vspace{-0.6cm}

\bea\nonumber
a_{44}&=&\frac{i (y_{1}^{-}-y_{2}^{+})
(y_{1}^{+}-y_{2}^{+}) }{2 (u_1-u_2+\frac{2i}{g})
(y_{2}^{-}-y_{1}^{+}) y_{1}^{+} y_{2}^{+} } \frac{\tilde{\eta}_1^2\tilde{\eta}_2}{\eta_1}\hspace{15cm}
\eea

\vspace{-0.6cm}

\bea\nonumber
a_{45}&=&a_{46}=a_{47}=a_{48}=0\hspace{15cm}
\eea

}

\subsubsection{Matrix form of invariant differential operators
$\Lambda_k$}

We use the basis of the two-particle bound state representation (\ref{basis2p}).
A  $\su(2)\oplus\su(2)$ invariant differential operator is
represented in the matrix form as a sum over symbols $E_{KILJ}$ which can be
equal to either $E_K{}^I \otimes E_L{}^J$ or to
$(-1)^{\epsilon_K\epsilon_L}\, E_K{}^I \otimes E_L{}^J$ (or
anything else one wants).
 \bea\nonumber \Lambda_1 &=& E_{1111}+\frac{1}
   {2}
   E_{1122}+\frac
   {1}{6}
   E_{1133}+\frac
   {1}{2}
   E_{1221}+\frac
   {1}{3}
   E_{1232}+\frac
   {1}{6}
   E_{1331}+\frac
   {1}{2}
   E_{2112}+\frac
   {1}{3}
   E_{2123}\\\nonumber
   &+&\frac
   {1}{2}
   E_{2211}+\frac
   {2}{3}
   E_{2222}+\frac
   {1}{2}
   E_{2233}+\frac
   {1}{3}
   E_{2321}+\frac
   {1}{2}
   E_{2332}+\frac
   {1}{6}
   E_{3113}+\frac
   {1}{3}
   E_{3212}+\frac
   {1}{2}
   E_{3223}\\\nonumber
   &+&\frac
   {1}{6}
   E_{3311}+\frac
   {1}{2}
   E_{3322}+E_{3333}
\hspace{15cm}
\eea

\vspace{-0.6cm}

\bea\nonumber
\Lambda_2 &=& \frac{1}{3}
   \left(E_{1133}
   -E_{1232}+E_{1
   331}-E_{2123}+E_{22
   22}-E_{2321}+E_{311
   3}-E_{3212}+E_{3311
   }\right)\hspace{15cm}
\eea

\vspace{-0.6cm}

\bea\nonumber
\Lambda_3 &=& \frac{1}{2}
   \left(E_{1122}
   +E_{1133}-E_{1
   221}-E_{1
   331}-E_{21
   12}+E_{22
   11}+E_{223
   3}-E_{233
   2}\right.\\\nonumber
   &~&~~~~~- \left. E_{3113
   }-E_{3223}+E_{
   3311}+E_{3322}\right)
\hspace{15cm}
\eea

\vspace{-0.6cm}

\bea\nonumber\Lambda_{4} &=& E_{5555}+\frac{1}{2}
   E_{5566}+\frac{1}{2}
   E_{5577}+\frac{1}{4}
   E_{5588}+\frac{1}{2}
   E_{5665}+\frac{1}{4}
   E_{5687}+\frac{1}{2}
   E_{5775}+\frac{1}{4}
   E_{5786} \\\nonumber
   &+&\frac{1}{4}
   E_{5885}+\frac{1}{2}
   E_{6556}+\frac{1}{4}
   E_{6578}+\frac{1}{2}
   E_{6655}+E_{6666}+\frac{1}{4} E_{6677}+\frac{1}{2}
   E_{6688}+\frac{1}{4}
   E_{6776} \\\nonumber
   &+&\frac{1}{4}
   E_{6875}+\frac{1}{2}
   E_{6886}+\frac{1}{2}
   E_{7557}+\frac{1}{4}
   E_{7568}+\frac{1}{4}
   E_{7667}+\frac{1}{2}
   E_{7755}+\frac{1}{4}
   E_{7766}+E_{7777} \\\nonumber
   &+&\frac{1}{2} E_{7788}+\frac{1}{4}
   E_{7865}+\frac{1}{2}
   E_{7887}+\frac{1}{4}
   E_{8558}+\frac{1}{4}
   E_{8657}+\frac{1}{2}
   E_{8668}+\frac{1}{4}
   E_{8756} \\\nonumber
   &+&\frac{1}{2}
   E_{8778}+\frac{1}{4}
   E_{8855}+\frac{1}{2}
   E_{8866}+\frac{1}{2}
   E_{8877}+E_{8888}
\hspace{15cm}
\eea

\vspace{-0.6cm}

\bea\nonumber
\Lambda_{5} &=& \frac{1}{4} \left(2
   E_{5566}+E_{5588}-2
   E_{5665}-E_{5687}+E
   _{5786}-E_{5885}-2
   E_{6556}-E_{6578}+2
   E_{6655}\right.\\\nonumber
   &+& \left.E_{6677}-E
   _{6776}+E_{6875}+E_
   {7568}-E_{7667}+E_{
   7766}+2
   E_{7788}-E_{7865}-2
   E_{7887}\right.\\\nonumber
   &-& \left.E_{8558}+E
   _{8657}-E_{8756}-2
   E_{8778}+E_{8855}+2
   E_{8877}\right)
\hspace{15cm}
\eea

\vspace{-0.6cm}

\bea\nonumber
\Lambda_{6} &=&\frac{1}{4} \left(2
   E_{5577}+E_{5588}+E
   _{5687}-2
   E_{5775}-E_{5786}-E
   _{5885}+E_{6578}+E_
   {6677}+2
   E_{6688}\right.\\\nonumber
   &-& \left.E_{6776}-E
   _{6875}-2
   E_{6886}-2
   E_{7557}-E_{7568}-E
   _{7667}+2
   E_{7755}+E_{7766}+E
   _{7865}\right.\\\nonumber
   &-& \left.E_{8558}-E_
   {8657}-2
   E_{8668}+E_{8756}+E
   _{8855}+2
   E_{8866}\right)\hspace{15cm}
\eea

\vspace{-0.6cm}

\bea\nonumber
\Lambda_{7} &=& \frac{1}{4}
   \left(E_{5588}-E_{5687}-E_{5786}+E_{5885}-E_{6578}+E_{667
   7}+E_{6776}-E_{6875
   }-E_{7568}\right.\\\nonumber
   &+& \left.E_{7667}
   +E_{7766}-E_{7865}+
   E_{8558}-E_{8657}-E
   _{8756}+E_{8855}\right)
\hspace{15cm}
\eea

\vspace{-0.6cm}

\bea\nonumber
\Lambda_{8} &=& E_{4444}
\hspace{15cm}
\eea

\vspace{-0.6cm}

\bea\nonumber
\Lambda_{9} &=& E_{1155}+E_{1166}+\frac{1}{3}
E_{1177}+\frac{1}{3}
   E_{1188}+\frac{ \sqrt{2}}{3}
   E_{1275}+\frac{ \sqrt{2}}{3}
   E_{1286}+\frac{ \sqrt{2}}{3}
   E_{2157}+\frac{ \sqrt{2}}{3}
   E_{2168}\\\nonumber
   &+&\frac{2}{3}
   E_{2255}+\frac{2}{3}
   E_{2266}+\frac{2}{3}
   E_{2277}+\frac{2}{3}
   E_{2288}+\frac{ \sqrt{2}}{3}
   E_{2375}+\frac{ \sqrt{2}}{3}
      E_{2386}+\frac{ \sqrt{2}}{3}
   E_{3257}\\\nonumber
   &+&\frac{ \sqrt{2}}{3}
   E_{3268}+\frac{1}{3}
   E_{3355}+\frac{1}{3}
   E_{3366}+E_{3377}+E
   _{3388}\hspace{15cm}
\eea

\vspace{-0.6cm}

\bea\nonumber
\Lambda_{10} &=& E_{5511}+\frac{2}{3}
   E_{5522}+\frac{1}{3}
   E_{5533}+\frac{1}{3} \sqrt{2}
   E_{5721}+\frac{1}{3} \sqrt{2}
   E_{5732}+E_{6611}+\frac{2}{3} E_{6622}
   \\\nonumber
   &+&\frac{1}{3}
   E_{6633}+\frac{1}{3} \sqrt{2}
   E_{6821}+\frac{1}{3} \sqrt{2}
   E_{6832}+\frac{1}{3} \sqrt{2}
   E_{7512}+\frac{1}{3} \sqrt{2}
   E_{7523}+\frac{1}{3}
   E_{7711}\\\nonumber
   &+&\frac{2}{3}
   E_{7722}+E_{7733}+\frac{1}{3} \sqrt{2} E_{8612}+\frac{1}{3}
   \sqrt{2} E_{8623}+\frac{1}{3}
   E_{8811}+\frac{2}{3}
   E_{8822}+E_{8833}
\hspace{15cm}
\eea

\vspace{-0.6cm}

\bea\nonumber
\Lambda_{11} &=& \frac{1}{3} \left(2 E_{1177}+2
   E_{1188}-\sqrt{2}
   E_{1275}-\sqrt{2}
   E_{1286}-\sqrt{2}
   E_{2157}-\sqrt{2}
   E_{2168}+E_{2255}+E
   _{2266}\right.\\\nonumber
   &+& \left. E_{2277}+E_
   {2288}-\sqrt{2}
   E_{2375}-\sqrt{2}
   E_{2386}-\sqrt{2}
   E_{3257}-\sqrt{2}
   E_{3268}+2
   E_{3355}+2
   E_{3366}\right)
\hspace{15cm}
\eea

\vspace{-0.6cm}

\bea\nonumber \Lambda_{12} &=& \frac{1}{3} \left(E_{5522}+2
   E_{5533}-\sqrt{2}
   E_{5721}-\sqrt{2}
   E_{5732}+E_{6622}+2
   E_{6633}-\sqrt{2}
   E_{6821}-\sqrt{2}
   E_{6832}\right.\\\nonumber
   &-& \left. \sqrt{2}
   E_{7512}-\sqrt{2}
   E_{7523}+2
   E_{7711}+E_{7722}-\sqrt{2} E_{8612}-\sqrt{2}
   E_{8623}+2
   E_{8811}+E_{8822}\right)
\hspace{15cm}
\eea

\vspace{-0.6cm}

\bea\nonumber
\Lambda_{13} &=& E_{1144}+E_{2244}+E_{3
   344}
\hspace{15cm}
\eea

\vspace{-0.6cm}

\bea\nonumber
\Lambda_{14} &=& E_{4411}+E_{4422}+E_{4
   433}
\hspace{15cm}
\eea

\vspace{-0.6cm}

\bea\nonumber
\Lambda_{15} &=& E_{4455}+E_{4466}+E_{4
   477}+E_{4488}
\hspace{15cm}
\eea

\vspace{-0.6cm}

\bea\nonumber
\Lambda_{16} &=& E_{5544}+E_{6644}+E_{7
   744}+E_{8844}
\hspace{15cm}
\eea

\vspace{-0.6cm}

\bea\nonumber
\Lambda_{17} &=& E_{1434}-E_{2424}+E_{3414}
\hspace{15cm}
\eea

\vspace{-0.6cm}

\bea\nonumber
\Lambda_{18}&=& E_{4143}-E_{4242}+E_{4341}
\hspace{15cm}
\eea

\vspace{-0.6cm}

\bea\nonumber
\Lambda_{19} &=&
E_{1538}-E_{1637}-E_{1736}+E_{1835}-E_{2528}+E_{262
   7}+E_{2726}-E_{2825
   }
\\\nonumber
   &+&
E_{3518}-E_{3617}
   -E_{3716}+E_{3815}
\hspace{15cm}
\eea

\vspace{-0.6cm}

\bea\nonumber
\Lambda_{20} &=&
E_{5183}-E_{5282}+E_{5381}-E_{6173}+E_{6272}-E_{637
   1}-E_{7163}+ E_{7262
   } \\\nonumber
   &-&E_{7361}+E_{8153}
   -E_{8252}+E_{8351}
\hspace{15cm}
\eea

\vspace{-0.6cm}

\bea\nonumber
\Lambda_{21} &=& E_{4548}-E_{4647}-E_{4746}+E_{4845}
\hspace{15cm}
\eea

\vspace{-0.6cm}

\bea\nonumber
\Lambda_{22} &=& E_{5484}-E_{6474}-E_{7464}+E_{8454}
\hspace{15cm}
\eea

\vspace{-0.6cm}

\bea\nonumber
\Lambda_{23} &=& \sqrt{2}
   E_{1526}+E_{1538}-\sqrt{2}
   E_{1625}-E_{1637}+E
   _{1736}-E_{1835}-\sqrt{2} E_{2516}+\sqrt{2}
   E_{2615}\\\nonumber
   &+&\sqrt{2}
   E_{2738}-\sqrt{2}
   E_{2837}-E_{3518}+E
   _{3617}-E_{3716}-\sqrt{2}
   E_{3728}+E_{3815}+\sqrt{2} E_{3827}
\hspace{15cm}
\eea

\vspace{-0.6cm}

\bea\nonumber
\Lambda_{24} &=& \sqrt{2} E_{5162}+E_{5183}-\sqrt{2} E_{5261}-E_{5381}-\sqrt{2}
   E_{6152}-E_{6173}+\sqrt{2}
   E_{6251}+E_{6371}\\\nonumber&+&E_{7163}+\sqrt{2}
   E_{7283}-E_{7361}-\sqrt{2} E_{7382}-E_{8153}-\sqrt{2}
   E_{8273}+E_{8351}+\sqrt{2} E_{8372}
\hspace{15cm}
\eea

\vspace{-0.6cm}

\bea\nonumber
\Lambda_{25} &=& E_{1441}+E_{2442}+E_{3
   443}
\hspace{15cm}
\eea

\vspace{-0.6cm}

\bea\nonumber
\Lambda_{26} &=& E_{4114}+E_{4224}+E_{4
   334}
\hspace{15cm}
\eea

\vspace{-0.6cm}

\bea\nonumber
\Lambda_{27} &=& \frac{1}{3} \left(\sqrt{2} E_{1572}+\sqrt{2}
   E_{1682}-2
   E_{1771}-2
   E_{1881}-E_{2552}+\sqrt{2}
   E_{2573}-E_{2662}+\sqrt{2} E_{2683}\right.\\\nonumber
   &+& \left.\sqrt{2}
   E_{2751}-E_{2772}+\sqrt{2}
   E_{2861}-E_{2882}-2
   E_{3553}-2
   E_{3663}+\sqrt{2}
   E_{3752}+\sqrt{2}
   E_{3862}\right)
\hspace{15cm}
\eea

\vspace{-0.6cm}

\bea\nonumber
\Lambda_{28} &=& \frac{1}{3} \left(\sqrt{2}
   E_{5127}-E_{5225}+\sqrt{2} E_{5237}-2
   E_{5335}+\sqrt{2}
   E_{6128}-E_{6226}+\sqrt{2} E_{6238}-2
   E_{6336}\right.\\\nonumber
   &-& \left.2
   E_{7117}+\sqrt{2}
   E_{7215}-E_{7227}+\sqrt{2} E_{7325}-2
   E_{8118}+\sqrt{2}
   E_{8216}-E_{8228}+\sqrt{2} E_{8326}\right)
\hspace{15cm}
\eea

\vspace{-0.6cm}

\bea\nonumber
\Lambda_{29} &=& -E_{5445}-E_{6446}-E_{
   7447}-E_{8448}
\hspace{15cm}
\eea

\vspace{-0.6cm}

\bea\nonumber
\Lambda_{30} &=& -E_{4554}-E_{4664}-E_{
   4774}-E_{4884}
\hspace{15cm}
\eea

\vspace{-0.6cm}

\bea\nonumber
\Lambda_{31} &=& E_{1551}+\frac{ \sqrt{2}}{3}
   E_{1572}+E_{1661}+\frac{ \sqrt{2}}{3}  E_{1682}+\frac{1}{3}
   E_{1771}+\frac{1}{3}
   E_{1881}+\frac{2}{3}
   E_{2552}+\frac{ \sqrt{2}}{3}
   E_{2573}\\\nonumber
   &+&\frac{2}{3}
   E_{2662}+\frac{ \sqrt{2}}{3}
   E_{2683}+\frac{ \sqrt{2}}{3}
   E_{2751}+\frac{2}{3}
   E_{2772}+\frac{ \sqrt{2}}{3}
   E_{2861}+\frac{2}{3}
   E_{2882}+\frac{1}{3}
   E_{3553}\\\nonumber
   &+&\frac{1}{3}
   E_{3663}+\frac{ \sqrt{2}}{3}
   E_{3752}+E_{3773}+\frac{ \sqrt{2}}{3}
   E_{3862}+E_{3883}
\hspace{15cm}
\eea

\vspace{-0.6cm}

\bea\nonumber
\Lambda_{32} &=& E_{5115}+\frac{ \sqrt{2}}{3}
   E_{5127}+\frac{2}{3}
   E_{5225}+\frac{ \sqrt{2}}{3}
   E_{5237}+\frac{1}{3}
   E_{5335}+E_{6116}+\frac{ \sqrt{2}}{3} E_{6128}+\frac{2}{3}
   E_{6226}\\\nonumber
   &+&\frac{ \sqrt{2}}{3}
   E_{6238}+\frac{1}{3}
   E_{6336}+\frac{1}{3}
   E_{7117}+\frac{ \sqrt{2}}{3}
   E_{7215}+\frac{2}{3}
   E_{7227}+\frac{ \sqrt{2}}{3}
   E_{7325}+E_{7337}+\frac{1}{3} E_{8118}\\\nonumber
   &+&\frac{ \sqrt{2}}{3}
   E_{8216}+\frac{2}{3}
   E_{8228}+\frac{ \sqrt{2}}{3}
   E_{8326}+E_{8338}\hspace{15cm}
\eea

\vspace{-0.6cm}

\bea\nonumber
\Lambda_{33} &=& \sqrt{2} E_{1546}-\sqrt{2}
   E_{1645}+E_{2548}-E
   _{2647}+E_{2746}-E_
   {2845}+\sqrt{2}
   E_{3748}-\sqrt{2}
   E_{3847}
\hspace{15cm}
\eea

\vspace{-0.6cm}

\bea\nonumber
\Lambda_{34} &=& \sqrt{2}
   E_{5164}+E_{5284}-\sqrt{2}
   E_{6154}-E_{6274}+E
   _{7264}+\sqrt{2}
   E_{7384}-E_{8254}-\sqrt{2} E_{8374}
\hspace{15cm}
\eea

\vspace{-0.6cm}

\bea\nonumber
\Lambda_{35} &=& \sqrt{2}
   E_{4516}+E_{4528}-\sqrt{2}
   E_{4615}-E_{4627}+E
   _{4726}+\sqrt{2}
   E_{4738}-E_{4825}-\sqrt{2} E_{4837}
\hspace{15cm}
\eea

\vspace{-0.6cm}

\bea\nonumber
\Lambda_{36} &=& \sqrt{2}
   E_{5461}+E_{5482}-\sqrt{2}
   E_{6451}-E_{6472}+E
   _{7462}+\sqrt{2}
   E_{7483}-E_{8452}-\sqrt{2} E_{8473}
\hspace{15cm}
\eea

\vspace{-0.6cm}

\bea\nonumber
\Lambda_{37} &=&
E_{1124}+E_{1234}-E_{2114}+E_{2334}-E_{3214}-E_{332
   4}
\hspace{15cm}
\eea

\vspace{-0.6cm}

\bea\nonumber
\Lambda_{38} &=&
E_{1142}-E_{1241}+E_{2143}-E_{2341}+E_{3243}-E_{334
   2}
\hspace{15cm}
\eea

\vspace{-0.6cm}

\bea\nonumber
\Lambda_{39} &=&
E_{1421}+E_{1432}-E_{2411}+E_{2433}-E_{3412}-E_{342
   3}
\hspace{15cm}
\eea

\vspace{-0.6cm}

\bea\nonumber
\Lambda_{40} &=&
E_{4112}+E_{4123}-E_{4211}+E_{4233}-E_{4321}-E_{433
   2}
\hspace{15cm}
\eea

\vspace{-0.6cm}

\bea\nonumber
\Lambda_{41} &=& -\sqrt{2} E_{1574}-\sqrt{2}
   E_{1684}+E_{2554}+E
   _{2664}-E_{2774}-E_
   {2884}+\sqrt{2}
   E_{3754}+\sqrt{2}
   E_{3864}
\hspace{15cm}
\eea

\vspace{-0.6cm}

\bea\nonumber
\Lambda_{42} &=& -\sqrt{2}
   E_{5147}+E_{5245}-\sqrt{2}
   E_{6148}+E_{6246}-E
   _{7247}+\sqrt{2}
   E_{7345}-E_{8248}+\sqrt{2} E_{8346}
\hspace{15cm}
\eea

\vspace{-0.6cm}

\bea\nonumber
\Lambda_{43} &=& -E_{5425}-\sqrt{2}
   E_{5437}-E_{6426}-\sqrt{2} E_{6438}+\sqrt{2}
   E_{7415}+E_{7427}+\sqrt{2}
   E_{8416}+E_{8428}
\hspace{15cm}
\eea

\vspace{-0.6cm}

\bea\nonumber
\Lambda_{44} &=& -E_{4552}-\sqrt{2}
   E_{4573}-E_{4662}-\sqrt{2} E_{4683}+\sqrt{2}
   E_{4751}+E_{4772}+\sqrt{2}
   E_{4861}+E_{4882}
\hspace{15cm}
\eea

\vspace{-0.6cm}

\bea\nonumber
\Lambda_{45} &=& -\sqrt{2} E_{1475}-\sqrt{2}
   E_{1486}+E_{2455}+E
   _{2466}-E_{2477}-E_
   {2488}+\sqrt{2}
   E_{3457}+\sqrt{2}
   E_{3468}
\hspace{15cm}
\eea

\vspace{-0.6cm}

\bea\nonumber
\Lambda_{46} &=& -\sqrt{2} E_{4157}-\sqrt{2}
   E_{4168}+E_{4255}+E
   _{4266}-E_{4277}-E_
   {4288}+\sqrt{2}
   E_{4375}+\sqrt{2}
   E_{4386}
\hspace{15cm}
\eea

\vspace{-0.6cm}

\bea\nonumber
\Lambda_{47} &=& -E_{5524}-\sqrt{2}
   E_{5734}-E_{6624}-\sqrt{2} E_{6834}+\sqrt{2}
   E_{7514}+E_{7724}+\sqrt{2}
   E_{8614}+E_{8824}
\hspace{15cm}
\eea

\vspace{-0.6cm}

\bea\nonumber
\Lambda_{48} &=& -E_{5542}+\sqrt{2}
   E_{5741}-E_{6642}+\sqrt{2} E_{6841}-\sqrt{2}
   E_{7543}+E_{7742}-\sqrt{2}
   E_{8643}+E_{8842}
\hspace{15cm}
\eea




\begin{thebibliography}{20}

\bibitem{MZ}
  J.~A.~Minahan and K.~Zarembo,
  ``The Bethe-ansatz for N = 4 super Yang-Mills,''
  JHEP {\bf 0303} (2003) 013, hep-th/0212208.

\bibitem{Serban:2004jf}
  D.~Serban and M.~Staudacher,
  ``Planar N = 4 gauge theory and the Inozemtsev long range spin chain,''
  JHEP {\bf 0406} (2004) 001, hep-th/0401057.

 \bibitem{BDS}
  N.~Beisert, V.~Dippel and M.~Staudacher,
  ``A novel long range spin chain and planar N = 4 super Yang-Mills,''
  JHEP {\bf 0407} (2004) 075, hep-th/0405001.

\bibitem{AFS}
  G.~Arutyunov, S.~Frolov and M.~Staudacher,
   ``Bethe ansatz for quantum strings,''
  JHEP {\bf 0410}, 016 (2004), hep-th/0406256.


\bibitem{S}
  M.~Staudacher,
  ``The factorized S-matrix of CFT/AdS,''
  JHEP {\bf 0505} (2005) 054, hep-th/0412188.


\bibitem{BS}
  N.~Beisert and M.~Staudacher,
  ``Long-range PSU(2,2$|$4) Bethe ansaetze for gauge theory and strings,''
  Nucl.\ Phys.\  B {\bf 727} (2005) 1, hep-th/0504190.



\bibitem{Gromov:2006cq}
  N.~Gromov and V.~Kazakov,
  ``Asymptotic Bethe ansatz from string sigma model on $S^3 \times R$,''
  Nucl.\ Phys.\  B {\bf 780} (2007) 143, hep-th/0605026.


\bibitem{B}
  N.~Beisert,
  ``The $\su(2|2)$ dynamic S-matrix,''
  hep-th/0511082.


\bibitem{MT}
  R.~R.~Metsaev and A.~A.~Tseytlin,
  ``Type IIB superstring action in AdS(5) x S(5) background,''
  Nucl.\ Phys.\  B {\bf 533} (1998) 109, hep-th/9805028.

\bibitem{AF04}
  G.~Arutyunov and S.~Frolov,
  ``Integrable Hamiltonian for classical strings on $\AdS$,''
  JHEP {\bf 0502} (2005) 059, hep-th/0411089.

\bibitem{FPZ}
  S.~Frolov, J.~Plefka and M.~Zamaklar,
  ``The $\AdS$ superstring in light-cone gauge and its Bethe
  equations,''
  J.\ Phys.\ A  {\bf 39} (2006) 13037, hep-th/0603008.



\bibitem{AFZ}
  G.~Arutyunov, S.~Frolov and M.~Zamaklar,
  ``The Zamolodchikov-Faddeev algebra for $\AdS$ superstring,''
  JHEP {\bf 0704} (2007) 002, hep-th/0612229.



\bibitem{Kazakov:2004qf}
  V.~A.~Kazakov, A.~Marshakov, J.~A.~Minahan and K.~Zarembo,
  ``Classical / quantum integrability in AdS/CFT,''
  JHEP {\bf 0405} (2004) 024, hep-th/0402207.

\bibitem{Janik}
  R.~A.~Janik,
  ``The $\AdS$ superstring worldsheet S-matrix and crossing symmetry,''
  Phys.\ Rev.\ D {\bf 73} (2006) 086006,
  hep-th/0603038.

\bibitem{BHL}
  N.~Beisert, R.~Hernandez and E.~Lopez,
  ``A crossing-symmetric phase for $\ads$ strings,''
  JHEP {\bf 0611} (2006) 070, hep-th/0609044.

\bibitem{BES}
  N.~Beisert, B.~Eden and M.~Staudacher,
  ``Transcendentality and crossing,''
  J.\ Stat.\ Mech.\  {\bf 0701} (2007) P021, hep-th/0610251.


\bibitem{AFPZ}
  G.~Arutyunov, S.~Frolov, J.~Plefka and M.~Zamaklar,
  ``The off-shell symmetry algebra of the light-cone $\ads$
  superstring,''
  J.\ Phys.\ A  {\bf 40} (2007) 3583, hep-th/0609157.




\bibitem{D}
  N.~Dorey,
  ``Magnon bound states and the AdS/CFT correspondence,''
  J.\ Phys.\ A  {\bf 39} (2006) 13119, hep-th/0604175.


  \bibitem{Bn}
  N.~Beisert,
  ``The Analytic Bethe Ansatz for a Chain with Centrally Extended
  $\su(2|2)$
  Symmetry,''
  J.\ Stat.\ Mech.\  {\bf 0701} (2007) P017, nlin.si/0610017.



  \bibitem{CDO}
  H.~Y.~Chen, N.~Dorey and K.~Okamura,
  ``The asymptotic spectrum of the N = 4 super Yang-Mills spin chain,''
  JHEP {\bf 0703} (2007) 005, hep-th/0610295.




\bibitem{AFTBA}
  G.~Arutyunov and S.~Frolov,
  ``On String S-matrix, Bound States and TBA,''
  JHEP {\bf 0712} (2007) 024
  [arXiv:0710.1568 [hep-th]].

\bibitem{Kulish:1981gi}
  P.~P.~Kulish, N.~Y.~Reshetikhin and E.~K.~Sklyanin,
  ``Yang-Baxter Equation And Representation Theory. 1,''
  Lett.\ Math.\ Phys.\  {\bf 5} (1981) 393.

\bibitem{Kazakov:2007fy}
  V.~Kazakov, A.~Sorin and A.~Zabrodin,
  ``Supersymmetric Bethe ansatz and Baxter equations from discrete Hirota
  dynamics,''
  Nucl.\ Phys.\  B {\bf 790} (2008) 345, hep-th/0703147.

\bibitem{Kazakov:2007na}
  V.~Kazakov and P.~Vieira,
  ``From Characters to Quantum (Super)Spin Chains via Fusion,''
  arXiv:0711.2470 [hep-th].



\bibitem{MM}
  M.~J.~Martins and C.~S.~Melo,
  ``The Bethe ansatz approach for factorizable centrally extended S-matrices,''
  Nucl.\ Phys.\  B {\bf 785} (2007) 246,
  hep-th/0703086.

\bibitem{Le}
  M.~de Leeuw,
  ``Coordinate Bethe Ansatz for the String S-Matrix,''
  J.\ Phys.\ A  {\bf 40} (2007) 14413
  [arXiv:0705.2369 [hep-th]].




\bibitem{Belitsky:2008wj}
  A.~V.~Belitsky,
  ``Fusion hierarchies for N = 4 superYang-Mills theory,''
  arXiv:0803.2035 [hep-th].


\bibitem{Beisert:2007ds}
  N.~Beisert,
  ``The S-Matrix of AdS/CFT and Yangian Symmetry,''
  PoS {\bf SOLVAY} (2006) 002
  [arXiv:0704.0400 [nlin.SI]].

\bibitem{Matsumoto:2007rh}
  T.~Matsumoto, S.~Moriyama and A.~Torrielli,
  ``A Secret Symmetry of the AdS/CFT S-matrix,''
  JHEP {\bf 0709}, 099 (2007)
  [arXiv:0708.1285 [hep-th]].

\bibitem{Matsumoto:2008ww}
  T.~Matsumoto and S.~Moriyama,
  ``An Exceptional Algebraic Origin of the AdS/CFT Yangian Symmetry,''
  arXiv:0803.1212 [hep-th].

\bibitem{Spill:2008tp}
  F.~Spill and A.~Torrielli,
  ``On Drinfeld's second realization of the AdS/CFT $su(2|2)$ Yangian,''
  arXiv:0803.3194 [hep-th].




\bibitem{sualg}
  A.~Torrielli,
  ``Classical r-matrix of the $\su(2|2)$
 SYM spin-chain,''
  Phys.\ Rev.\  D {\bf 75} (2007) 105020,
  hep-th/0701281.

\bibitem{Moriyama:2007jt}
  S.~Moriyama and A.~Torrielli,
  ``A Yangian Double for the AdS/CFT Classical r-matrix,''
  JHEP {\bf 0706} (2007) 083
  [arXiv:0706.0884 [hep-th]].

\bibitem{Beisert:2007ty}
  N.~Beisert and F.~Spill,
  ``The Classical r-matrix of AdS/CFT and its Lie Bialgebra Structure,''
  arXiv:0708.1762 [hep-th].

\bibitem{Leeuw}
  M.~de Leeuw,
  to appear.




\bibitem{magnon}
  G.~Arutyunov, S.~Frolov and M.~Zamaklar,
  ``Finite-size effects from giant magnons,''
  Nucl.\ Phys.\  B {\bf 778} (2007) 1,
  hep-th/0606126.


\bibitem{Klose:2008rx}
  T.~Klose and T.~McLoughlin,
  ``Interacting finite-size magnons,''
  arXiv:0803.2324 [hep-th].

\bibitem{Minahan:2008re}
  J.~A.~Minahan and O.~Ohlsson Sax,
  ``Finite size effects for giant magnons on physical strings,''
  arXiv:0801.2064 [hep-th].

\bibitem{HM}
  D.~M.~Hofman and J.~M.~Maldacena,
  ``Giant magnons,''
  J.\ Phys.\ A  {\bf 39} (2006) 13095, hep-th/0604135.


\bibitem{Luscher}
  M.~L\"uscher,
  ``Volume Dependence Of The Energy Spectrum In Massive Quantum Field Theories.
  1. Stable Particle States,''
  Commun.\ Math.\ Phys.\  {\bf 104} (1986) 177.


\bibitem{AJK}
  J.~Ambjorn, R.~A.~Janik and C.~Kristjansen,
  ``Wrapping interactions and a new source of corrections to the spin-chain  /
  string duality,''
  Nucl.\ Phys.\  B {\bf 736} (2006) 288, hep-th/0510171.


\bibitem{Janik:2007wt}
  R.~A.~Janik and T.~Lukowski,
  ``Wrapping interactions at strong coupling -- the giant magnon,''
  Phys.\ Rev.\  D {\bf 76} (2007) 126008
  [arXiv:0708.2208 [hep-th]].

\bibitem{Hatsuda:2008gd}
  Y.~Hatsuda and R.~Suzuki,
  ``Finite-Size Effects for Dyonic Giant Magnons,''
  arXiv:0801.0747 [hep-th].


\bibitem{Gromov:2008ie}
  N.~Gromov, S.~Schafer-Nameki and P.~Vieira,
  ``Quantum Wrapped Giant Magnon,''
  arXiv:0801.3671 [hep-th].

\bibitem{Heller:2008at}
  M.~P.~Heller, R.~A.~Janik and T.~Lukowski,
  ``A new derivation of Luscher F-term and fluctuations around the giant
  magnon,''
  arXiv:0801.4463 [hep-th].


\bibitem{M}
  J.~M.~Maldacena,
  ``The large N limit of superconformal field theories and supergravity,''
  Adv.\ Theor.\ Math.\ Phys.\  {\bf 2} (1998) 231
  [Int.\ J.\ Theor.\ Phys.\  {\bf 38} (1999) 1113], hep-th/9711200.



 \bibitem{Shastry} B.~S.~Shastry, ``Exact integrability of the
 one-dimensional Hubburd-model", Phys.Rev.Lett {\rm 56} (1986)
 2453.



\bibitem{Beisert:2008tw}
  N.~Beisert and P.~Koroteev,
  ``Quantum Deformations of the One-Dimensional Hubbard Model,''
  arXiv:0802.0777 [hep-th].


 \bibitem{KMRZ}
  T.~Klose, T.~McLoughlin, R.~Roiban and K.~Zarembo,
  ``Worldsheet scattering in $\ads$,''
  JHEP {\bf 0703} (2007) 094
  [arXiv:hep-th/0611169].


\bibitem{ArSok}
  G.~Arutyunov and E.~Sokatchev,
  ``Conformal fields in the pp-wave limit,''
  JHEP {\bf 0208} (2002) 014, hep-th/0205270.


\bibitem{kor}
  F.~G\"ohmann and V.~E.~Korepin,
  ``Solution of the quantum inverse problem,''
  J.\ Phys.\ A  {\bf 33} (2000) 1199, hep-th/9910253.

\bibitem{Dorey2}
  H.~Y.~Chen, N.~Dorey and K.~Okamura,
  ``On the scattering of magnon boundstates,''
  JHEP {\bf 0611} (2006) 035, hep-th/0608047.

\bibitem{Roiban}
  R.~Roiban,
  ``Magnon bound-state scattering in gauge and string theory,''
  JHEP {\bf 0704} (2007) 048,
  hep-th/0608049.


 \bibitem{AAF}
  L.~F.~Alday, G.~Arutyunov and S.~Frolov,
 ``New integrable system of 2dim fermions from strings on $\AdS$,''
  JHEP {\bf 0601} (2006) 078, hep-th/0508140.


\bibitem{AFlc}
  G.~Arutyunov and S.~Frolov,
  ``Uniform light-cone gauge for strings in $\AdS$: Solving
  $\su(1|1)$
  sector,''
  JHEP {\bf 0601} (2006) 055, hep-th/0510208.


  \bibitem{DHM}
  N.~Dorey, D.~M.~Hofman and J.~Maldacena,
  ``On the singularities of the magnon S-matrix,''
  hep-th/0703104.

\bibitem{CDO2}
  N.~Dorey and K.~Okamura,
  ``Singularities of the Magnon Boundstate S-Matrix,''
  JHEP {\bf 0803} (2008) 037
  [arXiv:0712.4068 [hep-th]].


\bibitem{AF06}
  G.~Arutyunov and S.~Frolov,
  ``On $\ads$ string S-matrix,''
  Phys.\ Lett.\ B {\bf 639} (2006) 378, hep-th/0604043.






\end{thebibliography}
\end{document}